\documentclass[bottom]{usbthesis}
\def\nn{\nonumber\\}
\usepackage{graphicx}
\usepackage{color}
\usepackage{bm}
\usepackage{amsmath}
\usepackage{amssymb}
\usepackage{setspace}
\usepackage{array}
\usepackage{tabularx}
\usepackage{multirow}
\usepackage{afterpage}
\usepackage{hhline}
\usepackage{url}
\usepackage{hyperref} % see the file hyperref.cfg for options
\usepackage{footmisc} % provides \footref and \mpfootnotemark commands
\usepackage{xspace}
\usepackage[numbers,sort&compress]{natbib}

\graphicspath{{vis_hydro/}{qgp_vis/}{phen/}}
\def\simge{%
    \mathrel{\rlap{\raise 0.511ex
    \hbox{$>$}}{\lower 0.511ex \hbox{$\sim$}}}}
\def\simle{%
    \mathrel{\rlap{\raise 0.511ex
    \hbox{$<$}}{\lower 0.511ex \hbox{$\sim$}}}}

\renewcommand{\vec}[1]{\boldsymbol{#1}}

\DeclareMathAccent{\ring}{\mathalpha}{operators}{"17}

\providecommand{\renewoperator}[3]{\renewcommand*{#1}{\mathop{#2}#3}}
\renewoperator{\Re}{\mathrm{Re}}{\nolimits}
\renewoperator{\Im}{\mathrm{Im}}{\nolimits}

\makeatletter
\providecommand*{\diff}{\@ifnextchar^{\DIfF}{\DIfF^{}}}
\def\DIfF^#1{\mathop{\mathrm{\mathstrut d}}\nolimits^{#1}\gobblespace}
\def\gobblespace{\futurelet\diffarg\opspace}
\def\opspace{%
    \let\DiffSpace\!%
    \ifx\diffarg(%
        \let\DiffSpace\relax
    \else
        \ifx\diffarg[%
            \let\DiffSpace\relax
        \else
            \ifx\diffarg\{%
                \let\DiffSpace\relax
            \fi\fi\fi\DiffSpace}

\def\N{{\mathcal N}}

\def\Re{{\rm Re}}
\def\Im{{\rm Im}}

\def\st{\begin{equation}}
\def\stp{\end{equation}}
\def\bg{\begin{eqnarray}}
\def\nd{\end{eqnarray}}
\def\Eq#1{Eq.~(\ref{#1})}

\def\Sect#1{Section~\ref{#1}}

\def\llangle{\left\langle}
\def\rrangle{\right\rangle}

\def\N{\mathcal{N}}

\def\gsim{\mbox{~{\protect\raisebox{0.4ex}{$>$}}\hspace{-1.1em}
	{\protect\raisebox{-0.6ex}{$\sim$}}~}}
\def\lsim{\mbox{~{\protect\raisebox{0.4ex}{$<$}}\hspace{-1.1em}
	{\protect\raisebox{-0.6ex}{$\sim$}}~}}

\def\nott#1{\setbox0=\hbox{$#1$}                % set a box for #1 
   \dimen0=\wd0                                 % and get its size
   \setbox1=\hbox{/} \dimen1=\wd1               % get size of /
   \ifdim\dimen0>\dimen1                        % #1 is bigger
      \rlap{\hbox to \dimen0{\hfil/\hfil}}      % so center / in box
      #1                                        % and print #1
   \else                                        % / is bigger
      \rlap{\hbox to \dimen1{\hfil$#1$\hfil}}   % so center #1
      /                                         % and print /
   \fi}                                         %

\def\baselinestretch{1.025}
\advance\parskip 1.9pt
\advance\voffset -0.2in

\def\st{\begin{equation}}
\def\stp{\end{equation}}
\def\bg{\begin{eqnarray}}
\def\nd{\end{eqnarray}}

\author{Kevin Dusling}%
\title{Hydrodynamic Description of Dilepton~Production}%

\month{August} \year{2008}%

\program{Physics}%

\director{Ismail Zahed}{Professor, Department of
 Physics and Astronomy}%
\chairman{Alexandre Abanov}{Professor, Department of Physics and Astronomy}%
\fstmember{Axel Drees}{Professor, Department of Physics and Astronomy}%
\outmember{Dmitri Kharzeev}{Senior Scientist, Brookhaven National Laboratory}%

\begin{document}

%\singlespacing
\doublespacing
\pagenumbering{roman}

\maketitle
\makeapproval

\begin{abstract}
    The first part of this thesis focuses on the production of thermal dileptons from a hadronic gas at finite temperature.  The rates are calculated by an expansion in Pion density and constrained by broken chiral symmetry and vacuum correlation functions, many of which have been measured by experiment.  We focus on emission processes having two Pions in the final state.  

Next, follows a separate discussion on viscous hydrodynamics and its effect on $p_\perp$ spectra and elliptic flow.  A non-central hydrodynamic model of Au-Au collisions in 2+1 dimensions is simulated.  Off-equilibrium corrections to the distribution can bring about large changes in the differential elliptic flow, especially at higher $p_\perp$.  Also discussed is the shear viscous correction to dilepton production in a quark-gluon plasma (QGP) emanating from q\={q} annihilation in the Born approximation.  It is argued that a thermal description is reliable for invariant masses less than $M_{max}\approx(2\tau_0 T_0^2)/(\eta/s)$. Shear viscosity leads to qualitative differences in dilepton $p_\perp$ spectrum, which could be used to extract information on the thermalization time, viscosity to entropy ratio and possibly the thermalization mechanism in heavy-ion collisions.      

Finally, the dilepton rates used in this work are integrated over the space-time evolution of the collision region and compared to the recent results from the NA60 experiment at CERN and the PHENIX experiment at RHIC.  The role played by chiral symmetry restoration in the hadronic phase and viscosity in the QGP phase is discussed. 

\end{abstract}

\begin{dedication}
    To my family.
\end{dedication}

\tableofcontents %
\listoffigures %
\listoftables %

\begin{acknowledgements}
    \vspace{-27pt}
First and foremost, I must thank Ismail Zahed.  Without his patience and guidance this thesis would not have been possible.  I am also grateful to Derek Teaney who also advised me on many of the topics in this thesis and played a role throughout.  Both Derek and Ismail, trained me thoroughly and played an intricate role in my development as a scientist.  I would also like to thank Gerry Brown and Edward Shuryak for their encouragement and support throughout my PhD. 

I am also grateful to have been a member of the Stony Brook graduate student community.  The high quality of graduate course work at Stony Brook was critical for my scientific development and intellectually enlightening.  I was also fortunate enough to study as an undergraduate at the The Cooper Union where I received the strong foundation necessary for any academic endeavour.  This was only possible because of Peter Cooper's conviction that ``education should be as free as water or air.''  

My fellow graduate classmates have also been a driving force in my studies.  I would like to thank Clint Young and Shu Lin for being backboards in which to bounce ideas from.

Finally, I am forever indebted to my wife Shaughnessy for introducing me to a life outside of Physics.

\end{acknowledgements}

\pagestyle{thesis}

%%%%%%%%%%%%%%%%%%%%%%%%%%%%%%%%%%%%%%%%%%%%%%%%%%%%%%%%%%%%%%%%%%%%%%%%%%%%%%%%

\newpage
\pagenumbering{arabic}

\chapter{Introduction}

\section{Basics of Quantum Chromodynamics}

Quantum chromodynamics (QCD) is believed to be the theory which describes the strong interactions of quarks and gluons which are found in hadrons.  The dynamics of both quarks and gluons are dictated by the QCD Lagrangian \cite{ZahedBook}
\begin{eqnarray}
\mathcal{L}_{QCD}&=&-\frac{1}{4}G^{a}_{\mu\nu}G^{\mu\nu}_a+\frac{\theta}{16\pi^2}\epsilon^{\mu\nu\alpha\beta}G^a_{\mu\nu}G^a_{\alpha\beta}\nn
&+&\sum_{f=1}^{N_f} \overline{q}^i_f\left(iD^{ij}_\mu\gamma^\mu-m_f\delta^{ij}\right)q_f^j
\end{eqnarray}
where we have used the following notation
\begin{eqnarray}
G_{\mu\nu}^a&=&\partial_\mu A^a_\nu-\partial_\nu A^a_\mu+igf^{abc}A^b_\mu A^c_\nu\nn
D_\mu^{ij}&=&\partial_\mu \delta^{ij}+ig\left(A^a_\mu T^a\right)^{ij}
\end{eqnarray}
and $A_\nu^a$ is the gluon field having color index $a$ and $q_f^i$ is a quark field having flavor index $f$ and color index $i$.  

Unfortunately there is no  calculational scheme which works well for all energies.  For large momentum transfers asymptotic freedom states that the coupling constant becomes small, therefore allowing for a perturbative treatment.  However, as the momentum transfer decreases the coupling becomes larger binding the quarks and gluons into hadrons.  This leads to the property of confinement, whereby the force required to separate two quarks increases with their relative distance.  

Since perturbative calculations are only allowable at high energies due to asymptotic freedom non-perturbative methods have been developed in order to gain insight into QCD.  One of the more well established non-perturbative approaches is lattice QCD. 
Only recently have first principal calculations by lattice QCD in the strong coupling regime become available, albeit with limitations.  Since the metric in lattice QCD is Euclidean the calculation is limited to static properties.  For example, it becomes very difficult, if not impossible, to calculate scattering amplitudes or transport coefficients.  

There have also been many non-perturbative methods developed based on effective theories.  In order for these theories to represent nature they should contain the same symmetries of QCD.  QCD with massless quarks and $N_f$ flavors has an exact global flavor symmetry $SU_L(N_f)\times SU_R(N_f)$ called chiral symmetry.  This symmetry is spontaneously broken generating three (for $SU(2)$ flavor) Goldstone bosons called the Pions.  In the real world the quarks are massive and electromagnetism is present, so the flavor symmetry is only approximate, leading to pseudo-Goldstone bosons having a small mass which can be calculated in the framework of chiral perturbation theory.  This explicit breaking gives rise to the partially conserved axial current (PCAC) hypothesis, $\partial^\mu A_\mu^a=f_\pi m_\pi^2 \pi^a$ with $A_\mu^a$ the axial-vector current.

\section{Heavy Ion Collisions}

Chiral symmetry which is spontaneously broken in the QCD vacuum is partially restored at finite temperature and/or density.  In addition, as the temperature is increased from zero, it is thermodynamically favorable for there to be a phase transition from a resonance gas of hadronic bound states to a quark gluon plasma.  One of the goals of the heavy ion collision program is to produce a quark-gluon plasma and study its properties.  It has already been accepted by many in the heavy ion community that a quark gluon plasma (QGP) has been triggered at RHIC consisting of a strongly interacting, low viscosity fluid.

In order to confirm these conclusions and quantify the properties of the QGP, a detailed study of heavy ion phenomenology is required.  One of the main experimental observations that led to the conclusion of the low viscosity nature of the QGP is the large amount of collective flow of the produced particles and its interpretation as coming from a hydrodynamic expansion.  Even though hydrodynamic behavior is able to explain a large amount of the available hadronic data it fails at a number of places, such as at high transverse momentum ($p_T$) and at forward rapidity.  It is believed that the deviations from ideal hydrodynamic behavior could be explained by dissipative effects.

In order to quantify these assertions, viscous relativistic hydrodynamic simulations have to be developed.  The first order Navier-Stokes theory is plagued with difficulties ({\em e.g.} the parabolic nature of the equations permit acausal signal propagation).  In order to correct for this unsatisfactory behavior a number of second order theories have been developed.  At this point in time there is still not a consensus in the heavy-ion community on which theory is appropriate to use.  A full study of viscous hydrodynamics is not only imperative for making quantitative predictions on the properties of the matter produced at RHIC but also helps in our theoretical understanding of kinetic theory results.  The importance of having a theoretical understanding of viscous relativistic hydrodynamics is not limited to heavy-ion collisions but is also necessary, for example, in cosmological simulations of the early universe.  

A second interesting phenomenological tool to study heavy ion collisions is electromagnetic probes.  In contrast to hadronic observables which interact strongly throughout the entire evolution of the heavy ion collision, electromagnetic probes leave the medium without further interaction and therefore carry direct information on the time evolution of the system \cite{457}.  This is in contrast to hadronic observables which thermalize after the collision and thus provide information only on the late stages of the evolution. 

In theory, the electromagnetic spectral function of the quark-gluon plasma could be extracted from thermal photon and dilepton emission, which would in turn permit one to learn about its properties ({\em e.g.} transport coefficients, presence of bound states, etc.)  In practice, however, this is not possible since the QGP dilepton yields are quenched by hadronic emission.  Therefore, in order to probe the QGP phase, there must be a solid theoretical understanding of the hadronic emission processes.

There is a long history of experimental dilepton measurements \cite{ShuryakBook} which we don't attempt to summarize here.  In regard to dilepton measurements from heavy-ion collisions there were three experiments prior to the recent results from NA60 and PHENIX.  These three past experiments were the NA45, HELIOS-3 and NA38/50, which were all performed at the CERN SPS collider, and focused respectively on low, intermediate and high mass dileptons.  All three experiments found an enhancement in the dilepton yields above expected hadronic sources (which is comprised of a {\em cocktail} designed to describe the measured dilepton spectra in p-p and p-Be collisions).  The quality of data however was limited.  Dilepton measurements in general are much more difficult then measuring hadronic observables.  Not only is there a large background which must be {\em rejected} but the cross sections involved are also relatively small.  These two facts together require high luminosity experiments in order to collect precision data.  

Let us discuss the low mass enhancement found at NA45.  For central S-Au and Pb-Au collisions NA45 found an enhancement by as much as a factor of 3-5 above known sources in the emission of di-electrons with invariant masses $0.2<M<0.6$ GeV.  A number of theoretical explanations were given for this phenomenon including {\em melting} of the $\rho$ due to chiral symmetry restoration \cite{BR,RW,paper1,paper2,paper3}.  The statistics were unfortunately too poor to confirm these predictions.  The upgrade from NA50 to NA60 consisted of a new vertex tracker, which now allows track matching in both coordinate and momentum space.  This leads to a considerable improvement in statistics and should allow one to discern between different theoretical approaches.  

Recently, the PHENIX experiment has also measured di-electron invariant mass spectra and found an enhancement by a factor as large as 7-8 above the {\em cocktail} for the most central collisions.  At first glance it appears that this result may be inconsistent with the measurements by NA60.  However, one must remember that the resultant yields must first be folded through the complicated detector acceptance which is specific to either PHENIX or NA60.

In order to support these interesting experimental programs it is necessary to generate realistic dilepton predictions.  In this direction we use a comprehensive set of rates for dilepton production taking into account the symmetries of QCD ({\em e.g.} broken chiral symmetry) at finite temperature and density integrated over the space-time history of relativistic viscous hydrodynamic simulations of the collision.  

\section{Outline of this thesis}

This thesis is separated into a number of self contained parts.  However, the last section on heavy ion phenomenology will rely on all the material presented throughout.  

The first part of this work focuses on the work done with my advisor, Ismail Zahed. In chapter two, we discuss the dilepton emission rates from a hadronic gas in thermal and chemical equilibrium.  The rates take into account broken chiral symmetry in a consistent manner and rely on experimental data as input.  The rates are treated in a density expansion and the effects of one and two pions in the final state are explored.  The new work consisted of evaluating the rates to second order in pion density, which include all hadronic processes involving two pions in the final state.  This work is currently unpublished. 

Chapter three is a separate discussion on viscous hydrodynamic simulations.  The work was done under the auspices of Derek Teaney and was published in \cite{Dusling:2007gi}.  We examine how shear viscosity changes the ideal hydrodynamic evolution and the effect it has on differential transverse momentum and elliptic flow spectra.  

Chapter four goes back to dilepton production, this time from a quark gluon plasma out of kinetic equilibrium.  In collaboration with Shu Lin \cite{Dusling:2008xj} we consider how shear viscosity modifies the leading order born $q\overline{q}$ dilepton production rates. 

In chapter five all the pieces are put together.  The dilepton rates are integrated over the space time evolution presented in chapter three.  Most of the work is done in kinetic equilibrium and was published in \cite{Dusling:2006yv,Dusling:2007kh,Dusling:2007su} and was done in collaboration with Ismail Zahed and with help from Derek Teaney regarding the hydrodynamic evolution and equation of state.  A final section in chapter five discusses the role of shear viscosity on dilepton emission from both the QGP and hadronic phases.

\chapter{\textquotedblleft Master Equation\textquotedblright Approach to Dilepton Production}

\section{Introduction}

It can be shown \cite{Bellac} that to lowest order in electromagnetic interactions and to all orders in strong coupling the differential rate for dilepton pair production can be expressed in terms of the correlation function of the hadronic electromagnetic current.

When lepton mass is ignored the rate is given by
\begin{equation}
\frac{dR}{d^4q}=\frac{4\alpha^2}{3(2\pi)^3}\frac{1}{q^4}\left(q^\mu q^\nu - q^2 g^{\mu\nu}\right)W_{\mu\nu}(q)\,,
\end{equation}
where
\begin{equation}
W_{\mu\nu}(q)=\int d^4x e^{-iqx} \langle J_\mu^{\text{em}}(x) J_\nu^{\text{em}}(0) \rangle_\beta\,.
\end{equation}
In the above equations $q$ is the time-like four-momentum of the lepton pair, $J_\mu^{\text{em}}$ is the hadronic part of the electromagnetic current and $\langle \cdots \rangle_\beta$ stands for the thermal averaging at a temperature $\beta\equiv 1/T$.  

In general, there are two ways in which the above thermal structure function, $W_{\mu\nu}(q)$, can be evaluated.  The first is by kinetic theory.  By inserting a complete set of states for each incoming component of the thermal density matrix the above equations can be shown to agree with relativistic kinetic theory reaction by reaction.  This method of evaluation is not only cumbersome but also relies on many approximations, such as the choice of Lagrangian and coupling constants.  

A second approach, which is used here, is to relate the thermal structure function directly to spectral functions as was first done by Z. Huang \cite{Huang:1995dd}.  From the spectral representation and symmetry the thermal structure function can be related to the absorptive part of the time ordered function
\begin{equation}
W_{\mu\nu}(q)=\frac{2}{1+e^{q_0/T}}\text{Im}\Pi_{\mu\nu}^{\text{em}}(q)\,,
\end{equation}
where
\begin{equation}
\Pi_{\mu\nu}^{\text{em}}(q)=i\int d^4x e^{iqx} \langle T\left( J_\mu^{\text{em}}(x) J_\nu^{\text{em}}(0) \right)\rangle_\beta\,.
\end{equation}

Let us now evaluate $\Pi_{\mu\nu}^{\text{em}}$ not reaction by reaction as done in kinetic theory but instead in a low temperature expansion as was first done by Dey, Eletsky and Ioffe \cite{Dey:1990ba}.  At low enough temperature the heat bath will be dominated by pions and therefore one can keep the first term in the expansion of the trace in the thermal averaging.  We quote the result and leave the details for the appendix.  To leading order in $(T/f_\pi)^2$ one finds
\begin{equation}
\Pi_{\mu\nu}(q,T) = \Pi_{\mu\nu}^{\text{em}}(q) - \epsilon\left( \Pi_{\mu\nu}^V(q) - \Pi_{\mu\nu}^A(q)\right)\,,
\end{equation}
where $\epsilon = T^2/6f_\pi^2$ and $\Pi^{A,V}$ are the axial-vector and vector correlators.  This is an example of how the dilepton production rate at finite temperature can be determined from measurable experimental data at zero temperature. The above result shows that the vector and axial-vector spectral densities mix at finite temperature.  Chiral symmetry makes the statement that $\Pi^V = \Pi^A$.  To leading order in temperature this occurs when $T = \sqrt{3}f_\pi \approx 160$ MeV.

Even though the above result was restricted to zero momentum pions it is general in the sense that it was derived from current algebra and PCAC alone.  As the number of soft-pion fields emitted or absorbed grows the current algebra formulation becomes increasingly difficult.  For this reason Weinberg \cite{Weinberg:1978kz} developed a method of calculating current algebra results using an effective Lagrangian formulation at tree level.  By renormalizing the tree level results one could obtain corrections to the soft-pion theorems.  A one loop calculation can still only describe data up to about 200 MeV above threshold.  Two loop calculations are intractable since over 100 new low energy constants appear.

A program that extends chiral symmetry consistently into the resonance region without the soft-pion restriction is discussed in the next section.
   
%\section{\textquotedblleft Master Equation\textquotedblright Approach}
\section{Chiral Reduction Formula}

The limitations of current algebra and chiral perturbation theory can be avoided by instead using an S matrix formalism.  H. Yamagishi and I. Zahed \cite{Yamagishi:1995kr} have derived a single equation (coined the \textquotedblleft Master Equation\textquotedblright) that contains all of the low energy theorems of current algebra.

The starting point for this program is an action {\bf I} with its kinetic part invariant under local $SU_L(2)\times SU_R(2)$ that is gauged with external sources.  Examples are two-flavor QCD or the nonlinear sigma model.

For two-flavor QCD the action is given as
\begin{eqnarray}
{\bf I}&=&\int d^4x \bar{q} \gamma^\mu\left(i\partial_\mu + G_\mu + v^a_\mu \frac{\tau^a}{2} + a_\mu^a\frac{\tau^a}{2}\gamma_5\right)q\nn
&-&\frac{m_q}{m_\pi^2}\int d^4x \bar{q}\left(m_\pi^2+s-i\gamma_5\tau^ap^a\right)q\nn
&-&\frac{1}{2g^2}\int d^4x \text{Tr}_C\left( G_{\mu\nu} G^{\mu\nu} \right)
\end{eqnarray}
where $G$ is the gluon field strength tensor defined by
\begin{equation}
G_{\mu\nu}=\partial_\mu G_\nu - \partial_\nu G_\mu -i\left[G_\mu, G_\nu\right]
\end{equation}

By Noether's theorem currents are defined by 
\bg
{\bf J}(x)=\frac{\delta{\bf I}}{\delta\phi(x)}
\nd

where ${\bf J} = \left({\bf V}, {\bf A}, f_\pi\sigma, f_\pi \pi\right)$ and $\phi=\left(v_\mu^a,a_\mu^a,s,p^a\right)$.  The currents must satisfy the Noether's equations in the presence of sources (also known as the Veltman--Bell \cite{VB} equations):
\bg
\nabla^\mu{\bf V}_\mu+\underline{a}^\mu {\bf A}_\mu+f_\pi\underline{p}\pi=0
\nd
\bg
\nabla^\mu{\bf A}_\mu + \underline{a}^\mu {\bf V}_\mu - f_\pi\left(m_\pi^2+s\right)\pi + f_\pi p\sigma=0
\nd
In the above equations $\nabla_\mu\equiv\partial_\mu{\bf 1}+\underline v_\mu$ is the vector covariant derivative and we have used the notation that $\underline{A}^{ac}\equiv\epsilon^{abc}A^b$.  Schwinger's quantum mechanical action principal
\bg
\delta\langle\beta_\text{out}|\alpha_\text{in}\rangle=i\langle\beta_\text{out}|\delta{\bf I}|\alpha_\text{in}\rangle
\nd
along with the completeness of asymptotic states leads to the Peierls-Dyson formula \cite{PD}
\bg
{\bf J}(x)=-i\mathcal{S}^\dagger\frac{\delta\mathcal{S}}{\delta\phi(x)}\,.
\nd
The Veltman-Bell equations can now be recast into the following form
\bg
\left({\bf X}_V+\underline{p}\frac{\delta}{\delta p}\right)\mathcal{S}&=&0\\
\left({\bf X}_A-(m_\pi^2+s)\frac{\delta}{\delta p}+p\frac{\delta}{\delta s}\right)\mathcal{S}&=&0\\
\label{eq:VB}
\nd
where we have defined
\bg
{\bf X}_V^a(x)=\nabla_\mu^{ac}\frac{\delta}{\delta v_\mu^c(x)}+\underline{a}_\mu^{ac}(x)\frac{\delta}{\delta a_\mu^c(x)}\\
{\bf X}_A^a(x)=\nabla_\mu^{ac}\frac{\delta}{\delta a_\mu^c(x)}+\underline{a}_\mu^{ac}(x)\frac{\delta}{\delta v_\mu^c(x)}
\nd
It can be shown that ${\bf X}_V$ and ${\bf X}_A$ are the generators of local $SU(2)\times SU(2)$.

So far we have not considered whether chiral symmetry is present or explicitly and/or spontaneously broken.  The spontaneous breaking of chiral symmetry is expressed in terms of the following asymptotic condition on the axial-vector field
\bg
{\bf A}_\mu^a(x)\to -f_\pi\partial_\mu\pi^{a}_{\text{out,in}}(x)\;\;\;\;\;\;\;\;\;\;x^0\to\pm\infty
%\pm\infnity
\nd
The above condition assumes the absence of any additional stable axial vector or pseudo-scalar resonances.  For explicit chiral symmetry breaking we must also impose
\bg
\partial^\mu{\bf A}_\mu^a(x)\to +f_\pi m_\pi^2 \pi^a_{\text{out,in}}(x) \;\;\;\;\;\;\;\;\;\;x^0\to\pm\infty
\nd

In order simplify the incorporation of the above boundary conditions into the approach a modified action $({\bf \hat{I}})$ and modified S-matrix $(\mathcal{\hat{S}})$ are introduced.
\bg
{\bf \hat{I}} = {\bf I}-f_\pi^2\int d^4x\left ( s(x)+\frac{1}{2}a^\mu(x)\cdot a_\mu(x)\right)
\nd
\bg
\mathcal{\hat{S}}=\mathcal{S}\exp\left[-i f_\pi^2\int d^4x\left ( s(x)+\frac{1}{2}a^\mu(x)\cdot a_\mu(x)\right)\right]
\nd
Also needed is a change of variables, $p=J/f_\pi-\nabla^\mu a_\mu$.  We now use $\hat{\phi}=\left(v_\mu^a, a_\mu^a, s, J^a\right)$ as independent variables and ${\bf \hat{J}}=\left({\bf j}_A, {\bf j}_V,f_\pi \hat{\sigma},\hat{\pi}\right)$ as modified current densities defined analogously as
\bg
{\bf \hat{J}}(x)=-i\mathcal{\hat{S}}^\dagger\frac{\delta\mathcal{\hat{S}}}{\delta\hat{\phi}(x)}
\nd

Under this new change of variables the Veltman-Bell equations \ref{eq:VB} can be integrated upon introduction of a retarded and advanced Green function $\mathcal{G}_{\text{R,A}}$ yielding a relation between the pion field and the other currents.  After some manipulation these relations can be written in the following form
\bg
\left[\pi_{\text{in}},\hat{S}\right]=\int{d^4y d^4z}e^{iky}\left(1+{\bf K}\mathcal{G}_\text{R}\right)^{ac}(y,z)\nn
\times\left( -i\hat{S}\left({\bf K}\pi_{\text{in}}\right)^c(z)+if_\pi J^c(z)\hat{S}-\frac{1}{f_\pi}\left( \nabla^\mu\mathcal{A}_\mu-J\right)^c(z)\frac{\delta \hat{S}}{\delta s(z)}+\frac{1}{f_\pi}{\bf X}_A^c(z)\hat{S} \right)\nn
\nd 

With the above {\em master equation} in hand it is easy to see a strategy in order to generate Ward identities.  After reducing out a pion from a scattering amplitude the commutator on the LHS can be replaced by an operator of many functional derivatives acting on the scattering matrix on the RHS.  These variations on the $S$ matrix can be written as time ordered products by
\bg
T^*\left( \mathcal{O}(x_1)\cdots \mathcal{O}(x_n)\right)=(-i)^n\hat{S}^\dagger\frac{\delta^n}{\delta\hat{\phi}(x_1)\cdots \delta\hat{\phi}(x_n)}\hat{S}
\nd
where $\mathcal{O}=-i\hat{S}^\dagger\delta\hat{S}/\delta\hat{\phi}$.

\section{Leading Order Lepton Emission: $I\to F\pi l^+l^-$}

The rate of dilepton emission per unit four volume for particles in 
thermal equilibrium at a temperature $T$ is related to the thermal 
expectation value of the electromagnetic current-current 
correlation function~\cite{McLerran, Weldon}.  For massless leptons 
with momenta $p_1$ and $p_2$, the rate per unit invariant momentum 
$q=p_1+p_2$ is given by:
\begin{equation}
\label{eq:rate1}
\frac{dR}{d^4q}=\frac{-\alpha^2}{3\pi^3 q^2}\frac{1}
{1+e^{ q^0/T}}\text{Im}{\bf W}^F(q)
\end{equation}
where $\alpha=e^2/4\pi$, $T$ is the temperature and 
\begin{equation}
\label{eq:WF1}
{\bf W}^{F}(q)= i\int{d^4x}\text{ } 
e^{iq\cdot x}\text{Tr}\left[e^{-({\bf H}-F)/T} 
T^* {\bf J}^\mu(x) {\bf J}_\mu(0) \right]
\end{equation}
where $e{\bf J}_\mu$ is the hadronic part of the electromagnetic current, 
{\bf H} is the hadronic Hamiltonian and $F$ is the 
free energy.  The trace is over a complete set of hadron states.

In order to take into account leptons with mass $m_l$ the right-hand side 
of Eq.~\ref{eq:rate1} is multiplied by
\begin{equation}
\label{eq:lep_mass1}
(1+\frac{2m^2_l}{q^2})(1-\frac{4m^2_l}{q^2})^{1/2}
\end{equation}

Even though there are various approaches to calculating production 
rates, they differ in the way in which the current-current correlation 
function in Eq.~\ref{eq:rate1} is approximated and evaluated.  
The approach taken here is to use a chiral reduction formalism in 
order to reduce the current-current correlation function in~\ref{eq:WF1} 
into a number of vacuum correlation functions which can be constrained 
to experimental $e^+e^-$ annihilation, $\tau$-decay, two-photon fusion 
reaction, and pion radiative decay experimental data.

For temperatures T $\leq m_\pi$ the trace in Eq.~(\ref{eq:WF1}) 
can be expanded in pion states.  
Keeping terms up to first order in pion density yields \cite{paper1}
\begin{equation}
\label{eq.exp1}
\text{Im} {\bf W}^F(q)=-3q^2
\text{Im} {\bf \Pi}_V(q^2)+\frac{1}{f^2_\pi}
\int{d\pi}{\bf W}^F_\pi(q,k)
\end{equation}
with the phase space factor
\begin{equation}
\label{eq:phase_space_a1}
d\pi=\frac{d^3k}{(2\pi)^3}\frac{1}{2E_k}\frac{1}{e^{E_k/T}-1}\,.
\end{equation}

The first term in~\ref{eq.exp1} is the transverse part of the isovector 
correlator $\langle 0|T^*{\bf VV}|0 \rangle$ which can be determined 
experimentally from electroproduction data and gives a result 
analogous to the resonant gas model.  At low and intermediate invariant 
mass the spectrum is dominated by the $\rho(770$ MeV) and $\rho'(1450$ MeV).

The term linear in pion density (the second term in Eq.~\ref{eq.exp1}) can 
be related to experimentally measured quantities via the  
chiral reduction formulas \cite{CRF}.  It is shown in~\cite{paper1} 
that the dominant contribution comes solely from the part involving 
two-point correlators which yields:
\bg
{\bf W}^F_\pi(q,k)&=&\frac{12}{f_\pi^2}q^2\text{Im} {\bf \Pi}_V(q^2)\nn
&-&\frac{6}{f_\pi^2}(k+q)^2\text{Im} {\bf \Pi}_A \left( (k+q)^2\right) + (q\to -q)\nn
&+&\frac{8}{f_\pi^2}\left( (k\cdot q)^2-m_\pi^2 q^2\right) \text{Im} {\bf \Pi}_V(q^2)\times\text{Re} \Delta_R(k+q)+(q\to-q)\nn
\label{eq:lin_in_meson1}
\nd
where $\text{Re}\Delta_R$ is the real part of the retarded pion propagator given by 
$1/(q^2-m_\pi^2+i\epsilon)$ and ${\bf \Pi}_A$ is the transverse part of 
the iso-axial correlator $\langle 0|T^*{\bf j}_A{\bf j}_A|0 \rangle$.  The 
spectral functions appearing in Eq.~(\ref{eq:lin_in_meson1}) can be 
related to both $e^+e^-$ annihilation as well as $\tau$-decay data 
as was compiled in \cite{Huang95}.  

\begin{figure}
\centering
\includegraphics[scale=.5]{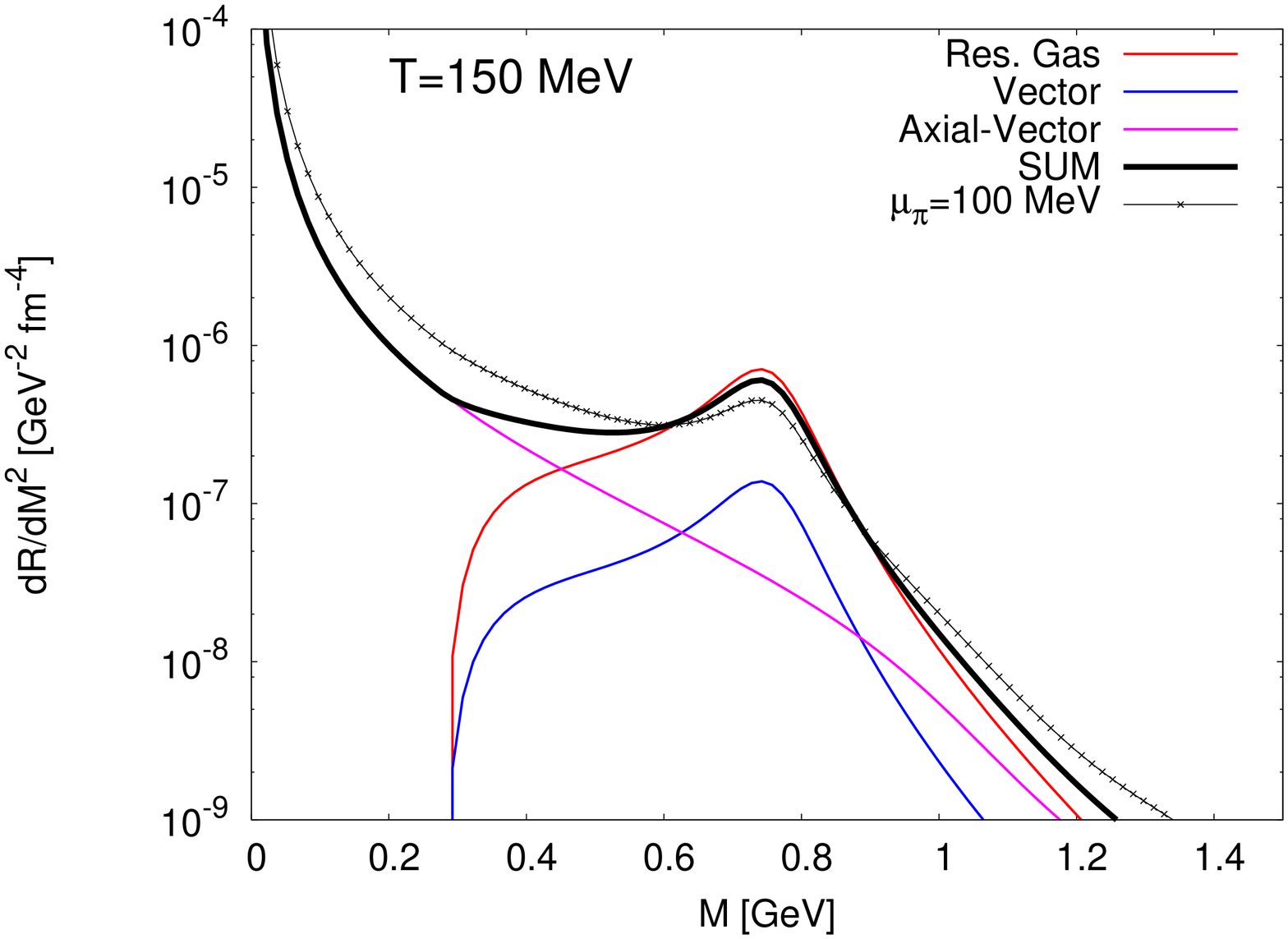}
\caption{(Color online) The total integrated dimuon rates from a pion gas at T=150 MeV.  
The curve labeled ``Res. Gas'' shows the analogue of the resonance gas contribution (the first term in 
Eq. \ref{eq.exp1}).  The curves labeled ``Vector'' and ``Axial-Vector''
show the contributions from the respective spectral functions 
in equation \ref{eq:lin_in_meson1}. } 
\label{fig:dimuon_rates_pion}
\end{figure}

It can be seen in Fig.~\ref{fig:dimuon_rates_pion} that the term 
linear in pion density decreases the rates from the resonance 
gas contribution for the mass region above the two pion threshold.  
However below the two pion threshold the only contribution to the 
rates come from the ${\bf \Pi}_A$ terms in Eq.~\ref{eq:lin_in_meson1}.  
This is because the axial spectral density is integrated over 
all momentum in the thermal averaging (Eq.~\ref{eq.exp1}), 
which weakens the $(k+q)^2$ factor in Eq.~\ref{eq:lin_in_meson1} 
allowing the $1/q^2$ term in Eq.~\ref{eq:rate1} to dominate at low $q^2$. 

\section{Second Order Lepton Emission: $I\to F\pi\pi l^+l^-$}
\subsection{Introduction}

We now keep terms up to second order in pion density
\begin{equation}
\label{eq:exp2}
\text{Im} {\bf W}^F(q)=-3q^2
\text{Im} {\bf \Pi}_V(q^2)+\frac{1}{f^2_\pi}
\int{d\pi}{\bf W}^F_\pi(q,k)
+\frac{1}{2f^4_\pi}
\int{d\pi_1 d\pi_2 }{\bf W}^F_{\pi\pi}(q,k_1,k_2)
\end{equation}
with the phase space factor
\begin{equation}
\label{eq:phase_space_a2}
d\pi=\frac{d^3k}{(2\pi)^3}\frac{1}{2E_k}\frac{1}{e^{E_k/T}-1}\,.
\end{equation}

The first two terms in the above density expansion were considered in the previous section.  As long as the system is sufficiently dilute ({\em i.e.} $\kappa\equiv n_\pi/2m_\pi f_\pi^2 \ll 1$) the expansion should converge rather quickly as long as no new thresholds open up.  What we will find is that the $2\pi$ contribution feeds into the low mass and low $p_\perp$ region where the zero and first order corrections do not contribute.  This will also enhance the real photon rate ($M^2=0$) at small energy.

\subsection{Result}

In this section we quote the full on-shell Ward identity for ${\bf W}_{\pi\pi}^F$.  We note that we correct a number of typographic errors from the result quoted in \cite{paper2}.  We also discuss in detail which terms are kept in the numerical calculations and argue which terms can be safely neglected.  First let us quote the terms included in the analysis:    
\begin{eqnarray}
\frac{1}{f_\pi^4}W^F_{\pi\pi}(q,k_1,k_2) &=& \nn
&=& \frac{2}{f_\pi^2}\left[g_{\mu\nu}-(2k_1+q)_\mu k_{1\nu}\text{Re}\Delta_R(k_1+q)\right]\text{Im}\mathcal{T}_{\pi\gamma}^{\mu\nu}\left(q,k_2\right)\nn
&+& (q\to -q) + (k_1\to -k_1) + (q,k_1 \to -q,-k_1)\label{eq:2pig}\\
\nn
&+& \frac{1}{f_\pi^2}k_1^\mu (2k_1+q)^\nu \text{Re}\Delta_R(k_1+q)\epsilon^{a3e}\epsilon^{e3g}\text{Im}\mathcal{B}^{ag}_{\mu\nu}(k_1,k_2)\nn
&-& \frac{1}{f_\pi^2}\left[g^{\mu\nu} - (k_1+q)^\mu(2k_1+q)^\nu\text{Re}\Delta_R(k_1+q)\right]\nn&\times&\epsilon^{a3e}\epsilon^{a3f}\text{Im}\mathcal{B}_{\mu\nu}^{ef}(k_1+q,k_2)\nn
&+& \frac{1}{f_\pi^2}(k_1+q)^\mu(k_1+q)^\nu(2k_1+q)^2\left[\text{Re}\Delta_R(k_1+q)\right]^2\nn&\times&\epsilon^{a3e}\epsilon^{a3f}\text{Im}\mathcal{B}_{\mu\nu}^{ef}(k_1+q,k_2) + (k_1\to-k_1)
\label{eq:2B}
\end{eqnarray}

Equation \ref{eq:2pig} contains the pion-spin averaged $\pi\gamma$ forward scattering amplitude ($i\mathcal{T}_{\pi\gamma}$).  This quantity can be constrained from measured photon fusion data by crossing and is discussed in section \ref{sec:ggpipi}.  We have defined the term $\mathcal{B}$ in equation \ref{eq:2B} as  
\begin{equation}
\mathcal{B}_{\mu\nu}^{ef}(k_1,k_2)\equiv i\int d^4x e^{ik_1 x}\langle\pi^b_{\text{out}}(k_2) | T^*\left( {\bf j}_{A\mu}^e(x) {\bf j}_{A\nu}^f(0) \right) | \pi^b_{\text{in}}(k_2)\rangle
\end{equation}
The pions in the above expression for $\mathcal{B}$ can be reduced out via the chiral reduction formula.  Since most of the strength will come from the vector and axial-vector spectral densities we keep these terms only.
\begin{eqnarray}
\text{Im}\mathcal{B}_{\mu\nu}^{ef}(k_1,k_2)&=&\frac{2}{f_\pi^2}\delta^{ef}\left[ g_{\mu\nu}(k_1+k_2)^2-(k_1+k_2)_\mu (k_1+k_2)_\nu\right]\text{Im}\Pi_V\left( (k_1+k_2)^2 \right)\nn &+& (k_2\to-k_2)\nn
&-&\frac{4}{f_\pi^2}\delta^{ef}\left[ g_{\mu\nu}k_1^2-k_{1\mu} k_{1\nu}\right]\text{Im}\Pi_A\left( k_1^2 \right)
\label{eq:Bfinal}
\end{eqnarray}
There are additional terms however which we should discuss. Most can be argued away by resonance saturation.  One term which we should quote which could appear at higher mass is the following four point function
\begin{equation}
\frac{1}{f_\pi^2}k_2^\alpha k_2^\beta i\int{d^4x d^4z_1 d^4z_2} e^{ik_1x}e^{ik_2z_2}e^{-ik_2z_1}\langle 0|T^*\left(j_{A\alpha}^b(z_2) j_{A\beta}^b(z_1) j_{A\mu}^e(x) j_{A\nu}^f(0)\right)|0\rangle_{\text{conn.}}\nn
\end{equation}
but we neglect it further in this analysis since we want to focus on the region below the $\rho$ mass.  Experimental information about this term could be extracted from $\pi-\pi$ scattering data \cite{ZahedBook}.
\begin{figure}
\label{fig:B}
\centering
\includegraphics[scale=0.8]{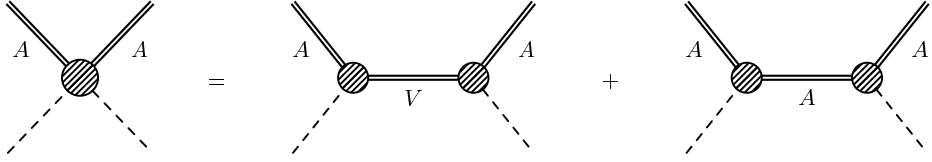}
\caption{Schematic representation of the chiral reduction of $\mathcal{B}$.  The dashed line is a pion.}
\end{figure}

For completeness we now quote the remaining terms of ${\bf W}_{\pi\pi}^F$.  These are not included in our analysis since they can be argued to be small in the kinematic regions which we are interested in.
\begin{eqnarray}
\label{eq:2Ba}
&+&\frac{1}{f_\pi^2}k_1^\alpha k_1^\beta (2k_1+q)^2\text{Re}\Delta_R(k_1+q)\text{Re}\Delta_R(k_1)\epsilon^{a3e}\epsilon^{g3e}\text{Im}\mathcal{B}_{\alpha\beta}^{ag}(k_1,k_2)\nn
&+& (q\to -q) + (k_1\to -k_1) + (q,k_1 \to -q,-k_1)\\
\nn
&+&\frac{1}{f_\pi}k_1^\alpha(2k_1+q)^\mu\epsilon^{a3e}\text{Im}\mathcal{A}^{ae}_{\alpha\mu}(k_1,q)\nn
&+& (q\to -q) + (k_1\to -k_1) + (q,k_1 \to -q,-k_1)\label{eq:2A}\\
\nn
&-&(2k_1+q)^2\left[\text{Im}\Delta_R(q+k_1)\right]^2\epsilon^{a3e}\epsilon^{a3f}\text{Im}\mathcal{T}_{\pi\pi}^{be\to bf}\left( (k_1+q),k_2\right)\nn
&+&(q\to -q)\label{eq:2pi}\\
\nn
\label{eq:2VVsig}
&+&\text{Im}\frac{3m_\pi^2}{f_\pi}\int d^4x d^4y e^{iq\cdot(x-y)}\langle\pi_{\text{out}}^b(k_2)|T^*{\bf V}_{\mu}^3(x){\bf V}^{3,\mu}(y)\hat{\sigma}(0)|\pi_{\text{in}}^b(k_2)\rangle\nn
\\
\label{eq:2C}
&-&\frac{1}{f_\pi^2}g^{\mu\nu}k_1^\alpha\epsilon^{a3e}\text{Im}i\mathcal{C}^{ea}_{\mu\nu\alpha}(q,k_1+q,k_1)+(k_1\to -k_1)\nn
&+&\text{Im}\frac{1}{f_\pi^2}(2k_1+q)^\mu(k_1+q)^\nu k_1^\alpha\text{Re}\Delta_R(k_1+q)\epsilon^{a3e}i\mathcal{C}^{ea}_{\mu\nu\alpha}(q,k_1+q,k_1)\nn
\\
\label{eq:2jjVV}
&-&\frac{1}{f_\pi^2}k_1^\alpha k_2^\beta\int d^4x d^4y d^4z e^{ik_1\cdot(y-x)}e^{iqy}\nn
&\times&\text{Im}i\langle\pi_{\text{out}}^b(k_2)|T^*{\bf j}_{A\alpha}^a(x) {\bf j}_{A\beta}^b(y) {\bf V}_\mu^3(z) {\bf V}^{3,\mu}(0)|\pi_{\text{in}}^b(k_2)\rangle\\
\nn
&-&\frac{2m_\pi^2}{f_\pi}(2k_1+q)^2\text{Re}\Delta_R(k_1+q)\left[\text{Re}\Delta_R(k_1+q)+\text{Re}\Delta_R(k_1)\right]\nn
&\times&\text{Im}\langle\pi_{\text{out}}^b(k_2)|\hat{\sigma}(0)|\pi_{\text{in}}^b(k_2)\rangle\nn
&+& (q\to -q) + (k_1\to -k_1) + (q,k_1 \to -q,-k_1)\label{eq:2sig}
\end{eqnarray}

We now discuss why the above terms are neglected.  First look at eq.~\ref{eq:2Ba}.  It is proportional to the principal value of the real part of the retarded pion propagator defined as,
\begin{equation}
\Delta_R(k)\equiv {\bf PP}\frac{1}{k^2-m_\pi^2}-i\pi\text{sgn}(k^0)\delta(k^2-m_\pi^2)\\.
\end{equation}
For on shell pions this term is proportional to ${\bf PP}\frac{1}{0}=0$ and therefore vanishes.  Now look at eq.~\ref{eq:2A} where we have defined $\mathcal{A}$ as:
\begin{eqnarray}
\mathcal{A}_{\alpha\mu}^{ae}(k_1,q)\equiv\int d^4x d^4y e^{ik_1 x}e^{iqy}\langle\pi_{\text{out}}^b(k_2)|T^*\left({\bf j}_{A\alpha}^a(x){\bf V}_\mu^3(y)\right)\pi_{\text{in}}^e(0)|\pi_{\text{in}}^b(k_2)\rangle\nn
\end{eqnarray}
Making use of the chiral reduction formula one can reduce the incoming pion $\pi_{\text{in}}^e(0)$ with the result:
\begin{eqnarray}
&&\left[\text{Im}\Delta_R(k_1+q)\right]^{-1}\text{Im}\mathcal{A}_{\alpha\mu}^{ae}(k_1,q)=\frac{1}{f_\pi}\epsilon^{e3g}\text{Im}\mathcal{B}_{\alpha\mu}^{ag}(k_1,k_2)\nn
&+&\frac{1}{f_\pi}\epsilon^{e3g}\left(2k_1+q\right)_\mu k_1^\beta\text{Re}\Delta_R(k_1)\text{Im}\mathcal{B}_{\alpha\beta}^{ag}(k_1,k_2)\nn
&+&\frac{1}{f_\pi}\epsilon^{age}\int d^4x e^{iqx}\text{Im}i\langle\pi_{\text{out}}^b(k_2)|T^*{\bf V}_{\mu}^3(x){\bf V}_\alpha^g(0)|\pi_{\text{in}}^b(k_2)\rangle\nn
&-&\frac{1}{f_\pi}(k_1+q)^\beta \int d^4x d^4y e^{iqx} e^{ik_1 y}\nn&\times&\text{Im}i\langle\pi_{\text{out}}^b(k_2)|T^*{\bf V}_{\mu}^3(x){\bf j}_{A\alpha}^a(y){\bf j}_{A\beta}^e(0)|\pi_{\text{in}}^b(k_2)\rangle\nn
&+&\frac{1}{f_\pi}\delta^{ae}(2k_1+q)_\alpha\int d^4x e^{ik_1x}\nn&\times&\text{Im}\langle\pi_{\text{out}}^b(k_2)|T^*{\bf V}_{\mu}^3(x){\bf \hat{\sigma}}(0)|\pi_{\text{in}}^b(k_2)\rangle
\end{eqnarray}
%\begin{eqnarray}
%\text{Im}\mathcal{A}_{\alpha\mu}^{ae}(k_1,q)&=&\frac{1}{f_\pi}\text{Im}\Delta_R(k_1+q)\left[\epsilon^{e3g}\text{Im}\mathcal{B}_{\alpha\mu}^{ag}(k_1,k_2)-\frac{1}{f_\pi}\epsilon^{a3e}\text{Im}T_{\pi\gamma}^{\alpha\mu}(k_1,k_2)\right]
%\end{eqnarray}
From reducing out the incoming pion we find that $\mathcal{A}\propto \text{Im}\Delta_R(k_1+q)$.  The term only contributes when the pion from the heat bath has the kinematics specified by the delta function.  Due to the small amount of phase space this term will be suppressed compared to the terms in \ref{eq:2pig} and \ref{eq:2B}.  The same argument can be made for neglecting eq.~\ref{eq:2pi}. 

\begin{figure}
\centering
\includegraphics[scale=1]{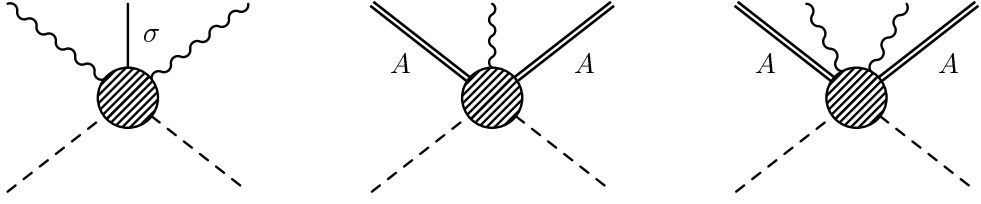}
\caption{Diagrams of the matrix elements in eqns \ref{eq:2VVsig}, \ref{eq:2C} and \ref{eq:2jjVV} (left to right).  The dashed line is a pion and the wavy line denotes a photon.}
\label{fig:W2ignorea}
\end{figure}

The matrix elements appearing in terms \ref{eq:2VVsig}, \ref{eq:2C} and \ref{eq:2jjVV} are shown in figure~\ref{fig:W2ignorea} where we have defined the term $\mathcal{C}$ in eq. \ref{eq:2C} as
\begin{eqnarray}
\mathcal{C}_{\mu\nu\alpha}^{ea}(q,k_1+q,k_1)&\equiv&\int d^4x d^4y e^{i(k_1+q)\cdot x}e^{-ik_1y}\nn
&\times&\langle\pi_{\text{out}}^b(k_2)|T^*{\bf V}_\mu^3(0){\bf j}_{A\nu}^e(x) {\bf j}_{A\alpha}^a(y)|\pi_{\text{in}}^b(k_2)\rangle
\end{eqnarray}
It turns out that these three contributions can be argued to be small.  First note that eqn.~\ref{eq:2VVsig} vanishes in the chiral limit.  Furthermore, after reducing out the incoming pions the remaining vacuum spectral functions will mostly consist of $\langle 0| VV\sigma|0\rangle$ and $\langle 0|  AA\sigma|0\rangle$ for which the resonance saturation is small.  Equation \ref{eq:2C} depends on $\mathcal{C}$ which after chiral reduction mostly reduces to the three correlators; $\langle 0|  VAA|0\rangle$, $\langle 0|  VVA|0\rangle$ and $\langle 0|VVV|0\rangle$ for which there is no s-channel cut through resonance saturation.  Finally the matrix element in eq.~\ref{eq:2jjVV} will mostly contribute in the four and six $\pi$ range and higher.  Since we are focusing our attention near the $2\pi$ threshold it is safe to say that the above processes can be neglected since the correlators contribute at higher mass.  In addition these processes are Boltzmann suppressed in comparison to the resonance and one $\pi$ final state reactions.

The final expression, eq.~\ref{eq:2sig}, is a direct consequence of the way chiral symmetry is broken through $\langle 0|\hat{\sigma}|0\rangle$.  The term can be related directly to the scalar form factor through $\text{Im}\langle \pi^b(k_2)|\hat{\sigma}|\pi^b(k_2)\rangle=\text{Im}{\bf F}_S(t=0)=0$ and therefore vanishes.

\subsection{Pion Compton Scattering Amplitude}
\label{sec:ggpipi}

We consider the reaction $\gamma(q_1) + \pi^\pm(k_1) \to \gamma(q_2) + \pi^\pm(k_2)$ and define the Mandelstam variables to be
\begin{eqnarray}
s&=&(k_2+q_2)^2\nn
t&=&(k_1-k_2)^2\nn
u&=&(k_q-q_2)^2
\end{eqnarray}

Let us express the total Compton scattering matrix element $\mathcal{M}$ as
\begin{equation}
\mathcal{M}=e^2\epsilon_1^\mu(q_1)\epsilon_2^\nu(q_2)\mathcal{T}_{\mu\nu}
\end{equation}

\begin{figure}
\centering
\includegraphics[scale=0.8]{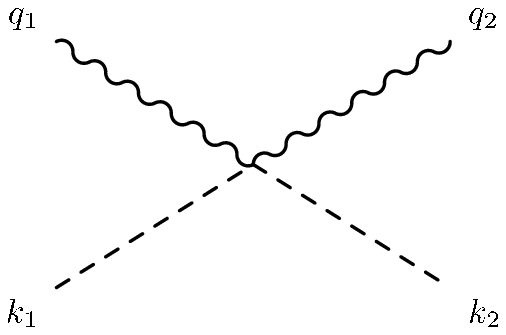}
\hspace{5pt}
\includegraphics[scale=0.8]{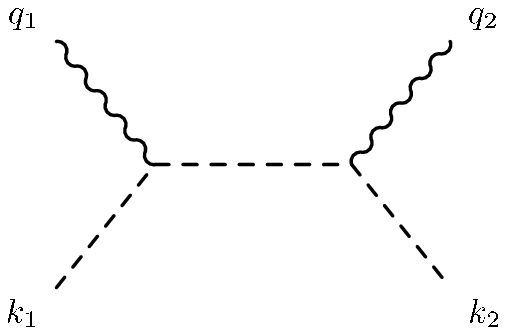}
\hspace{5pt}
\includegraphics[scale=0.8]{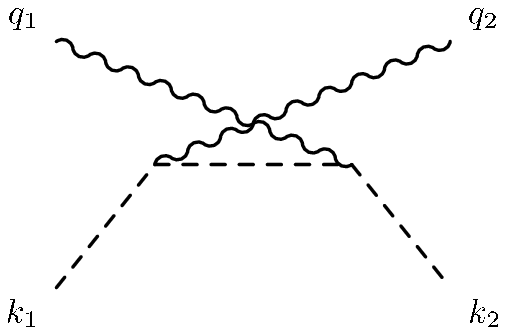}
\caption{Tree level contribution to the $\pi-\gamma$ scattering amplitude.}
\label{fig:compton}
\end{figure}

The pion compton scattering amplitude in the born approximation is by now a textbook example \cite{GreinerBook}.  The three Feynman diagrams of figure~\ref{fig:compton} contribute which evaluate for forward scattering ({\em i.e.} $q_1=q_2=q$ and $k_1=k_2=k$) to
\begin{equation}
\mathcal{T}_{\mu\nu}^{\pi\gamma}(q,k)=\frac{2}{3}\left[\frac{(2k+q)_\nu (2k+q)_\mu}{s-m_\pi^2}+\frac{(2k-q)_\nu (2k-q)_\mu}{u-m_\pi^2} - 2g_{\mu\nu}\right]
\end{equation}
where
\begin{eqnarray}
s&=&(k+q)^2\nn
u&=&(k-q)^2
\end{eqnarray}

\begin{figure}
\centering
\includegraphics[scale=0.8]{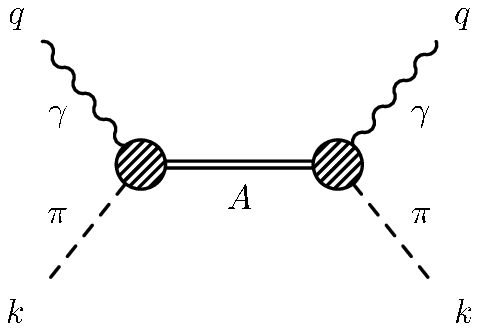}
\hspace{5pt}
\includegraphics[scale=0.8]{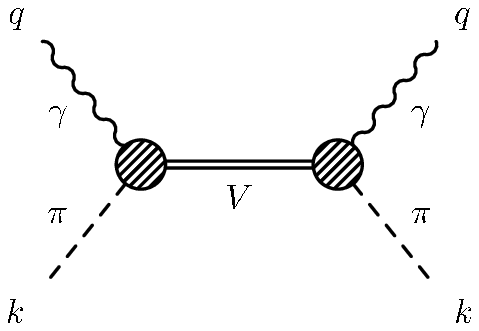}
\caption{Schematic representation of the higher order correction from the chiral reduction of the $\gamma-\pi$ scattering amplitude.}
\label{fig:compton2}
\end{figure}

However, since we are always below threshold the amplitude is always real and does not contribute to the imaginary part of the amplitude.  To go further we make use of the {\em master formula approach} of the $\gamma\pi\to\gamma\pi$ reaction \cite{Yamagishi:1995kr}.  The pion compton scattering process for real photons ($q^2=0$) was examined in \cite{Chernyshev:1995pq}.  The dominant contributions from the chiral reduction of the $\pi\gamma^*\to\pi\gamma^*$ process is   
\begin{eqnarray}
\mathcal{T}_{\mu\nu}^{\pi\gamma}(q,k) &=& 
\frac{1}{f_\pi^2}(2k+q)_\mu\left(-q^2k_\nu + (k\cdot q) q_\nu)\epsilon^{a3e}\epsilon^{eb3}\text{Re}\Delta_R(k+q)\right)\Pi_V(q^2)\nn
&+&(q\to-q) + (k\to-k) + (q,k\to -q,-k)\nn
&+&\frac{2}{f_\pi^2}\epsilon^{a3e}\epsilon^{be3}\left( -g_{\mu\nu}q^2+q_\mu q_\nu\right)\Pi_V(q^2)\nn
&-&\frac{1}{f_\pi^2}\epsilon^{a3e}\epsilon^{be3}\left( -g_{\mu\nu}(k+q)^2+(k+q)_\mu (k+q)_\nu\right)\Pi_A\left((k+q)^2\right)\nn
&+&(k\to-k)\nn
\end{eqnarray}
and corresponds to the diagrams in figure \ref{fig:compton2}.
Finally, we average over isospin ({\em i.e.} $\times 1/3\delta^{ab}$) obtaining
\begin{eqnarray}
\text{Im}\mathcal{T}_{\mu\nu}^{\pi\gamma}(q,k) &=& 
\frac{2}{3f_\pi^2}(2k+q)_\mu\left(-q^2k_\nu + (k\cdot q) q_\nu)\text{Re}\Delta_R(k+q)\right)\text{Im}\Pi_V(q^2)\nn
&+&(q\to-q) + (k\to-k) + (q,k\to -q,-k)\nn
&+&\frac{4}{3f_\pi^2}\left( g_{\mu\nu}q^2-q_\mu q_\nu\right)\text{Im}\Pi_V(q^2)\nn
&-&\frac{2}{3f_\pi^2}\left( g_{\mu\nu}(k+q)^2-(k+q)_\mu (k+q)_\nu\right)\text{Im}\Pi_A\left((k+q)^2\right)\nn
&+&(k\to-k)\nn
\label{eq:Tpigfinal}
\end{eqnarray}

\subsection{Soft Limit}

It is instructive to look at the above result for soft pions ($k_1,k_2\to 0$) in the chiral limit ($m_\pi=0$).  To first order in pion density the result is given in \cite{paper1}
\begin{equation}
W^F_{\pi}(q,0,0) = 12q^2\left[\text{Im}\Pi_V(q^2)-\text{Im}\Pi_A(q^2)\right]\,.
\end{equation}

To second order in pion density there are two terms which remain in the soft limit.  They are the terms proportional to $g_{\mu\nu}$ in equations~\ref{eq:2pig} and~\ref{eq:2B}.  In this limit we have
\begin{eqnarray}
\frac{1}{f_\pi^4}W^F_{\pi\pi}(q,0,0)&=&\frac{2}{f_\pi^2}g_{\mu\nu}\left[\text{Im}\mathcal{T}_{\pi\gamma}^{\mu\nu}(q,0)+\frac{1}{2}\epsilon^{a3e}\epsilon^{a3f}\text{Im}\mathcal{B}^{\mu\nu}(q,0)\right]\nn&=&\frac{2}{f_\pi^4}q^2\left[4-\frac{1}{2}24\right]\left(\text{Im}\Pi_V(q^2)-\text{Im}\Pi_A(q^2)\right)
\end{eqnarray}

We can now substitute the above result into~\ref{eq:exp2} and noting that each phase space integral leads to a factor of $T^2/24$ we find 
\begin{equation}
\text{Im}\Pi_{\text{em}}(q^2,T)=\text{Im}\Pi_{\text{em}}(q^2,0)-\left(\epsilon-\frac{\epsilon^2}{2}\right)\left[\text{Im}\Pi_V(q^2)-\text{Im}\Pi_A(q^2)\right]
\end{equation}
where 
\begin{equation}
\epsilon = \frac{T^2}{6f_\pi^2}
\end{equation}
consistent with the result of Dey, Eletsky and Ioffe \cite{Dey:1990ba,Eletsky:1994rp}.

\subsection{Dilepton Rates}

The phase space integrals in the rate equations were evaluated numerically.  As discussed earlier we expect the dominate contributions to come from eqns.~\ref{eq:2pig} and~\ref{eq:2B} where $\mathcal{T}_{\pi\gamma}$ and $\mathcal{B}$ are given by equations \ref{eq:Tpigfinal} and~\ref{eq:Bfinal} respectively.  

\begin{figure}[h!]
\centering
\includegraphics[scale=.4]{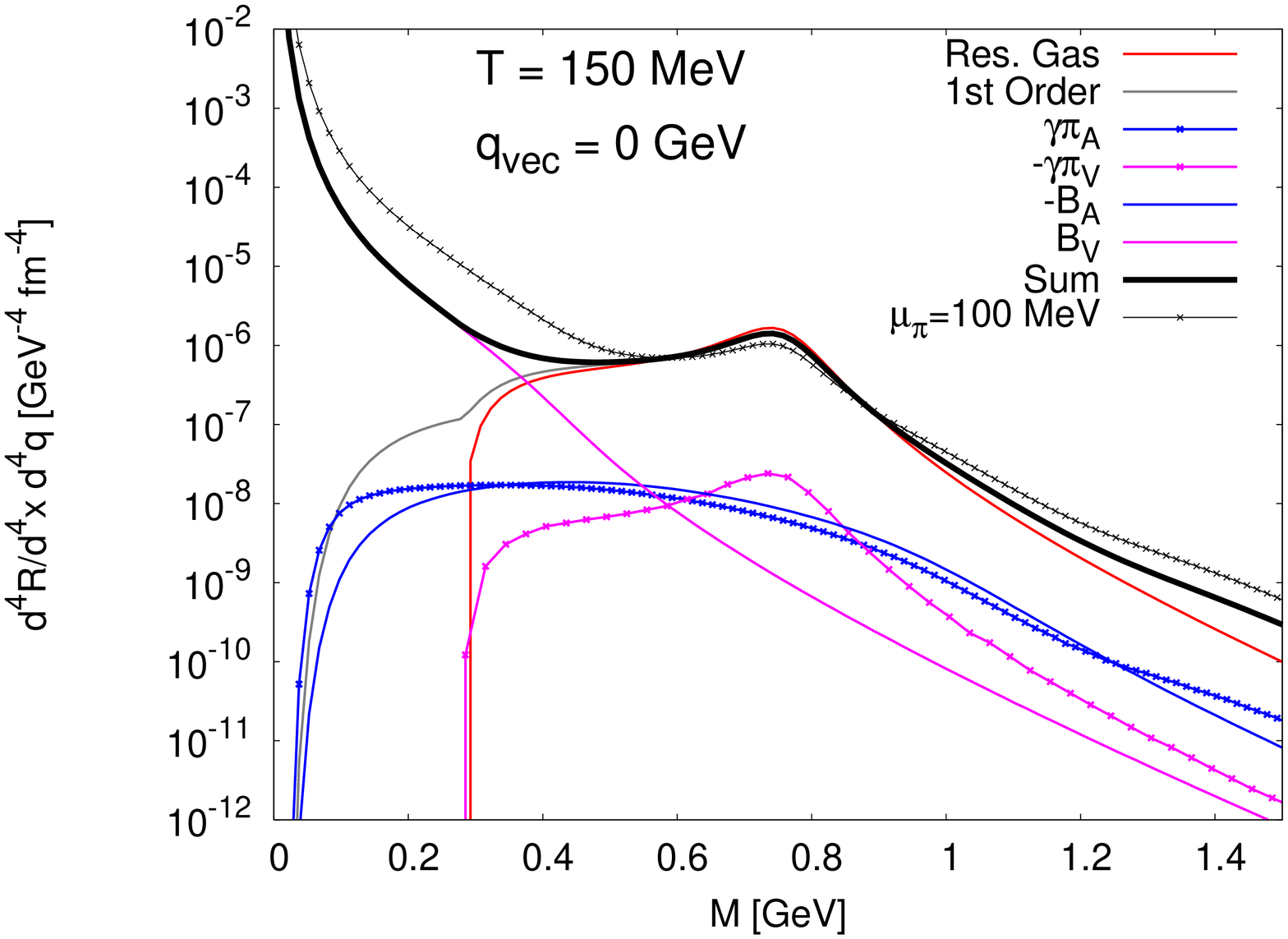}
  \vspace{9pt}
\includegraphics[scale=.4]{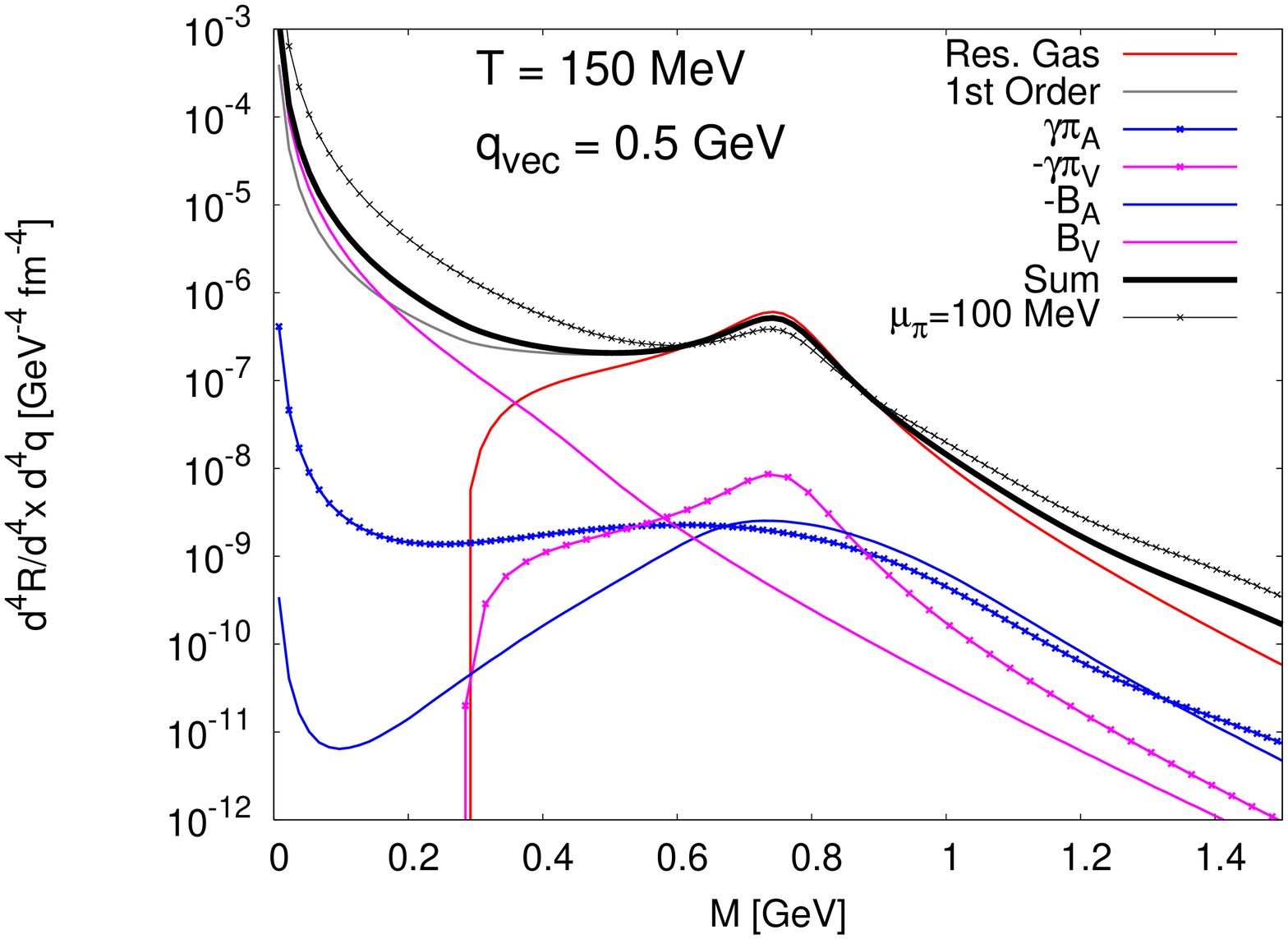}
\caption{(Color online) Differential virtual photon rates at $\vec{q}=0$ and $0.5$ GeV.  The curve labeled ``Res. Gas'' is the zeroth order contribution.  The first order curve is the sum of the resonance gas and first order, $W_\pi^F$, contributions.  The remaining terms show the contributions from various parts of $W^F_{\pi\pi}$.  $B_{V(A)}$ and $\gamma\pi_{V(A)}$ shows the vector (axial-vector) contribution of $\mathcal{B}$ and $\mathcal{T}_{\gamma\pi}$.  (Note that $B_A$ and $\gamma\pi_V$ are negative thus absolute values are shown.)  The solid (dashed) black lines shows the total rate evaluated at $\mu_\pi=0$ MeV ($\mu_\pi=100$ MeV).} 
\label{fig:ratespipi}
\end{figure}

We now discuss the results which are shown in figure \ref{fig:ratespipi}.  The upper (lower) figure shows the virtual photon rates for $\vec{q}=0$ $(0.5)$ GeV.  The red curve labeled ``Res Gas'' and the gray curve labeled ``1st order'' correspond to the first and second terms of equation \ref{eq:exp2}.  These are the results originally found in \cite{paper1} and correspond to all processes with zero and one pion in the final state.  At $\vec{q}=0.5$ GeV there is a huge enhancement in the low mass region, especially below the two pion threshold, and can be thought of as arising from processes of the type $I\to F\pi+e^+e^-$.  However, this enhancement gradually disappears at lower values of $\vec{q}$ and is negligible at $\vec{q}=0$. 

Let us now discuss the new results of this work, which are contributions from processes of the type $I\to F\pi\pi+e^+e^-$.  The magenta curves labeled ``$B_V$'' and ``-$\gamma\pi_V$'', show the contribution from the vector spectral densities in the expressions for $\mathcal{B}$ and $\mathcal{T}_{\pi\gamma}$ respectively.  We should note that the $\gamma\pi_V$ term is negative throughout and therefore its absolute value is shown.  The blue curves labeled ``$-B_A$'' and ``$\gamma\pi_A$'' show the corresponding contributions from the axial-vector spectral density.  In this case ``$B_A$'' is negative throughout and its absolute value is shown.  This first thing to note is that the contributions from $B_A$ and $\gamma\pi_{A,V}$ are well below the first order contribution in all kinematic regions.  Furthermore, there is even some cancellation between $B_A$ and $\gamma\pi_A$ which makes these contributions even less significant.  It should be pointed out that these terms cannot be neglected in the soft pion limit if one wants to be consistent with current algebra as discussed in the previous section.  However, at finite temperature they can be neglected.

The dileptons emanating from the $B_V$ term cannot be neglected however, and lead to a large enhancement below the two pion threshold at small momentum where the first order term is suppressed.  More precisely, at $\vec{q}=0$ GeV this term makes up for all of the emission below the two pion threshold.  By $\vec{q}=1$ GeV the second order term can be neglected since the first order term dominates.     
\begin{figure}[h!]
\centering
\includegraphics[scale=.8]{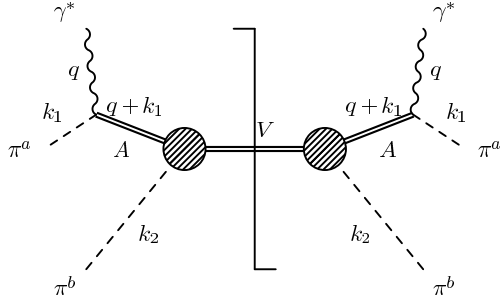}
\caption{Example of one of the matrix elements contributing to $\mathcal{B}$.}
\label{fig:Bfinal}
\end{figure}

In order to gain some physical insight into where the low mass and low $\vec{q}$ enhancement is coming from we show in figure~\ref{fig:Bfinal} one of the main contributors to the terms in eq.~\ref{eq:2B}.  The density expansion blurs the relation between the spectral densities and kinetic theory so it is not possible to make a mapping from a given term in the $\chi RF$ to the relevant physical reaction.  Regardless, one can think of the imaginary part of the diagram as coming from the thermal decay of $\rho,\rho^\prime, \rho^{\prime\prime}\cdots\to \pi h_1, \pi b_1, \pi a_1\cdots \to 2\pi \gamma^*$.  The two final state Pions carry away most of the momentum leaving behind a low momentum photon. 

\section*{Appendix: Hadronic Correlators at Finite Temperature}
\addcontentsline{toc}{section}{Appendix: Hadronic Correlators at Finite Temperature}
\chaptermark{Appendix}

Let us consider the vector spectral function at finite temperature when the heat bath is dominated by pions.  Expanding the thermal trace one finds
\begin{eqnarray}
\Pi^{ij}_{V\mu\nu}(q,T) &=& i\int d^4x e^{iqx} \text{Tr}\left( e^{-\beta(H-F)} T\left( V^i_\mu(x) V^j_\nu(0) \right) \right)
\approx \Pi^{ij}_{V\mu\nu}(q,T=0) \nonumber\\ 
&+& \int d^4x e^{iqx}\int\frac{d^3k}{(2\pi)^3 2k_0} n_B(k_0/T)\langle \pi^m(k)|T \left( V^i_\mu(x) V^j_\nu(0)\right)|\pi^m(k)\rangle\nonumber\\
\label{eq:specV}
\end{eqnarray}

Let's start with the following matrix element:
\begin{equation}
M^{ij}_{\mu\nu}=\langle \pi^a(k)|T \left( V^i_\mu(x) V^j_\nu(0)\right)|\pi^b(k^\prime)\rangle
\end{equation}
Using the LSZ reduction formula one out-going pion can be extracted yielding
\begin{equation}
M^{ij}_{\mu\nu}=-i(k^2-m_\pi^2)\int d^4z e^{ikz} \langle 0|T \left( \pi^a(z) V^i_\mu(x) V^j_\nu(0) \right) |\pi^b(k^\prime)\rangle\,.
\end{equation}
Using the PCAC hypothesis,
\begin{equation}
\pi^a(z)=\frac{1}{m_\pi^2 f_\pi}\partial^\alpha A_\alpha^a(z)\,,
\end{equation}
the matrix element becomes
\begin{equation}
M^{ij}_{\mu\nu}=-i\frac{(k^2-m_\pi^2)}{m_\pi^2 f_\pi}\int d^4z e^{ikz} \langle 0|T \left( \partial^\alpha A^a_\alpha(z) V^i_\mu(x) V^j_\nu(0) \right) |\pi^b(k^\prime)\rangle\,.
\end{equation}

We will use the following form of the divergence of the time ordered product:
\begin{eqnarray}
&& \partial_z^\alpha \langle 0|T \left(A^a_\alpha(z) V^i_\mu(x) V^j_\nu(0) \right) |\pi^m(k^\prime)\rangle \nonumber\\
&=& \langle 0|T \left(\partial^\alpha A^a_\alpha(z) V^i_\mu(x) V^j_\nu(0) \right) |\pi^b(k^\prime)\rangle\nonumber\\
&+& \delta(z_0-x_0)\langle 0|T \left(\left[ A^a_0(z), V^i_\mu(x)\right] V^j_\nu(0) \right) |\pi^b(k^\prime)\rangle\nonumber\\
&+& \delta(z_0-0)\langle 0|T \left(\left[ A^a_0(z), V^j_\nu(0)\right] V^i_\mu(x) \right) |\pi^b(k^\prime)\rangle
\end{eqnarray}

Note that the term on the left hand size of the equality will vanish after integrating by parts, neglecting the surface term and finally taking the soft pion limit.  The commutators in the above expression are simplified by the current algebra relations, which hold for equal times ($x_0=y_0$):
\begin{eqnarray}
\left[ A^a_0(x), V^b_\mu(y)\right] &=& i\epsilon^{abc}A_\mu^c(x)\delta^3(x-y)\nn
\left[ A^a_0(x), A^b_\mu(y)\right] &=& i\epsilon^{abc}V_\mu^c(x)\delta^3(x-y)\nn
\left[ V^a_0(x), V^b_\mu(y)\right] &=& i\epsilon^{abc}V_\mu^c(x)\delta^3(x-y)\nn
\left[ V^a_0(x), A^b_\mu(y)\right] &=& i\epsilon^{abc}A_\mu^c(x)\delta^3(x-y)
\end{eqnarray}

Performing the integrals in the matrix element one is left with
\begin{eqnarray}
f_\pi M^{ij}_{\mu\nu}&=&\epsilon^{ail}\langle 0|T\left(A_\mu^l(x)V_\nu^j(0)\right)|\pi^b(k^\prime)\rangle e^{ikx}\nonumber\\
&=&\epsilon^{ajl}\langle 0|T\left(V_\mu^i(x)A_\nu^l(0)\right)|\pi^b(k^\prime)\rangle
\end{eqnarray}
The incoming pion in the above matrix element can be reduced in a similar fashion.  Once both pions are reduced the matrix element can be substituted back into eq.~\ref{eq:specV}
\begin{eqnarray}
\Pi^{V}_{\mu\nu}(q,T) &=& \Pi^{V}_{\mu\nu}(q,T=0) \nn 
&+& \int d^4x e^{iqx}\left(\int\frac{d^3k}{(2\pi)^3 2k_0} n_B(k_0/T)\right)\nn
&\times& \frac{4}{f_\pi^2}\left[ \langle 0|T \left( A_\mu(x) A_\nu(0)\right)|0\rangle - \langle 0|T \left( V_\mu(x) V_\nu(0)\right)|0\rangle\right] \nonumber\\
\end{eqnarray}

The result for the integral in parenthesis is
\begin{equation}
\int\frac{d^3k}{(2\pi)^3 2k_0} n_B(k_0/T) = \frac{T^2}{24}\,.
\end{equation}

The final result can be written as
\begin{equation}
\Pi^V_{\mu\nu}(q,T) = \left(1-\epsilon\right)\Pi^V_{\mu\nu}(q,T=0) + \epsilon\Pi^A_{\mu\nu}(q,T=0) 
\end{equation}

where $\epsilon=\frac{T^2}{6f_\pi^2}$.

\section*{Appendix: Collisional Broadening of the $\rho$ Meson}
\addcontentsline{toc}{section}{Appendix: Collisional Broadening of the $\rho$ Meson}
\chaptermark{Appendix}
\label{sec:coll}

In this appendix we briefly discuss how the $\rho$ meson propagator is modified at finite temperature in the resonance saturation approximation.  This procedure was carried out in \cite{Gao:1998mn,Gao:1999ux} but we include it in this appendix for completeness of this work.  It will be interesting to compare how {\em standard} collisional broadening of the rho meson compares to the spectral function approach based on the chiral reduction formula discussed in full in the previous section.  

The total width of the $\rho$ meson is the decay plus collisional width 
\begin{eqnarray}
\Gamma_{\text{tot}}=\Gamma_{\text{decay}}+\Gamma_{\text{coll}}
\end{eqnarray}

The decay width is given by the imaginary part of the rho self-energy at one loop while the collisional width is given at the two loop level.  Employing an effective Lagrangian for the $\rho-\pi$ interaction the decay width is given as \cite{Kapusta:2006pm}
\begin{eqnarray}
\Gamma_{\text{decay}}=\frac{g_{\rho\pi\pi}^2}{48\pi\omega^2}\left(\omega^2-4m_\pi^2\right)^{3/2}\left[2n_\pi\left(\frac{\omega}{2}\right)+1\right]
\end{eqnarray}

An effective Lagrangian can also be used to calculate the two loop level $\rho$ self energy.  However, the various couplings can not always be measured precisely and instead we resort to resonance saturation in order to calculate the $\pi\rho$ cross section.  The thermal collision rate is 
\begin{eqnarray}
\Gamma_{\text{coll}}=\frac{g_\pi g_\rho}{n_\rho}\int_{s_0}^{\infty}\frac{T}{2(2\pi)^4\sqrt{s}}\lambda(s,m_\pi^2,m_\rho^2)\text{K}_1(\sqrt{s}/T)\sigma_{\pi\rho}(s)
\end{eqnarray}

In the above equation $g_\pi,g_\rho$ is the spin/iso-spin degeneracy factor of the $\pi,\rho$ mesons.  $n_\rho$ is the number density of $\rho$ mesons and $\lambda(x,y,z)=x^2-2x(y+z)+(y-z)^2$ is the kinematical triangle function.  We use the standard Breit-Wigner formula for the $\pi\rho$ cross section
\begin{equation}
\sigma_{\rho\pi}=\frac{\pi}{q^2}\sum_R F_s F_i \frac{B_R\Gamma_R^2}{\left(\sqrt{s}-m_R^2\right)^2+\Gamma_R^2/4}
\end{equation}  

where 'R' refers to intermediate resonances (here we include $\phi$, $a_1$, $a_2$, $\omega^\prime$, and $\pi(1670)$ with properties taken from \cite{Eidelman:2004wy}) and $q$ is the three-momentum of the $\rho$ meson
\begin{equation}
q=\frac{1}{2\sqrt{s}}\lambda^{1/2}(s,m_\pi^2,m_\rho^2)
\end{equation} 

\begin{figure}
\centering
\includegraphics[scale=.5]{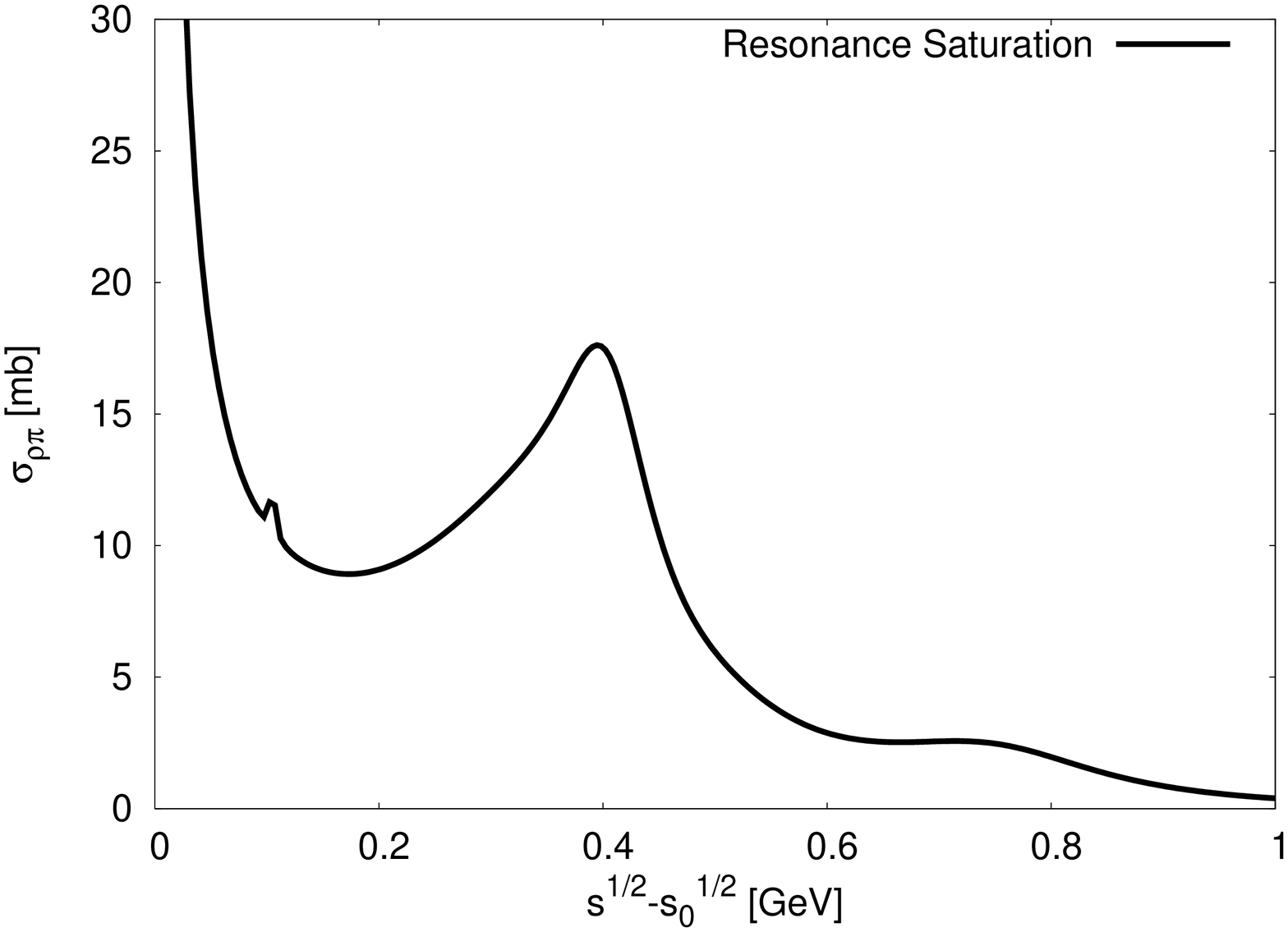}
\caption{The $\pi\rho\to\pi\rho$ cross section.}
\label{fig:cross}
\end{figure}

\chapter{Viscous Hydrodynamics}
\section{Introduction}
\subsection{Motivation}
One of the first and most exciting observations from the 
Relativistic Heavy Ion Collider (RHIC) was the very strong 
elliptic flow in non-central collisions \cite{Adams:2005dq,Adcox:2004mh}. 
The elliptic flow is quantified by the anisotropy of
particle production with respect to the 
reaction plane $v_2$
\st
      v_2 = \llangle \frac{p_x^2 - p_y^2}{p_x^2 + p_y^2} \rrangle,
\stp
and can be measured as a function of $p_T$, rapidity, centrality 
and particle type. 

The adopted interpretation of the $v_2$ measurements is that the medium
responds as a fluid to the differences in pressure gradients in the $x$
and $y$ directions. The fluid then  expands preferentially in the
reaction plane and establishes the observed momentum space anisotropy.
This hydrodynamic interpretation is supported by the qualified success
of ideal hydrodynamic models in describing a large variety of data over
a range of colliding systems and energies \cite{Hirano:2004en,Teaney:2001av,Kolb:2000fh,Huovinen:2001cy,Nonaka:2006yn}.  Nevertheless, the
hydrodynamic interpretation of the flow results is not unassailable.  A
back of the envelope estimate of viscous corrections to hydrodynamic
results \cite{Danielewicz:1984ww} suggests that viscous corrections are actually rather
large, {\em i.e.,} the mean free path is comparable to the system size
\cite{Drescher:2007cd}.
 These estimates are best conveyed in terms of the shear
viscosity to entropy ratio, $\eta/s$.  The conditions for partial
equilibrium at RHIC are so unfavorable that at unless $\eta/s$ is small
(say $0.5$ or less), it is difficult to imagine that the medium would
participate in a coordinated  collective flow.

From a theoretical perspective, it is difficult to  reliably estimate
$\eta/s$ in the vicinity of the QCD phase transition where the system
is strongly coupled.  Lattice QCD measurements of transport are hard
(perhaps impossible \cite{Petreczky:2005nh,Aarts:2002cc}) though recent efforts have lead to estimates which
are not incompatible with the hydrodynamic interpretation of RHIC
results \cite{Aarts:2007wj,Meyer:2007ic}.  In a strict perturbative setting (where the
quasi-particle picture is exact) $\eta/s$ is large $\sim 1/g^4$.
Nevertheless an extrapolation of weak coupling results to moderate
coupling also leads to an $\eta/s$ which is perhaps reconcilable with
the hydrodynamic interpretation \cite{Arnold:2003zc,Baymetal}.  
Finally, these perturbative estimates
should be contrasted with $\N=4$ Super Yang Mills at strong coupling,
where $\eta/s$ is $1/4\pi$ \cite{Policastro:2001yc,Kovtun:2004de}.  Although $\N=4$  SYM is not QCD,  the
calculation was important because it showed that there is at least one theory where $\eta/s$ is
sufficiently small that collective phenomena would be observed under
conditions similar to those produced at RHIC.

From a phenomenological perspective one of the most compelling evidences
for the hydrodynamic interpretation of RHIC flow results is the fact
that the deviations  from hydrodynamics are qualitatively reproduced by
kinetic theory \cite{Molnar:2001ux,Xu:2004mz}. In particular, kinetic theory calculations generically
reproduce the flattening of $v_2(p_T)$ at higher $p_T$, and the
reduction of elliptic flow at large impact parameters.  Some aspects of
these kinetic theory results can be understood by considering the first
viscous corrections to the thermal distribution function \cite{Teaney:2003kp}. These
estimates  motivated full viscous hydrodynamic simulations of the
elliptic flow which will be performed in this work.  Recently such
viscous simulations have been performed by two other groups \cite{Heinzv2,RRv2}.  A brief
discussion of the history surrounding viscous relativistic
hydrodynamics is given below.

\subsection{Viscous Hydrodynamics}
\label{viscoushydro}

The Navier-Stokes equations describe viscous corrections to ideal fluid
flow by keeping  terms up to first order in gradients of ideal quantities \cite{LL}.
%From the kinetic point of view one expands the collision term using the
%Chapman-Enskog procedure ({\em i.e.} a systematic expansion in
%$\lambda_{mfp}/L$) and keeps terms up to first order.  The relativistic
%generalization of the Navier-Stokes was first done by Eckart
%\cite{eckart} and Landau and Lifshitz \cite{LL}.   
The resulting equations are parabolic which permit acausal signal propagation \cite{Lindblom:1985}.  For instance, the stress tensor
instantaneously adjusts to any thermodynamic force, $\partial_iu_j$.  
This is, of
course, an unphysical picture  since the stress tensor should relax 
to the  thermodynamic forces over a typical collision timescale.

One would therefore like a phenomenological theory that explains this
relaxation correctly.  Much work has been done in this direction but
there is still no completely satisfactory theory.  
Probably the most used model is that of 
Israel and Stewart \cite{IS,Israel:1979wp}, but there are also others by Lindblom and
Geroch \cite{LG}, Pav\'on, Jou and Casas-V\'asquez \cite{PJC} and also
by \"{O}ttinger and Grmela \cite{OG,Ottinger} which is used in this work. In fact a wide class of models was developed by Lindblom and Geroch in two separate papers \cite{LG,LG2} 
% All of the
%above theories (except the one by \"{O}ttinger and Grmela) are
%restricted to small deviations from equilibrium.

All of the above theories have the same behavior: they relax on small
time scales to the first-order relativistic Navier-Stokes equations
and have some generalized entropy which increases as a function of
time.
It was shown by Lindblom \cite{Lindblom} that for a large class 
class of these second order theories, the physical fields should be
indistinguishable from the simple Navier-Stokes form.  To paraphrase
Lindblom; any measurement of the stress energy tensor or particle
current on a time scale larger than the microscopic time scale will be
indistinguishable from the Navier-Stokes theory.  
The differences between the causal theories 
and the acausal Navier-Stokes equations 
are indicative of the corrections quantitatively captured by the full kinetic theory.
Nevertheless, the causal theories provide
a qualitative guide to the magnitude of these corrections 
 \cite{Forster}.  However, the form of these corrections implicitly 
assumes a good quasi-particle description which may not exist in
a strongly coupled plasma \cite{Teaney:2006nc}.

\section{The Hydrodynamic Model}

In the following section we outline the equations of motion for the
hydrodynamical model used in the following simulations.  We start by
summarizing the well known first-order Navier-Stokes theory.  Then we
outline the equations required for a second-order causal description of
dissipative fluid dynamics.  This is done assuming a boost invariant
expansion as first proposed by Bjorken \cite{Bjorken:1982qr}, where the equations
of motion are expressed in terms of the proper time
$\tau=\sqrt{t^2-z^2}$ and the spatial rapidity
$\eta_s=\frac{1}{2}\text{ln}\frac{t+z}{t-z}$. The cartesian coordinate z
denotes the position along the beam axis while x,y label positions
transverse to the beam axis.

\subsection{1$^{st}$ Order Viscous Hydrodynamics - Navier Stokes}

Viscous hydrodynamics was originally formulated in the first-order
Navier-Stokes approximation where the energy momentum tensor and baryon
flux is a sum of their ideal and dissipative parts:
\bg
T^{\mu\nu} &=& \epsilon u^{\mu} u^{\nu} + (p + \Pi) \Delta^{\mu\nu} + \pi^{\mu\nu}\, , \\
n^\mu&=&n u^\mu+j_d^\mu\, ,
\nd
where $p, \epsilon, n$ and $u^\mu=(\gamma,\gamma{\bf \text{v}})$ are
the pressure, energy density, baryon density and four-velocity of the
fluid.  We use the convention that \newline $g^{\mu\nu}=\text{diag}(-1,+1,+1,+1)$ and therefore $u^\mu u_\mu=-1$.
The dissipative terms, $\pi$ and $j_d$ depend on the definition of the
local rest frame (LRF) of the fluid.  A specific form of $\pi^{\mu\nu}$
and $v^\mu$ can be found using the Landau-Lifshitz definition \cite{LL}
of the LRF ($u_\mu \pi^{\mu\nu}=0$), constraining the the entropy to
increase with time and by working within the Navier-Stokes
approximation (keeping terms to first order in gradients only) resulting in
\bg
\pi^{\mu\nu}&=&-\eta(\nabla^\mu u^\nu+\nabla^\nu u^\mu-\frac{2}{3}\Delta^{\mu\nu}\nabla_\beta u^\beta)\, , \\
\Pi&=&-\zeta \nabla_\beta u^\beta\, , \\
j_d^\mu&=&-\kappa(\frac{nT}{\epsilon+p})^2\nabla^\mu(\frac{\mu}{T})\, ,
\label{eq:pimunu}
\nd
where $\kappa, \eta$ and $\zeta$ are the heat conduction, shear and bulk
viscosities of the fluid with temperature $T$ and chemical potential
$\mu$.  The viscous tensor is constructed with the differential
operator $\nabla^\mu=\Delta^{\mu\nu}d_\nu$ where
$\Delta^{\mu\nu}=g^{\mu\nu}+u^\mu u^\nu$ is the local three-frame
projector, $d_\mu u^\nu=\partial_\mu
u^\nu+\Gamma^\nu_{\gamma\mu}u^\gamma$ is the covariant derivative and $\Gamma^\nu_{\gamma\mu}\equiv 1/2 g^{\nu\alpha}(\partial_\mu g_{\alpha\gamma}+\partial_\gamma g_{\alpha\mu}-\partial_\alpha g_{\gamma\mu})$ are the Christoffel symbols.   

The transport coefficients in a quark-gluon plasma and also in the
hadronic gas were studied in Refs.~\cite{Danielewicz:1984ww,Prakash:1993bt,Baymetal,Arnold:2003zc}.  It was found that the dominate
dissipative mechanism was shear viscosity in both the QGP and hadronic
gas. Bulk viscosity may however dominate in the transition region 
\cite{Kharzeev:2007wb}.  
Heat transport can be ignored in the limit that $\mu_B\ll T$ which
is the limit taken here.  
   
In the following work we will consider viscous effects in a quark-gluon
plasma phase only.  
For this purpose we  
consider a constant shear to entropy ratio, $\eta/s=\mbox{const}$ 
and a massless gas $p=1/3 \epsilon$. Future work will discuss viscosity in the mixed
and hadronic phases.  From this point on we will neglect
the thermal conductivity.  We keep the bulk viscosity in the
equations for consistency, but always set $\zeta=0$ in any
calculations. 

\subsection{2$^{nd}$ order Viscous Hydrodynamics}
\label{GenericSect}

In order to render a second order theory it is necessary to introduce
additional  variables.  These variables will relax on very short time
scales to the standard thermodynamic quantities in the first order
theory, but an evolution equation for them is still required in order
to avoid acausal signal propagation.  One such theory that has been
used in a number of works was introduced by Israel and Stewart
\cite{IS}.  Instead we use a theory developed by \cite{OG,Ottinger} due
to its appealing structure when implemented numerically.  However, as discussed above, all
of these theories should agree ({\em i.e.,} they all relax on short
time scales to the same the first-order equations).  

We now summarize the evolution equations used in the current analysis
following the mathematical structure outlined in Ref.~\cite{Ottinger}.  We use a
simplified version of the model for deviations of the stress energy
tensor close to equilibrium.  The new dynamical variable that is
introduced is the tensor variable $c_{\mu\nu}$ which will later be
shown to be closely related to the velocity gradient tensor,
$\pi_{\mu\nu}$.   The tensor variable $c_{\mu\nu}$ is conveniently
defined to have the property
\bg
c_{\mu\nu}u^\nu=u_\mu\, ,
\nd 
and the energy momentum tensor is given by
\bg
T^{\mu\nu}=(\epsilon-u_\alpha \mathbb{P}^{\alpha\beta}u_\beta)u^\mu u^\nu+\mathbb{P}^{\mu\nu}\, .
\label{eq:Tmunu}
\nd
The explicit form of the stress tensor $\mathbb{P}^{\mu\nu}$ is given
in \cite{Ottinger} and has a fairly complicated form.  The discussion
is simplified by considering small deviations from local thermal
equilibrium and working in the local rest frame so that the stress tensor
can be approximated as
\bg
T^{ij}_{LRF}=p(\delta^{ij}-\alpha c^{ij})\, ,
\nd
where $\alpha$ is a small parameter related to the relaxation time.  The equations of motion are dictated by
conservation of energy and momentum which is given by $d_\mu T^{\mu\nu}=0$.
In addition, an evolution equation for the generalized mechanical force
tensor is needed and is given by \cite{Ottinger}
\bg
u^\lambda(\partial_\lambda c_{\mu\nu}-\partial_\mu c_{\lambda\nu}-\partial_\nu c_{\mu\lambda})=\frac{-1}{\tau_0}\overline{c}_{\mu\nu}-\frac{1}{\tau_2}\mathring{c}_{\mu\nu}\, ,
\label{eq:cevol}
\nd
where $\overline{c}$ and $\mathring{c}$ are defined as the isotropic and traceless parts of the tensor variable $c_{\mu\nu}$ defined as
\bg
\overline{c}_{\mu\nu}=\frac{1}{3}(c^{\lambda}_\lambda-1)(\eta_{\mu\nu}+u_\mu u_\nu)\, , \\
c_{\mu\nu}+u_\mu u_\nu=\mathring{c}_{\mu\nu}+\overline{c}_{\mu\nu}\, .
\nd

In the limit that the relaxation times ($\tau_0, \tau_2$) are very small the evolution equation yields
\bg
c^{ij}=\tau_2( \partial_i u^j+\partial_j u^i-\frac{2}{3}\delta^{ij}\partial_k u^k)+\frac{2}{3}\tau_0\delta^{ij}\partial_k u^k\, .
\label{cijequ}
\nd
Substituting the above equation into $T^{ij}_{LRF}$ and comparing the result to the Navier-Stokes equation (\ref{eq:pimunu}) the bulk and shear viscosities can be identified as
\bg
\eta=\tau_2 p \alpha \nonumber\, , \\
\zeta=\frac{2}{3}\tau_0 p \alpha\, .
\nd
In the model proposed by \"{O}ttinger \cite{Ottinger}  
the quantity $\alpha$ is related
to the equation of state,  but in the
linearized version it is simply treated as a constant parameter related
to the relaxation time. 
We fix $\alpha=0.7$ in all calculations, which
then fixes the relaxation times ($\tau_2, \tau_0$) as a function of
$\eta$ and $\zeta$.  

It is natural to ask what is the effect of the relaxation time on the
theory.  In some sense this was already answered by Lindblom
\cite{Lindblom}.  
He showed that
the physical fluid must relax to a state that is indistinguishable from the
Navier-Stokes form.  
Therefore we expect the
physical velocity gradients to agree with those given by the auxiliary
tensor variable $c^{\mu\nu}$ as in \Eq{cijequ} .  
We expect higher order
gradient terms to be necessary when there are large deviations between
any observable computed using the physical fields or the auxiliary
field $c^{\mu\nu}$.  This will be used as a gauge in order to find the
limit of applicability of any hydrodynamic calculations.

\subsubsection{1+1 Dimensions}

We now outline the equations of motion for the stress-energy tensor and
the generalized mechanical force tensor assuming a boost-invariant
expansion as well as azimuthal symmetry with arbitrary transverse
expansion.  It is easiest to work in polar coordinates
$(\tau,r,\phi,\eta)$ and since there is no dependence on $\phi$ or
$\eta$ the four-velocity can be expressed as $u^\mu=(\gamma,\gamma
v_r,0,0)$ where $\gamma=\frac{1}{\sqrt{1-v_r^2}}$.  In this coordinate
system the metric tensor is given by
$g^{\mu\nu}=\text{diag}(-1,1,1/r^2,1/\tau^2)$.

The first two equations of motion are given by the conservation of energy and momentum, $d_\mu T^{\mu\nu}=0$ for $\nu=\tau$ and $\nu=r$.  (Due to boost invariance and azimuthal symmetry the $\nu=\eta$ and $\nu=\phi$ equations are trivial.)
\bg
\partial_\tau T^{00}+\partial_r T^{01}=\frac{-1}{\tau}( T^{00}+\Tilde{P}^{33})-\frac{1}{r} T^{01}  \\
\partial_\tau T^{01}+\partial_r T^{11}=\frac{-1}{\tau} T^{01} -\frac{1}{r}( T^{11}-\Tilde{P}^{22}) 
\label{eq:eqom1dT}
\nd
where $\Tilde{P}^{22}=r^2 P^{22}$ and $\Tilde{P}^{33}=\tau^2 P^{33}$.  The evolution equations for the generic mechanical force tensor $c^{\mu\nu}$ are: 
\bg
\partial_\tau c^{11}+v\partial_r c^{11}-\frac{2}{\gamma}[(1-c^{11})\partial_r u^1+c^{01}\partial_r u^0]=\frac{-1}{\gamma\tau_0}\overline{c}^{11}-\frac{1}{\gamma\tau_2}\mathring{c}^{11}   \nn
\partial_\tau \Tilde{c}^{22}+v\partial_r \Tilde{c}^{22}+\frac{2v}{r}(\Tilde{c}^{22}-c^{11})+\frac{2}{r}c^{10}=\frac{-1}{\gamma\tau_0}\overline{\Tilde{c}}^{22}-\frac{1}{\gamma\tau_2}\mathring{\Tilde{c}}^{22}   \nn
\partial_\tau \Tilde{c}^{33}+v\partial_r \Tilde{c}^{33}+\frac{2}{\tau}(\Tilde{c}^{33}+c^{00})-\frac{2v}{\tau}c^{10}=\frac{-1}{\gamma\tau_0}\overline{\Tilde{c}}^{33}-\frac{1}{\gamma\tau_2}\mathring{\Tilde{c}}^{33} \nn
\label{eq:eqom1dc}
\nd
where $\Tilde{c}^{22}=r^2 c^{22}$ and $\Tilde{c}^{33}=\tau^2 c^{33}$.

\subsubsection{1+2 Dimensions}

We now consider the 1+2 dimensional case without azimuthal symmetry but
still having longitudinal boost invariance and use a coordinate system
whereby the coordinates transverse to the beam axis are cartesian,
$(\tau,x,y,\eta)$.  Since there is no dependence on $\eta$ the
four-velocity can be expressed as $u^\mu=\gamma(1, v_x,v_y,0)$ where
$\gamma=\frac{1}{\sqrt{1-v_x^2-v_y^2}}$.  In this coordinate system the
metric tensor is given by $g^{\mu\nu}=\text{diag}(-1,1,1,1/\tau^2)$.

In this coordinate system the first three equations of motion are given by the $\nu=\tau$, $x$, and $y$ components of the conservation law $d_\mu T^{\mu\nu}=0$:
\bg
\label{eq:2dstress}
\partial_\tau T^{00}+\partial_x T^{01}+\partial_y T^{02}=\frac{-1}{\tau}( T^{00}+\tau^2 P^{33} )  \\
\partial_\tau T^{10}+\partial_x T^{11}+\partial_y T^{12}=\frac{-1}{\tau}T^{10}\\
\partial_\tau T^{20}+\partial_x T^{21}+\partial_y T^{22}=\frac{-1}{\tau} T^{20}
\nd
The evolution equations for the generalized mechanical force tensor are:
\bg
(\partial_\tau+v_x\partial_x +v_y\partial_y)c^{11}+2[(c^{11}-1)\partial_x v_x+c^{12}\partial_x v_y]=\frac{-1}{\gamma\tau_0}\overline{c}^{11}-\frac{1}{\gamma\tau_2}\mathring{c}^{11}   \nn
(\partial_\tau+v_x\partial_x +v_y\partial_y)c^{22}+2[(c^{22}-1)\partial_y v_y+c^{21}\partial_y v_x]=\frac{-1}{\gamma\tau_0}\overline{c}^{22}-\frac{1}{\gamma\tau_2}\mathring{c}^{22}   \nn
(\partial_\tau+v_x\partial_x +v_y\partial_y)\Tilde{c}^{33}+\frac{2}{\tau}(\Tilde{c}^{33}-1)=\frac{-1}{\gamma\tau_0}\overline{\Tilde{c}}^{33}-\frac{1}{\gamma\tau_2}\mathring{\Tilde{c}}^{33} \nn
(\partial_\tau+v_x\partial_x +v_y\partial_y)c^{12}+c^{12}(\partial_x v_x + \partial_y v_y)+(c^{22}-1)\partial_x v_y+(c^{11}-1)\partial_y v_x\nonumber\\=\frac{-1}{\gamma\tau_0}\overline{c}^{12}-\frac{1}{\gamma\tau_2}\mathring{c}^{12}  \nn
\label{eq:2dcs}
\nd

\subsubsection{Initial Conditions}
\label{sec:IC}

The hydrodynamic simulation is a $2+1$ dimensional boost invariant model with an ideal gas equation
of state $p = \frac{1}{3}\,\epsilon$. 
The temperature is
related to the energy density with the $N_f=3$ ideal
QGP equation of state.  
We have chosen this
extreme equation of state because the resulting radial
and elliptic flow are too large relative to data on
light hadron production. Thus,
this equation of state will estimate the largest elliptic
flow possible for a given shear viscosity.  We note that for any non-central collision we have chosen a default impact parameter of b=6.5 fm.   

Aside from the equation of state, the hydrodynamic model is based upon
reference \cite{Teaney:2001av}.
At an initial time $\tau_0=1$ fm/c,
the entropy is distributed in the transverse plane
according to the distribution of participants
for a Au-Au collision.
Then one parameter, $C_s$,
is  adjusted to set the initial temperature and total particle yield.
Specifically the initial entropy density in the transverse plane is
\st
    s(x, y,\tau_0) = \frac{C_{s}}{\tau_0}\,\frac{dN_{p}}{dx\,dy},
\stp
where $\frac{dN_p}{dx\,dy}$ is the number of participants per
unit area.
The value $C_{s}=15$ closely corresponds to the
results of full hydrodynamic simulations \cite{Teaney:2001av,Kolb:2000fh,Huovinen:2001cy} and  corresponds to a maximum initial
temperature of $T_{0} = 420\,\mbox{MeV}$ at impact
parameter $b=0$. 
With the entropy density specified the energy density can be 
determined. This requires inverting the equation of 
state.

In a viscous formulation we must also specify the viscous 
fields, $i.e.$ the  $c^{\mu\nu}$ in the second order setup.
Following the general philosophy outlined in \Sect{viscoushydro} we
will choose the $c^{\mu\nu}$ such that the stress tensor
deviations are
\st
 \pi_{\mu\nu} = -\eta\llangle \nabla_{\mu} u_{\nu} \rrangle  
\qquad  \Pi  = -\zeta \nabla_{\mu} u^{\mu} = 0 
\stp
Since at time $\tau_o$ the transverse flow velocity and the longitudinal
flow velocity is Bjorken this means that
at mid rapidity  
\st
\pi_{xx} = \pi_{yy} =  -\frac{1}{2} \,\pi_{zz}= \frac{2}{3}\, \eta\, \partial_z u^{z}  \qquad \Pi = 0
\stp

To achieve this condition we first rewrite the flow equations 
for small $c_{\mu\nu}$ and vanishing transverse flow. The 
$c_{ij}$ equations become
\bg
   \partial_\tau c^{11}  &=&  
- \frac{\bar{c}^{11}}{\tau_0}
- \frac{\mathring{c}^{11}}{\tau_2}\, , \\
   \partial_\tau c^{22}  &=&  
- \frac{\bar{c}^{22}}{\tau_0}
- \frac{\mathring{c}^{22}}{\tau_2}\, , \\
   \partial_\tau c^{33}  - \frac{2}{\tau} &=&  
- \frac{\bar{c}^{33}}{\tau_0}
- \frac{\mathring{c}^{33}}{\tau_2}\, . 
\nd
In writing this we have used the fact that for small velocity
$c^{00} \approx  -u^{0} u^{0}$ .
Then looking for the quasi stationary state we set the time derivatives
to zero, and  use the relations $\bar{c}^{ij} = \frac{1}{3} c^l_l\,\delta^{ij}$ and 
$c^{ij} = \mathring{c}^{ij} + \bar{c}^{ij}$ to find that 
\bg
  c^{11}&=& \frac{2}{3} \frac{\tau_0}{\tau} - \frac{2}{3} \frac{\tau_2}{\tau}\, , \\
  c^{22}&=& \frac{2}{3} \frac{\tau_0}{\tau} - \frac{2}{3} \frac{\tau_2}{\tau}\, ,\\
  c^{33}&=& \frac{2}{3} \frac{\tau_0}{\tau} + \frac{4}{3} \frac{\tau_2}{\tau}\, .
\nd

\section{Hydrodynamic Results}
\label{sec:HydroResults}
The equations outlined in the previous two sections were integrated
numerically using the initial conditions described above.  In this section we now show the results of the
simulation.  Before showing the results of the  2+1 dimensional
simulation we outline some of the main physics points using results
from the 1+1 dimensional case.

\begin{figure}[hbtp]
\centering
\includegraphics[scale=0.75]{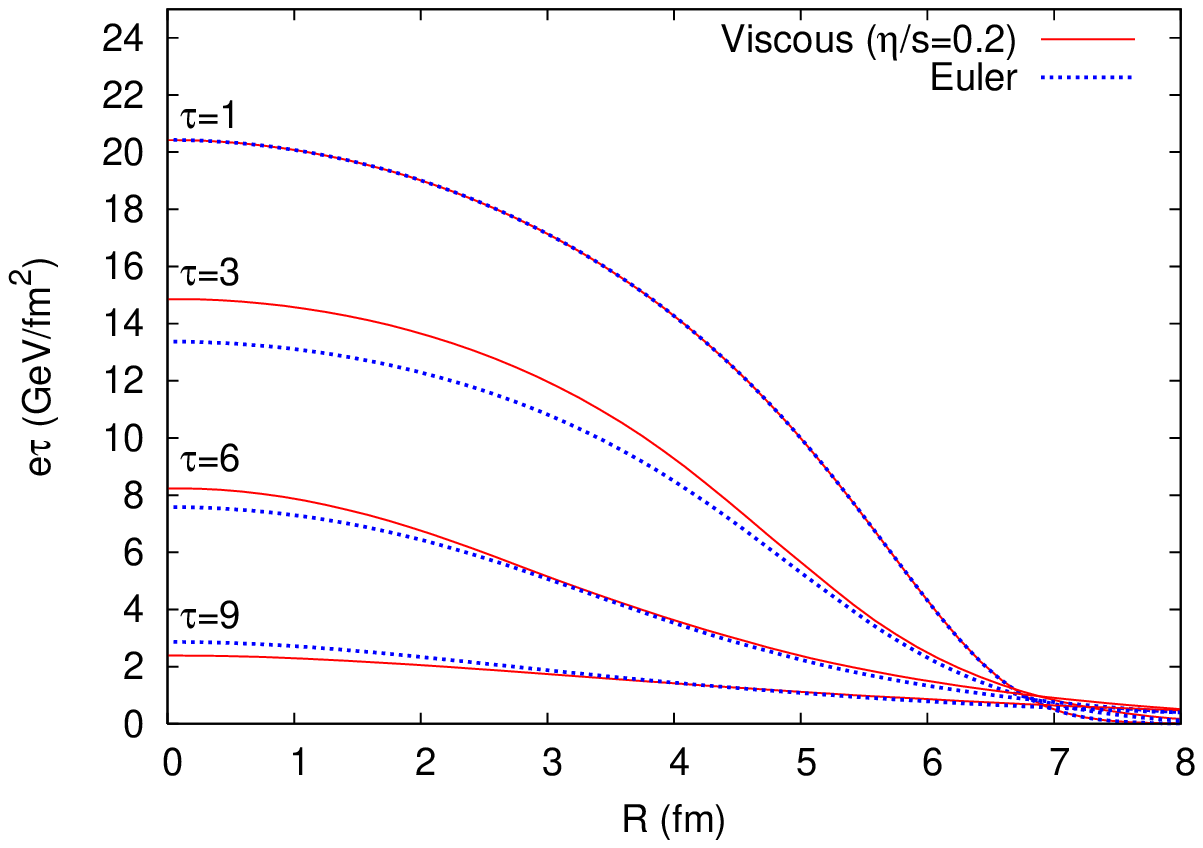}
\vspace{9pt}
\includegraphics[scale=0.75]{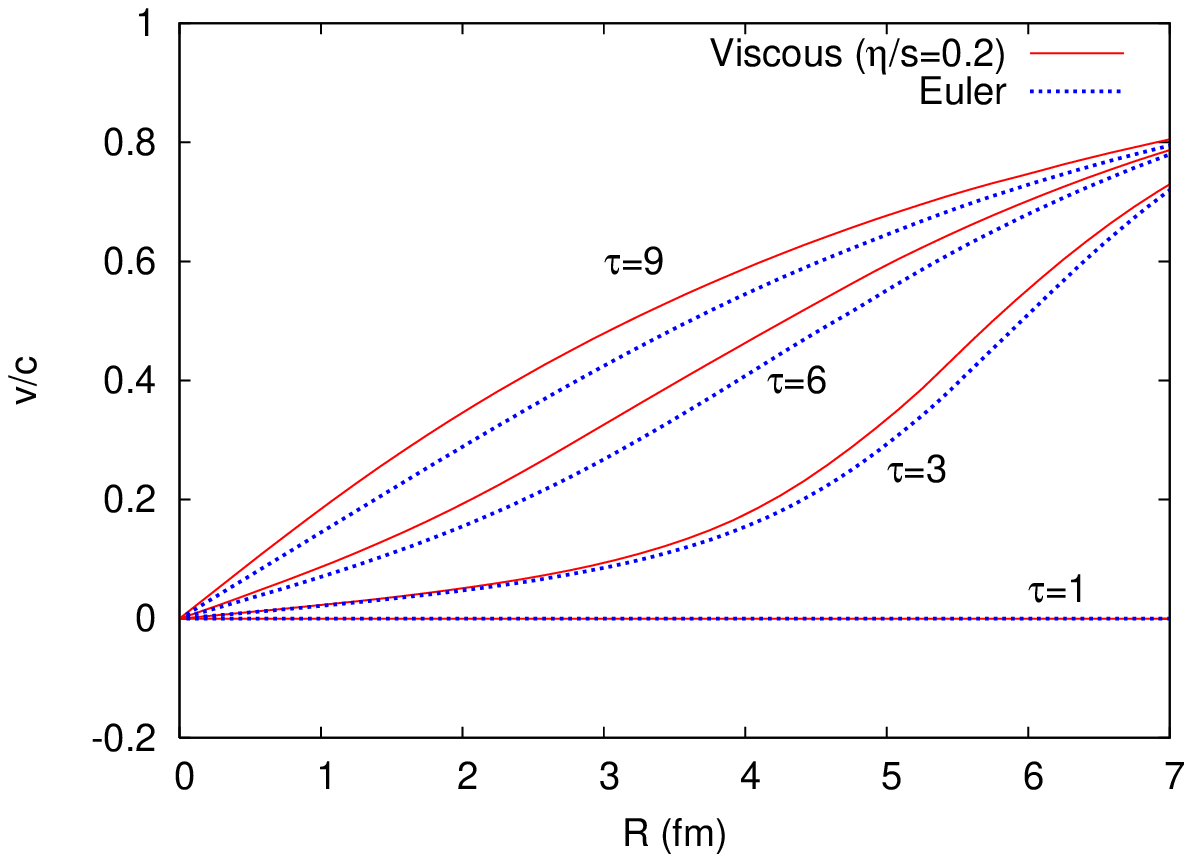}
\caption{(Color online) Plot of the energy density per unit rapidity (top) and of the
transverse velocity (bottom) at times of $\tau=1,3,6,9$ fm/c, for
$\eta/s=0.2$ (solid red line) and for ideal hydrodynamics (dotted blue line).}
\label{fig:v1d}
\end{figure}

Fig.~\ref{fig:v1d} shows the energy density per unit rapidity (upper)
and the transverse velocity (bottom) at various times for both ideal
hydrodynamics and for finite viscosity ($\eta/s=0.2$).  
The effect of viscosity is twofold. The longitudinal 
pressure is initially reduced and the viscous case does less longitudinal $pdV$ work as in the simple Bjorken expansion \cite{Danielewicz:1984ww}.  This means that at early times the energy per rapidity decreases
more slowly in the viscous case.
The reduction of longitudinal pressure is 
accompanied by a larger transverse pressure.
This causes the transverse velocity
to grow more rapidly.  The larger transverse velocity causes the energy
density to deplete faster at later times in the viscous case.  The net result is that a
finite viscosity (even as large as $\eta/s=0.2$) does not integrate to
give major deviations from the ideal equations of motion. A preliminary account of this effect was given long ago \cite{DT1}.

\begin{figure}[hbtp]
\centering
  \vspace{9pt}
  \centerline{\hbox{ \hspace{0.0in} 
\includegraphics[scale=1.1]{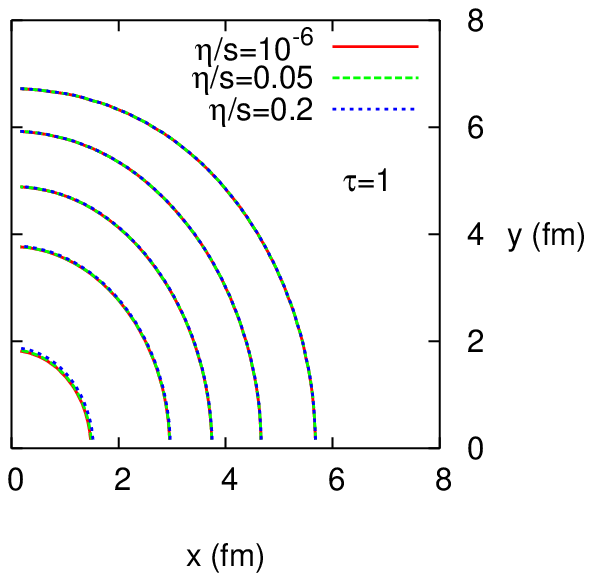}
    \hspace{0.1in}
\includegraphics[scale=1.1]{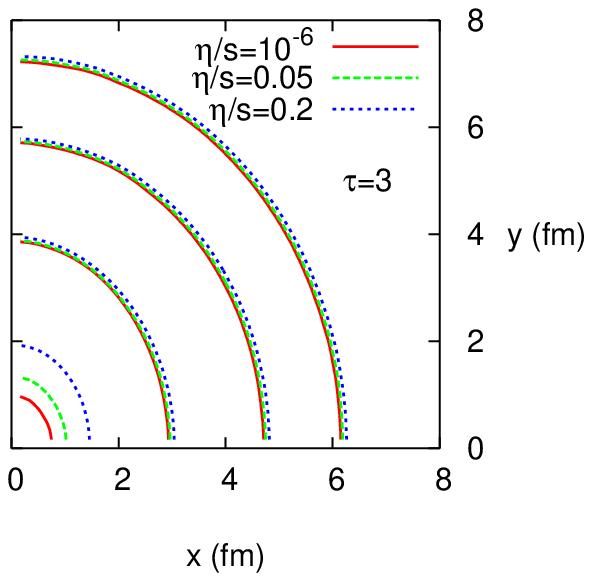}
    }
  }
  \vspace{9pt}
  \centerline{\hbox{ \hspace{0.0in}
\includegraphics[scale=1.1]{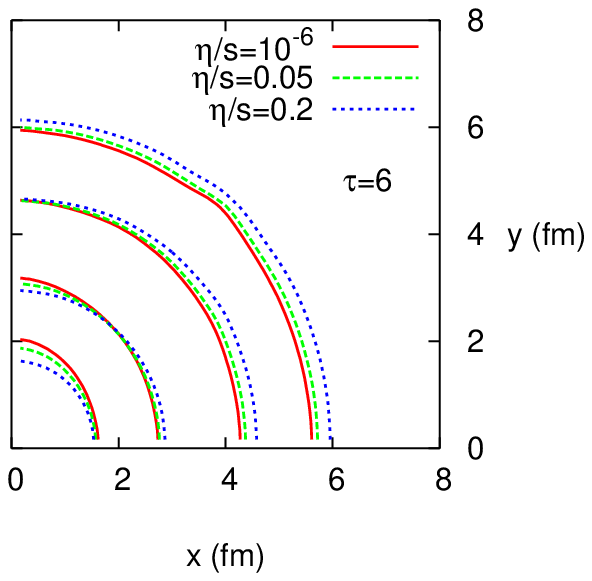}
    \hspace{0.1in}
\includegraphics[scale=1.1]{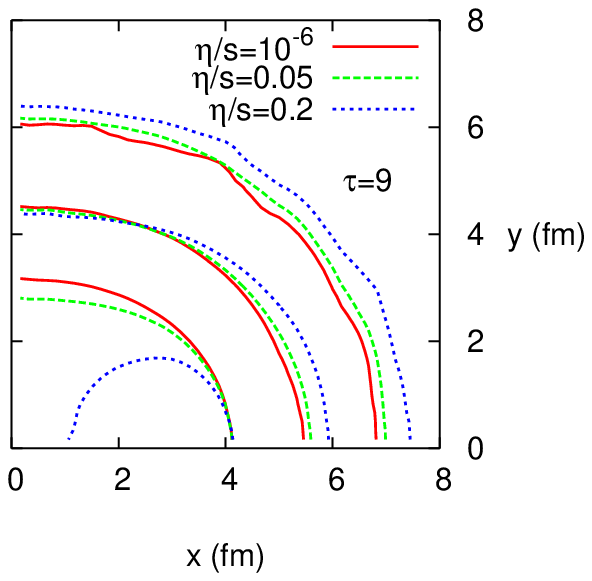}
    }
  }
  \vspace{9pt}
  \caption{(Color online) Contour plot of energy density per unit
rapidity in the transverse plane.  The contour values working outward
are for $\tau=1$ fm/c: 15, 10, 5, 1, 0.1, for $\tau=3$ fm/c: 10, 5, 1, 0.1, for $\tau=6$ fm/c: 3,
2, 1, 0.5 and for $\tau=9$ fm/c: 0.5, 0.375, 0.25, in units of GeV/fm$^2$.  }
  \label{fig:enedens}
\end{figure}

We now present results of the 2+1 dimensional boost invariant
hydrodynamic model.  Fig.~\ref{fig:enedens} shows contour plots of the
energy density per unit rapidity in the transverse plane at proper
times of $\tau=1, 3, 6, 9$ fm/c.  The initial conditions ($\tau=1$) is taken
from the Glauber model discussed before.

\begin{figure}[hbtp]
\centering
  \vspace{9pt}
  \centerline{\hbox{ \hspace{0.0in} 
\includegraphics[scale=1.1]{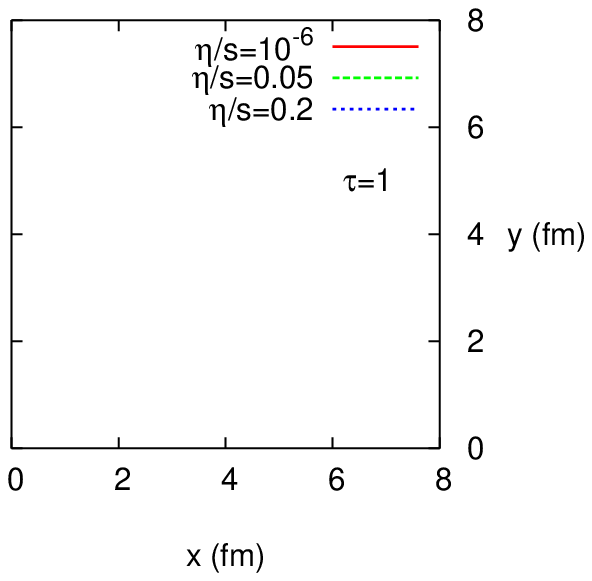}
    \hspace{0.1in}
\includegraphics[scale=1.1]{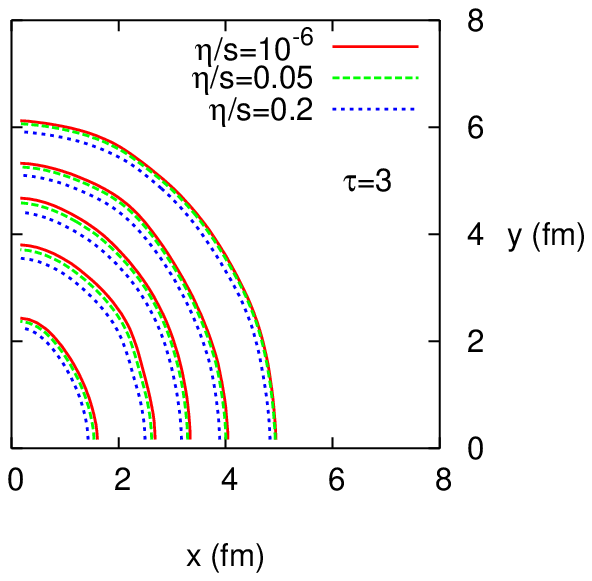}
    }
  }
  \vspace{9pt}
  \centerline{\hbox{ \hspace{0.0in}
\includegraphics[scale=1.1]{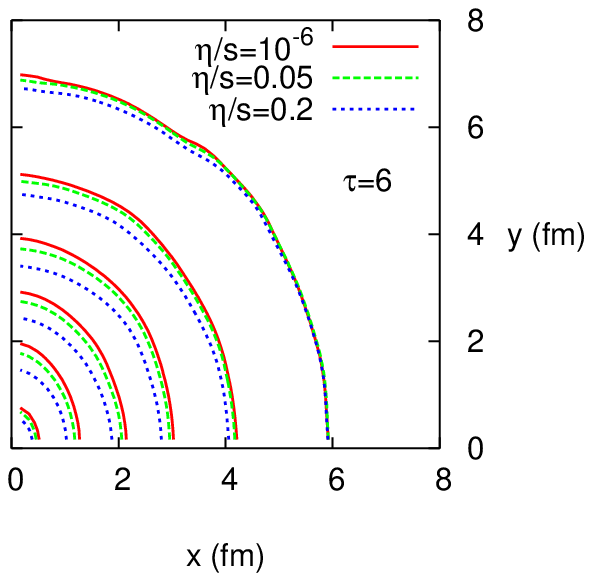}
    \hspace{0.1in}
\includegraphics[scale=1.1]{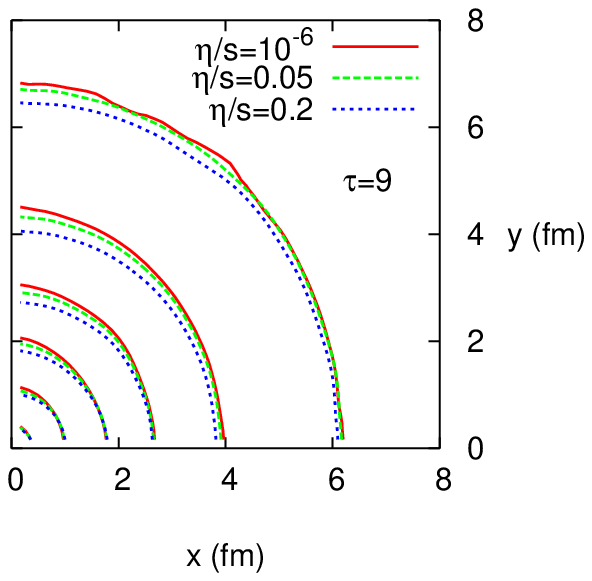}
    }
  }
  \vspace{9pt}
  \caption{(Color online) Contour plot of transverse velocity, $v_\perp=\sqrt{v_x^2+v_y^2}$.  The inner most contour is for $v_\perp=0.1$ and increases in steps of $\Delta v_\perp = 0.15$. }
  \label{fig:vt}
\end{figure}

Fig.~\ref{fig:vt} shows contour plots of the transverse velocity at the
same times of $\tau=$1, 3, 6, 9 fm/c.  At $\tau=1$ the figure is blank since
the velocity in the transverse plane is zero as set by the initial
conditions.  By looking at the contours of constant $v/c$ one can see
that a finite viscosity increases the transverse velocity.

\begin{figure}[hbtp]  
\centering
\includegraphics[scale=.7]{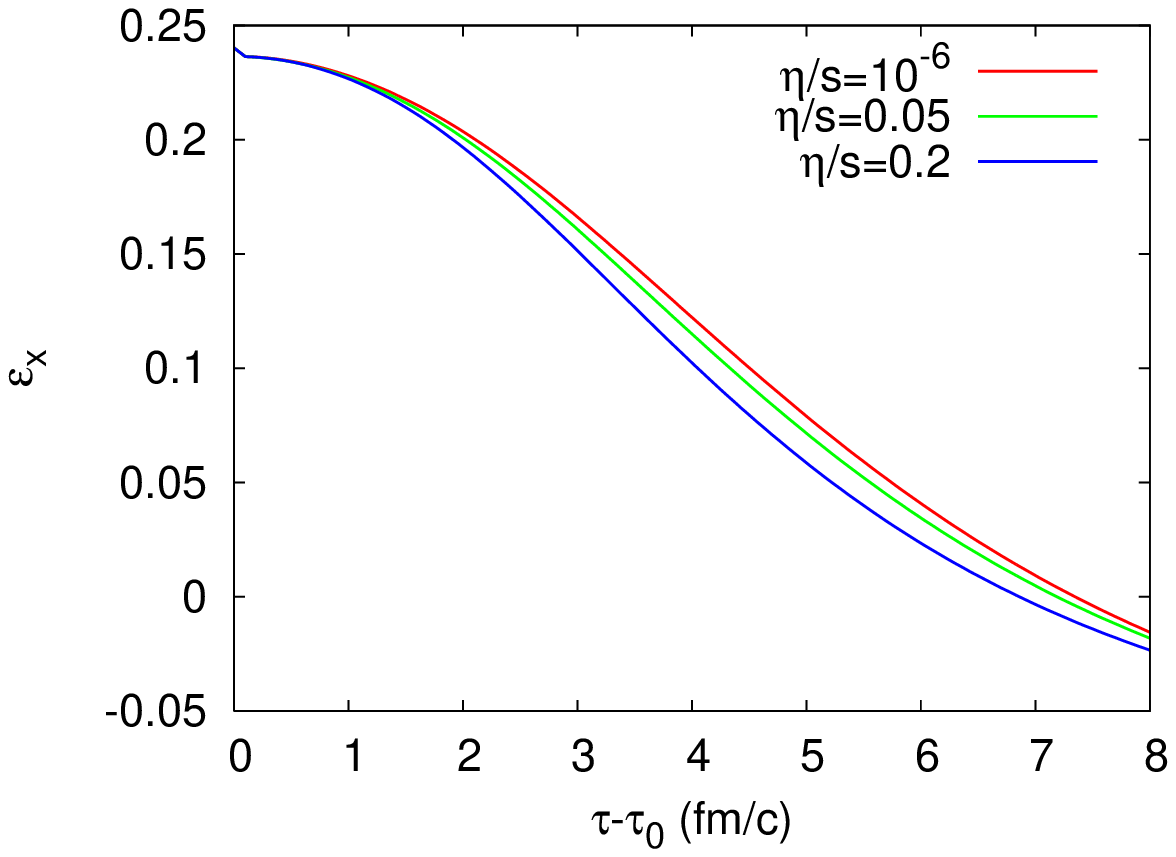}
 \vspace{9pt}
\includegraphics[scale=.7]{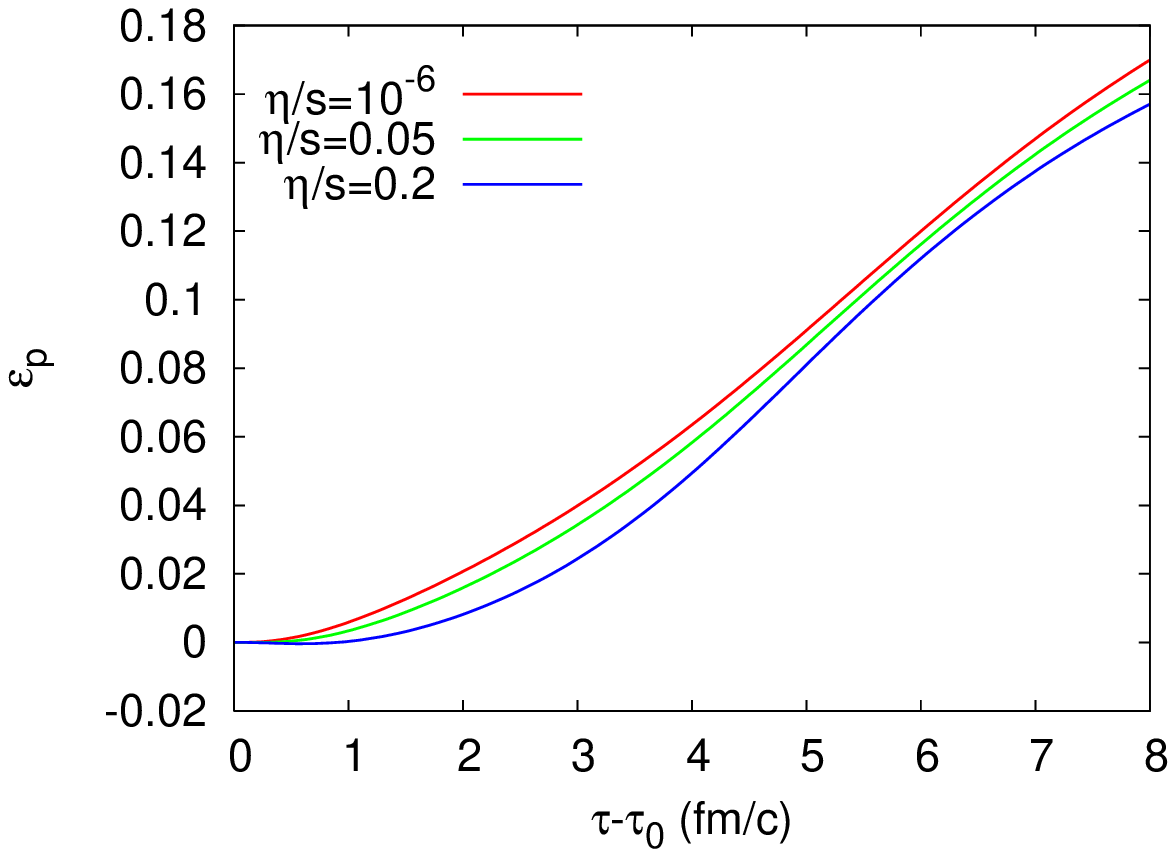}
\vspace{9pt}
\includegraphics[scale=.7]{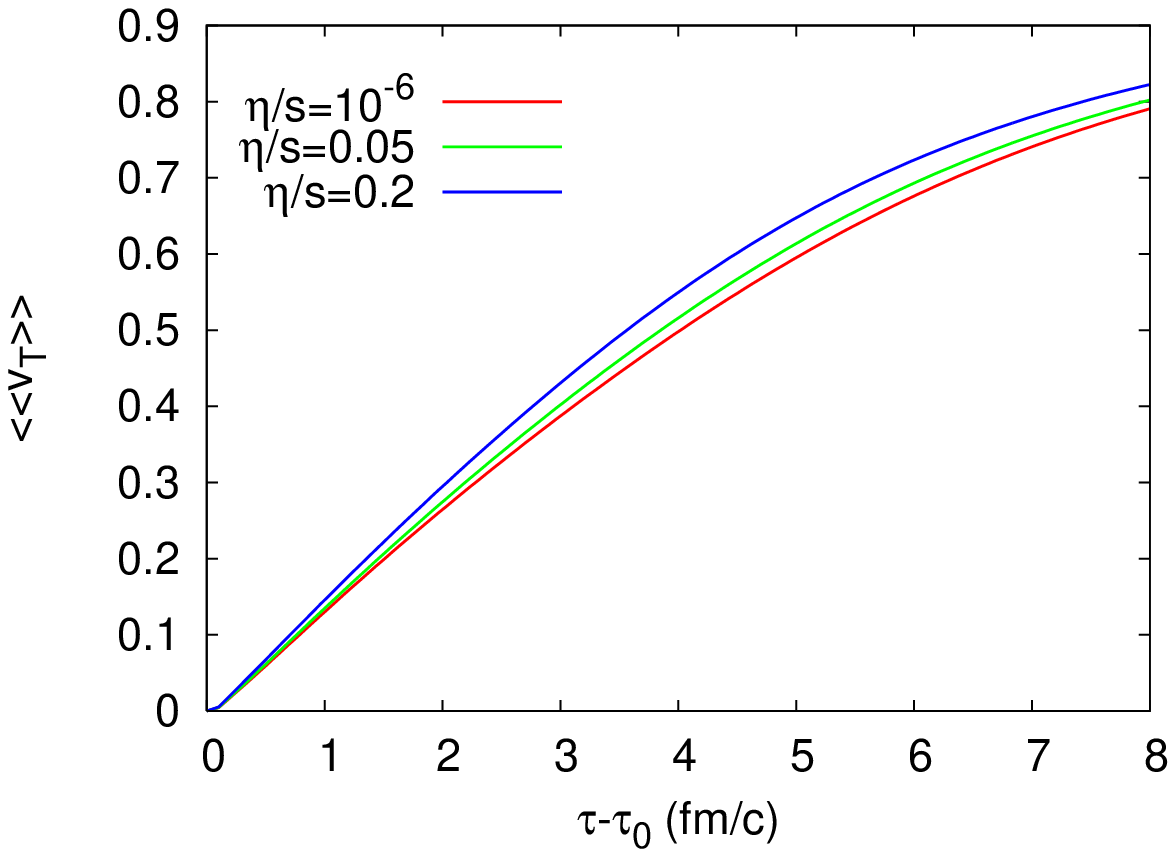}
  \vspace{9pt}
  \caption{(Color online) Time evolution of the spatial ellipticity
$\epsilon_x$, the momentum anisotropy $\epsilon_p$, and the energy
density weighted transverse flow $\langle\langle v_\perp
\rangle\rangle$, see Eq.~\ref{eq:anis}. }
  \label{fig:anis}
\end{figure}

Since we are interested in elliptic flow which originates from the
initial spatial anisotropy of the collision region it is useful to see
how the spatial and momentum anisotropy develop in time.  We therefore
look at the following three quantities \cite{KSH}:
\bg
\epsilon_x=\frac{\langle\langle y^2-x^2 \rangle\rangle}{\langle\langle y^2+x^2 \rangle\rangle} \nonumber\\
\epsilon_p=\frac{\langle\langle T^{xx}-T^{yy} \rangle\rangle}{\langle\langle T^{xx}+T^{yy} \rangle\rangle} \nonumber\\
\langle\langle v_T \rangle\rangle=\frac{\langle\langle \gamma \sqrt{v_x^2+v_y^2} \rangle\rangle}{\langle\langle \gamma \rangle\rangle}
\label{eq:anis}
\nd
where the double angular bracket $\langle\langle \cdots \rangle\rangle$
denote an energy density weighted average.  The spatial ellipticity
($\epsilon_x$) is a measure of the spatial anisotropy as a function of
time.  The spatial anisotropy is what drives the momentum anisotropy
($\epsilon_p$).  This quantity can be thought of as characterizing the
$p_T^2$ weighted integrated elliptic flow \cite{OllitraultSph}. 
 The final quantity $\langle\langle v_T \rangle\rangle$ is the average
radial flow velocity.  All three of these quantities are plotted in
fig.~\ref{fig:anis} for $\eta/s$=0.2, 0.05 and $10^{-6}$.

As already shown in the 1+1 dimensional case the finite viscosity case
does less longitudinal work. The longitudinal pressure is reduced
while the transverse pressure is uniformly increased in the radial
direction, {\em i.e.} gives no addition $v_2$ component.
This causes the transverse velocity (as seen in
$\langle\langle v_T \rangle\rangle$ and fig.~\ref{fig:vt}) to grow more
rapidly while $\epsilon_p$ lags behind the ideal case. 
Furthermore, the larger radial symmetric transverse velocity 
causes a faster decrease in
the spatial anisotropy. This further frustrates the build-up 
of the momentum anisotropy $\epsilon_p$.
 We therefore expect to see a decrease in the integrated
$v_2$ as the viscosity is increased.  This is indeed the case as will
be shown.  However, this effect is small compared to the change in
$v_2$ from use of the off-equilibrium distribution function.

\section{Freezeout}
\label{sec:fo}

As discussed in the introduction, ideal hydrodynamics is
applicable when $\lambda_{mfp}\ll L$ where $L$ denotes the typical
system size.  When dissipative corrections are included, one must
remember that the Navier Stokes equations are derived assuming that the
relaxation time $\tau_R$ is much 
smaller than the inverse expansion rate, 
$\tau_R \partial_{\mu} u^{\mu}\ll 1$.
Therefore, in the simulations we determine the freezeout surface by monitoring the expansion rate relative to the relaxation time using a generalization of the freezeout criteria first proposed in \cite{Bondorf:1978kz, Heinz:1987ca} and later in \cite{Hung:1997du}.

Specifically, freezeout is signaled when\footnote{
In actual simulations we take 
$(\eta/p)\,\partial_{\mu} u^{\mu} = 0.6$ for most runs (see below).
}
\st
     \frac{\eta}{p} \partial_{\mu} u^{\mu} \sim \frac{1}{2}\,.
\stp
This combination  of parameters can be motivated from the kinetic 
theory estimates \cite{Reif}.  The pressure is $p\sim \epsilon
\llangle v_{\rm th}^2\rrangle  $  with $\llangle v_{\rm th}^2\rrangle $ the 
typical quasi-particle velocity and $\epsilon$ the energy density.
The viscosity is of order $\eta \sim \epsilon \llangle v_{\rm th}^2 \rrangle \tau_{R}$ 
with $\tau_R$ the relaxation time.  Thus the freezeout condition is
simply
\st
    \frac{\eta}{p} \partial_{\mu} u^{\mu} \sim \tau_R \partial_{\mu} u^{\mu} \sim \frac{1}{2}\,.
\stp
In the model we are considering $\eta/p=\alpha \tau_{2}$ with 
$\alpha =0.7$ as described in \Sect{GenericSect}.

The value of $\frac{1}{2}$ can be considered as a parameter chosen to
be smaller than one.  The point is that as the above quantity becomes
large the Navier Stokes approximation is no longer applicable and the
simulation should freezeout.  At this point one would need to include
further higher order corrections in the gradients or switch to a
kinetic approach.

%For the Bjorken geometry we are considering the derivative is:
%\st
%   \partial_{\mu} u^{\mu} = \partial_{\tau} u^{\tau} + \frac{u^{\tau}}{\tau} + \partial_{x} u^{x} + \partial_y u^{y}  
%\label{eq:grad}
%\stp
%\st
%  \eta \sim e \, v_{\rm th}^2\,  \tau_{\scriptscriptstyle R}   \qquad p \sim e \, v_{\rm th}^2
%\stp

It is also convenient to have a definition for an analogous freezeout
surface in the case of ideal hydrodynamics.  One can think of keeping
the freezeout surface fixed as $\eta/s$ is taken to zero.  Dividing the
freezeout criterion by $\eta/s$ and using $s=(\epsilon+p)/T\sim 4p/T$
we define
\st
\chi=\frac{4}{T}\partial_\mu u^\mu\,,
\stp
which involves only quantities in the ideal simulation.
This is a separate freezeout parameter independent of the viscosity.  We should point out that the ideal freezeout conditions becomes more complicated in a hadronic resonance gas phase.  For example, the simple temperature dependence in $\chi$ is modified due to the rho resonance peak in the $\pi-\pi$ cross section \cite{Eskola:2007zc}.  This will be considered in more detail in a future work. 

\begin{figure}[hbtp]
\centering
\includegraphics[height=85mm]{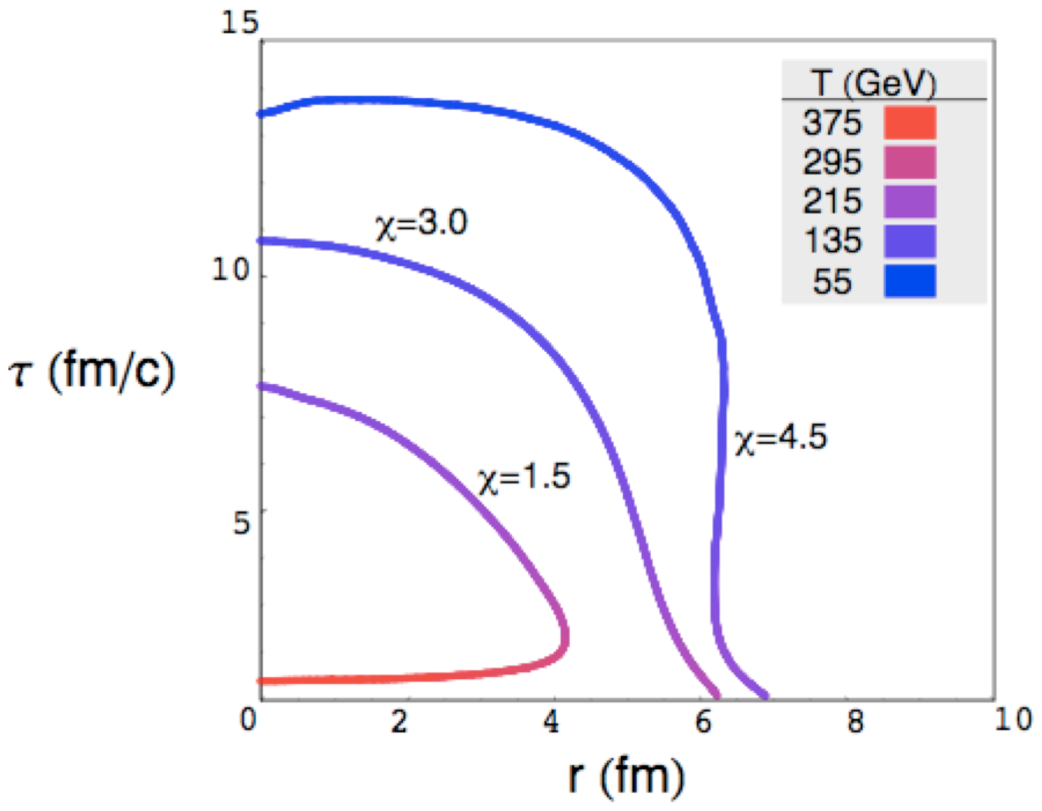}
\vspace{9pt}
\includegraphics[height=85mm]{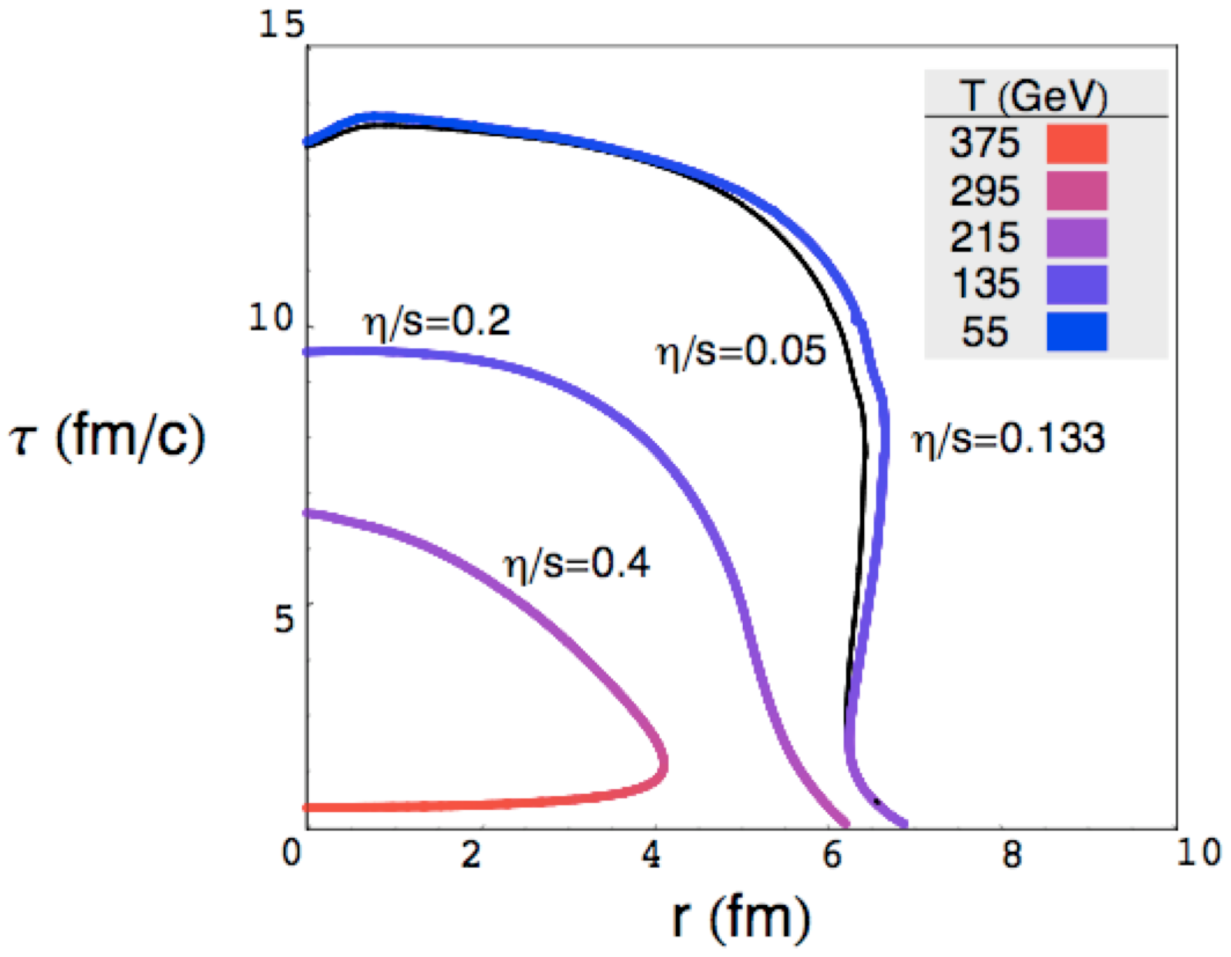}
\caption{(Color online) Contour plot of various freezeout surfaces for central Au-Au
collisions.  Top: Surfaces from ideal hydrodynamics where the freezeout
condition is set by the parameter $\chi$=1.5, 3 and 4.5.  Bottom:
Corresponding viscous solution where $\eta/s$ was fixed by the
condition $\frac{\eta}{p}\partial_\mu u^\mu=0.6$.  The thin solid black
curve shows the contour set by $\frac{\eta}{p}\partial_\mu u^\mu=0.225$
for comparison.  }
\label{fig:fosurf}
\end{figure}

We show in fig.~\ref{fig:fosurf} contour plots of the freezeout
surface for fixed $\chi$ from both ideal (upper plot) and viscous
hydrodynamics (lower plot).  For fixed $\chi$ the freezeout surfaces
remain approximately the same in both cases.  The freezeout surface from now on
will be specified by $\chi$ in order to facilitate a comparison between
the ideal and viscous cases when comparing spectra.

We have typically chosen $\chi$ and $\eta/s$ in order that
$\frac{\eta}{p}\partial_\mu u^\mu=0.6$. 
Thus in Table~\ref{tabFO} for $\eta/s=0.2$ we have $\chi=3.0$
and $\frac{\eta}{p}\partial_\mu u^\mu=0.6$. 
However, for $\eta/s=0.05$ the
freezeout parameter is $\chi=12$ giving an unphysically large
surface. 
This would normally not be the case in a more
realistic model with a phase transition present, since in the hadronic
phase the viscosity goes like $\eta\sim\frac{T}{\sigma_0}$.  The change
in scaling with temperature would cause the system to freezeout soon
after hadronization.  We plan on quantifying this statement in a future
work.  We therefore use $(\eta /p)\partial_\mu u^\mu=0.225$ when
$\eta/s=0.05$ giving $\chi=4.5$.  The thin solid curve in the lower 
plot of fig.~\ref{fig:fosurf} shows this particular freezeout contour.     
In table~\ref{tabFO} we summarize the freezeout parameters used
throughout this work.  For a given $\eta/s$ the most physical
choice of freezeout parameter $\chi$ is selected such that $(\eta /p)\partial_\mu u^\mu\approx0.6$.               
However, if the viscosity becomes so
small that the volume becomes unphysically large (such as for $\eta/s=0.05$) we
set $\chi=4.5$ as a maximum.  These three physically motivated 
parameter sets are given in bold  in the table.

We should stress that the freezeout surface taken in this work is
different from the typical constant temperature surface
used in many hydrodynamic simulations.  From fig.~\ref{fig:fosurf}, one
can see from the temperature map that the surface is not an
isotherm and actually spans a very wide range of temperatures.  
The freezeout surface is understood by examining the expansion rate in Bjorken geometry
\st
   \partial_{\mu} u^{\mu} = \partial_{\tau} u^{\tau} + \frac{u^{\tau}}{\tau} + \partial_{x} u^{x} + \partial_y u^{y}  \, .
\label{eq:grad}
\stp
The
resulting surface is due to a competition 
between the first two terms
in \ref{eq:grad} at early times and the last two terms at later times.

%The surface is somewhat unrealistic without the presence of the
%QGP/hadron phase transition.  $\eta/s$ in the hadronic phase
%scales differently with temperature compared to  the QGP.  This will
%modify the freezeout surface in the hadronic phase. This discussion
%is reserved for future work. 

\begin{table}[hbtp]
\centering
\begin{tabular}{c|c|c}
\hline
$\eta/s$ & $\frac{\eta}{p}\partial_\mu u^\mu$ & $\chi$\\
\hline
0.05 & 0.6 & 12.0 \\ 
{\bf 0.05} & {\bf 0.225} & {\bf 4.5} \\
0.05 & 0.15 & 3.0 \\
\hline
0.2 & 0.9 & 4.5 \\
{\bf 0.2} & {\bf 0.6} & {\bf 3.0} \\
\hline
{\bf 0.133} & {\bf 0.6} & {\bf 4.5} \\ 
\hline
\end{tabular}
\caption{Freezeout parameters used throughout this work.  For a given $\eta/s$ the most physical
choice of freezeout parameter $\chi$ is selected such that $(\eta /p)\partial_\mu u^\mu\approx0.6$. 
However, if the viscosity becomes so small (such as for $\eta/s=0.05$) that the volume becomes unphysically large (see text for discussion) we
set $\chi=4.5$ as a maximum.  These three physically motivated
parameter sets are in bold.}
\label{tabFO}
\end{table}

\section{Spectra}

\subsection{Anisotropy}

Before computing the differential spectrum we will compute the 
momentum anisotropy as a function of time. The momentum 
anisotropy $A_2$ (which differs from $v_2$ by the placement 
of averages) is defined as
\bg
A_2=\frac{\langle p_x^2\rangle-\langle p_y^2\rangle}{\langle p_x^2\rangle+\langle p_y^2\rangle}= \frac{S_{11}-S_{22}}{S_{11}+S_{22}}\, ,
\label{eq:A2} 
\nd
where $S^{ij}$ is the sphericity tensor and can be related to the 
hydrodynamics fields (i.e. $u^{\mu}$ ,$\pi^{\mu\nu}$, $\Pi$) 
and moments of the ideal particle distribution function. 
From a theoretical perspective, $A_2$ is preferred because it is
almost independent of the details of the particle content of the theory
\cite{OllitraultSph}.

We plot $A_2$ in the following manner.  At a given proper time we integrate over the surface of constant $\chi$, 
which has developed by time $\tau$.  The remaining part of the surface is fixed by integrating
over the matter which has not frozen out ($\chi < \chi_{f.o.}$) at fixed
proper time.  This can be thought of as a freezeout surface with a flat top at time $\tau$. As time
moves forward eventually all of the matter is frozen out over a surface set by constant $\chi$ yielding a constant $A_2$.

\begin{figure}[hbtp]
\centering
\includegraphics[scale=0.85]{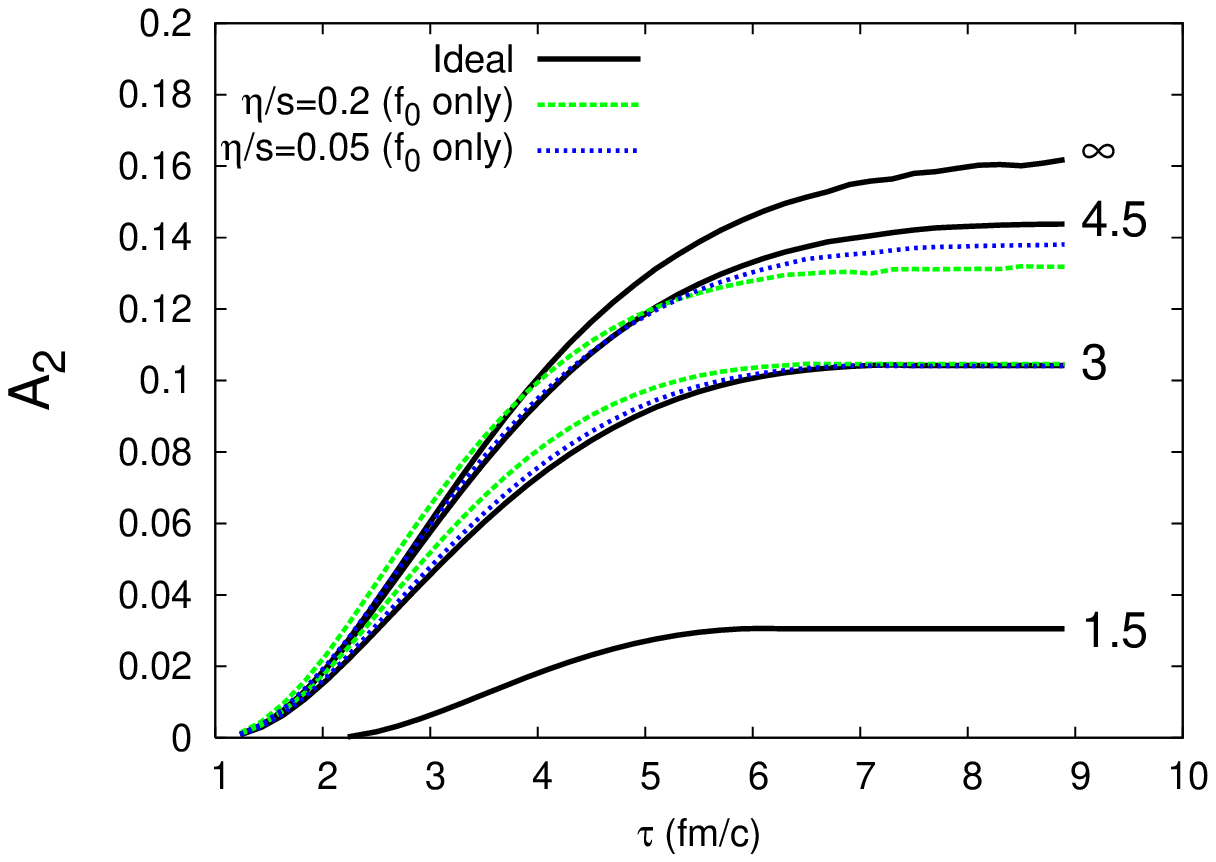}
  \vspace{9pt}
\includegraphics[scale=0.85]{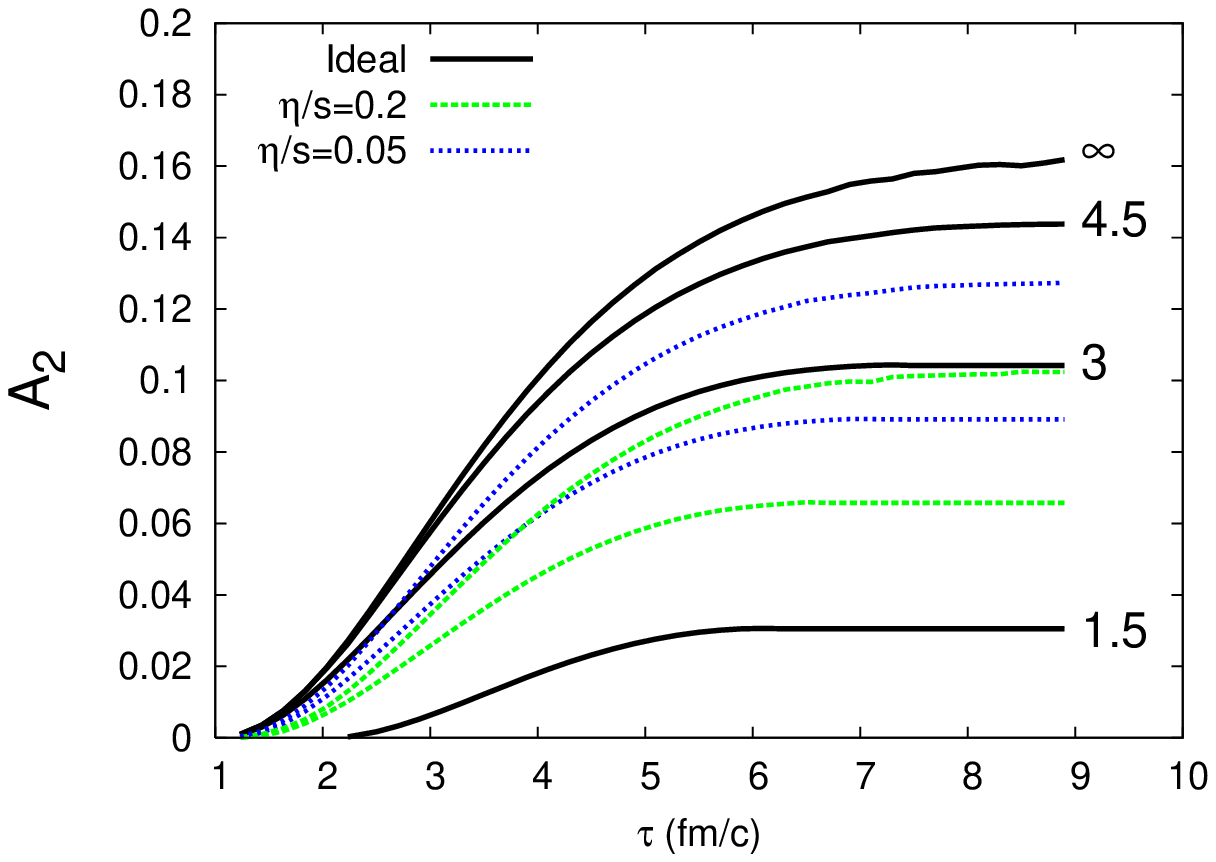}
\caption{(Color online) $A_2$ (defined in Eq.~\ref{eq:A2}) as a function of $\tau$.  The solid black
lines show the ideal result for $\chi=$1.5, 3.0, 4.5 and $\infty$.
Also shown in the bottom and top figures respectively are the viscous results with 
and without including the viscous correction to the distribution
function, for $\chi=3.0$ and $4.5$ and $\eta/s=0.2$ (dashed green curve) and for $\eta/s$=0.05
(dotted blue curve).} 
\label{fig:A2}
\end{figure}

Figure~\ref{fig:A2} shows $A_2$ for four different freezeout surfaces.
The figure on the top shows the results using only the ideal
contribution to the sphericity (regardless of if viscosity is present).
This will be analogous to using only the ideal particle distribution
function when generating the spectrum.  
First look at the solid black curves which
are generated using ideal hydrodynamics and a specified $\chi$.  
For a larger value of $\chi$
a larger space-time region is evolved by 
hydrodynamics producing a larger elliptic flow or $A_2$. 
The
true ideal case  where hydrodynamics is universally applicable  
is given by $\chi=\infty$.  We see that for $\chi=4.5$ most of
the elliptic flow is reproduced.

In order to assess the role of viscosity we first look at the figure on
the top.  The dashed curves show $A_2$ 
for $\eta/s=0.05$ and $\eta/s=0.2$ without including viscous
corrections to the distribution function. (For clarity, these curves are shown only for $\chi=3.0$ and $\chi=4.5$.)  
Without the corrections to the distribution function the viscous 
corrections to $A_2$ are modest.
The lower figure shows the analogous
plot, this time including the viscous corrections to the distribution
function.  The corrections are much larger and we therefore expect the
viscosity to decrease the integrated elliptic flow.    

\subsection{Spectra}

The thermal $p_T$ and differential $v_2$ spectra of particles are
generated using the Cooper-Frye formula \cite{CF} given by
\st
E\frac{d^3N}{d^3p}=\frac{g}{2\pi^3}\int_\sigma f(p_\mu u^\mu,T) p^\mu d\sigma_\mu\, .
\label{eq:CF}
\stp
The thermal distribution function used in the Cooper-Frye formula above also needs to include corrections due to finite viscosity.  We therefore write $f = f_o + \delta f$ where $f_o$ is the ideal particle distribution and $\delta f$ is the viscous correction given by
\st
\delta f = \frac{1}{2(e + p) T^2}\, f_o(1 + f_o) \, 
p^{\mu} p^{\nu} \left[\pi_{\mu\nu} + \frac{2}{5} \Pi \Delta_{\mu\nu}  \right].
\label{eq:df}
\stp
For boltzmann statistics $f_o(1+f_o)$ is replaced by $f_o$.  
The elliptic flow is defined as the weighted average of the yields with $\cos(2\phi)$:
\st
v_2(p_T)=
\langle \cos(2\phi) \rangle_{p_T}=\frac{\int_{-\pi}^\pi d\phi \cos(2\phi) \frac{dN}{dy p_T dp_T d\phi}}{\int_{-\pi}^\pi d\phi\frac{dN}{dy p_T dp_T d\phi}}\, ,
\stp
where $\phi$ is the angle between the decaying particle's momentum
(${\bf p}_T$) and the azimuthal angle of the collision region.

\begin{figure}
\centering
\includegraphics[scale=1]{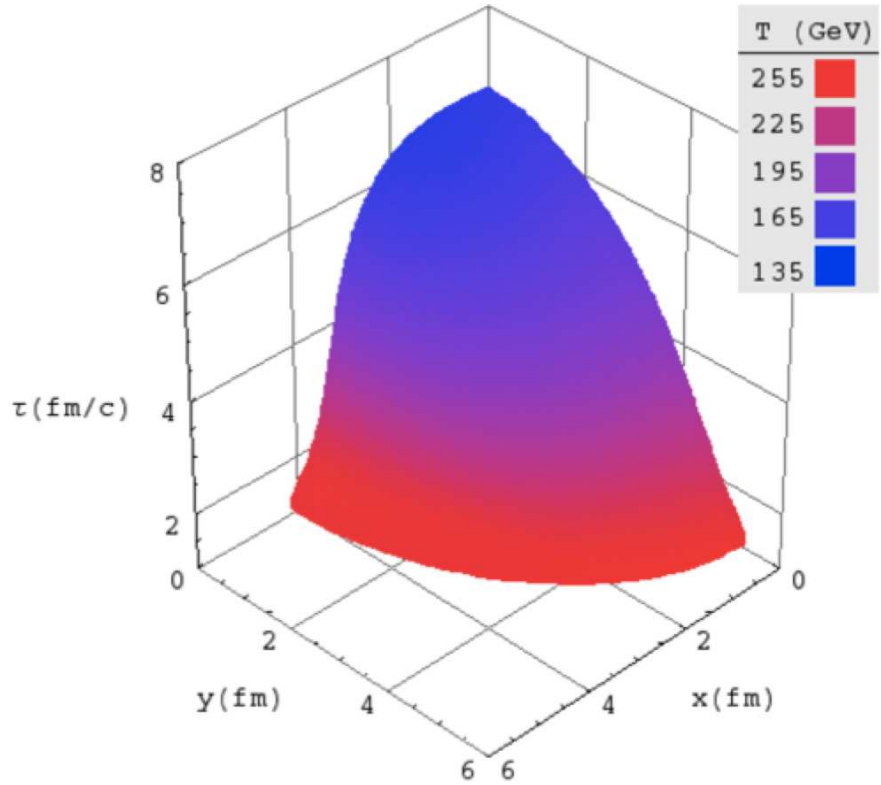}
\caption{(Color online) Freezeout surface for semi-central (b=6.5) Au-Au collisions
for $\eta/s=0.2$ and $\chi=3.0$.}
\label{fig:FOSurf1}
\end{figure}

A typical freezeout surface for $\chi=3$ at an impact parameter b=6.5
is shown in fig.~\ref{fig:FOSurf1}.  Color gradients show the
temperature profile on the freezeout surface and as noted before the
surface is not necessarily an isotherm.

\begin{figure}[hbtp]
\centering
\includegraphics[scale=0.85]{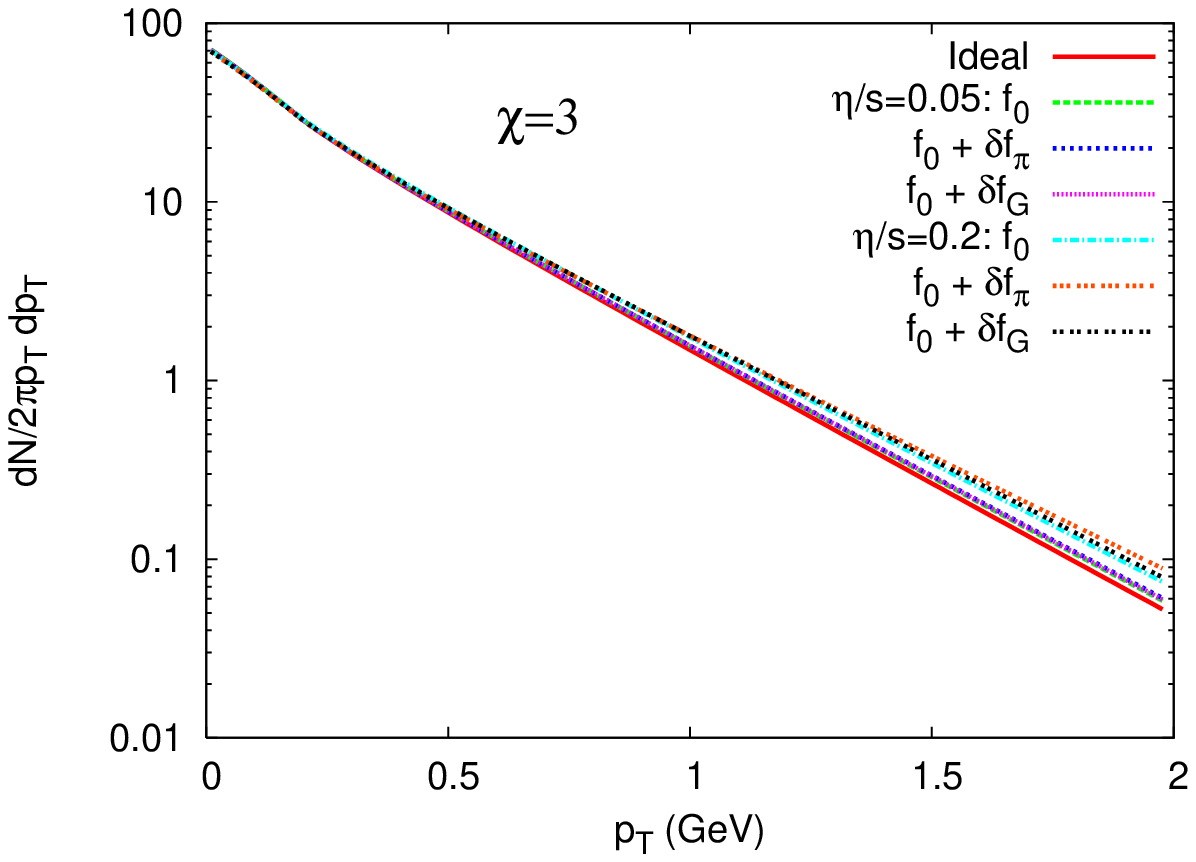}
  \vspace{9pt}
\includegraphics[scale=0.85]{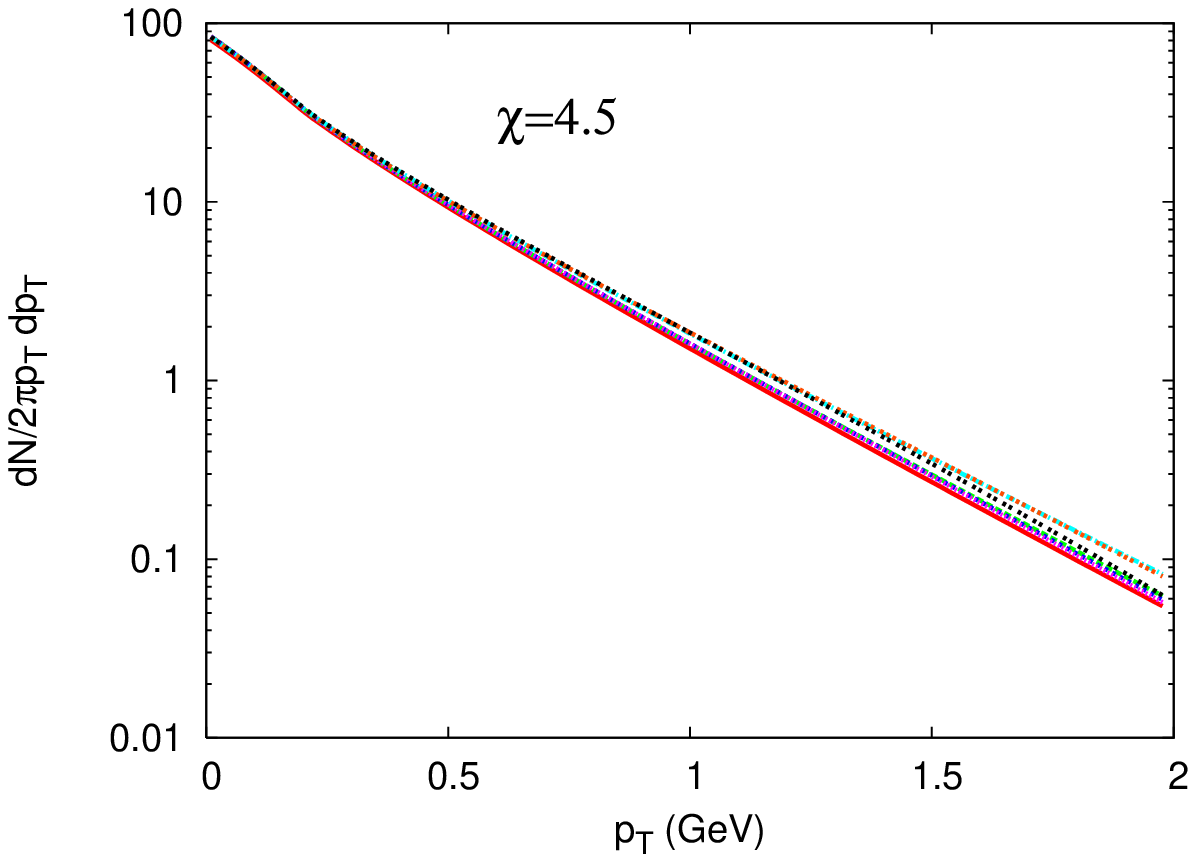}
\caption{(Color online) Differential transverse momentum spectra for Au-Au collisions at b=6.5 fm.  The upper plot is for freeze-out parameter $\chi=3$ and the bottom for $\chi=4.5$.  In both plots the ideal case is shown by the solid red curve.  Then the viscous case is shown without including the viscous corrections to the distribution function and is denoted by $f_o$.  The addition of the viscous correction to the distribution function is generated in two different ways.  $\delta f_\pi$ is calculated using the auxiliary tensor $c^{\mu\nu}$ while $\delta f_G$ is calculated using the physical gradients {\em i.e., $\pi^{\mu\nu}=-\eta\langle \partial^\mu \partial^\nu\rangle$}.    }
\label{fig:ptsp}
\end{figure}

Differential $p_T$ spectra for massless particles are shown in
fig.~\ref{fig:ptsp} for two different freezeout surfaces: $\chi=3.0$
(top) and $\chi=4.5$ (bottom).  In both plots the ideal case is shown
by the solid red line.  First we discuss changes to the spectra
brought about by modifications to the equations of motion by looking at
the spectra generated with the ideal particle distribution ($f_o$
only).  For both values of viscosity and both freezeout choices a
hardening of the spectra is observed.  This is expected since viscosity
tends to increase the transverse velocity. 

The effect from the viscous corrections to the distribution function
are more subtle.  At earlier times the transverse flow has not fully
developed and the longitudinal pressure is reduced while the 
transverse pressure is increased \cite{Teaney:2003kp}. This 
is a consequence of the fact that the shear tensor is traceless.
The increase in transverse pressure 
leads to a hardening of the spectrum after integration
over the space-time freezeout surface. 
This is the case for $\chi=3$
even though the corrections are small.  At later times the larger 
transverse
flow alleviates some of the longitudinal shear. When the hydro is
finally in a full 3D expansion, the viscous correction tends to 
reduce the transverse pressure. This 
changes the sign of the viscous correction term.
This is seen for $\chi=4.5$ where the viscous corrections soften the
spectrum slightly. 

As discussed above, any observable created by using the auxiliary
variable $c^{\mu\nu}$ should agree with the results using the physical
velocity fields.  Therefore we  also show the viscous corrections
calculated using the physical gradients (denoted by $\delta f_G$), {\em i.e.,}
in the local rest frame the $\pi^{ij}$ is approximated by
\st
    \pi^{ij} = -\eta(\partial^i u^j + \partial^j u^i - \frac{2}{3}\delta^{ij} \partial_l u^{l})\, , 
\stp 
when computing $\delta f$.

Overall, the corrections to the spectra are small so it is hard to see
any differences between the two calculations.  This will not be the
case for the differential elliptic flow where this comparison will be
more important.

\begin{figure}[hbtp]
\centering
\includegraphics[scale=0.85]{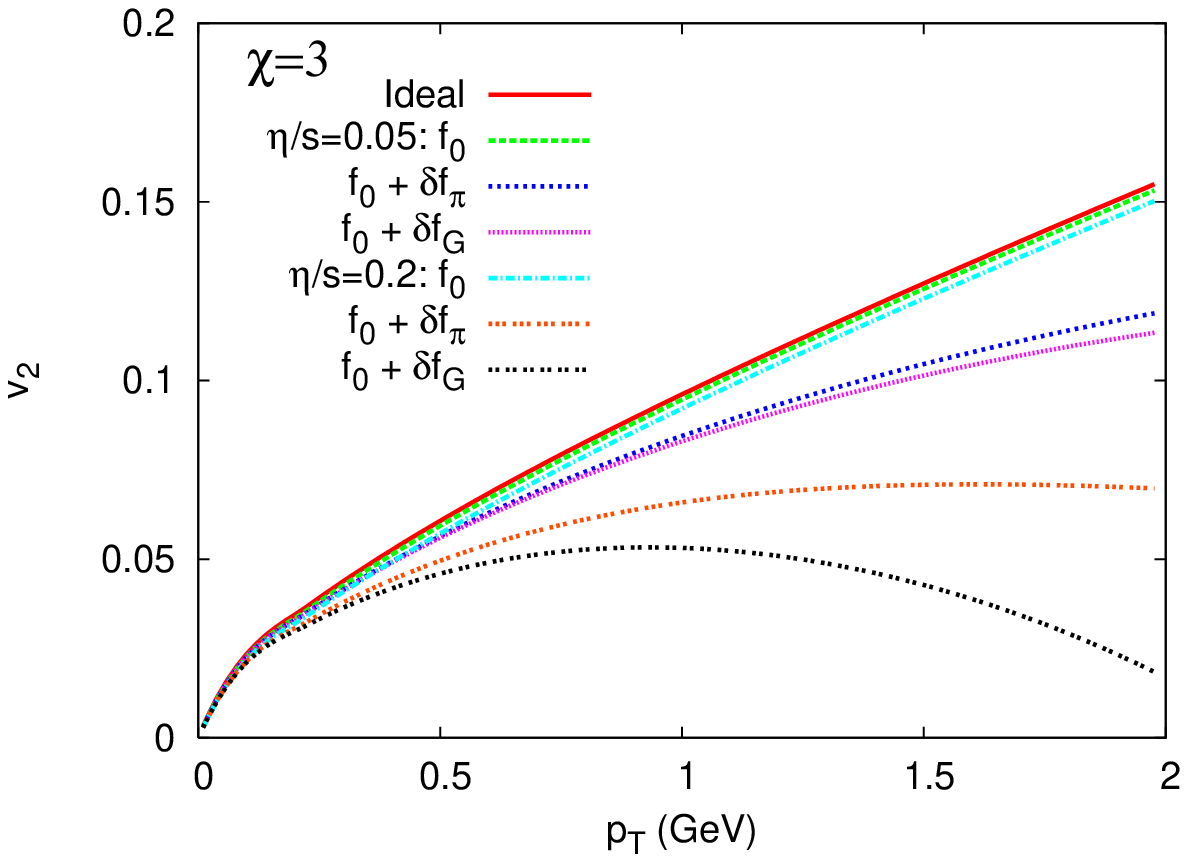}
  \vspace{9pt}
\includegraphics[scale=0.85]{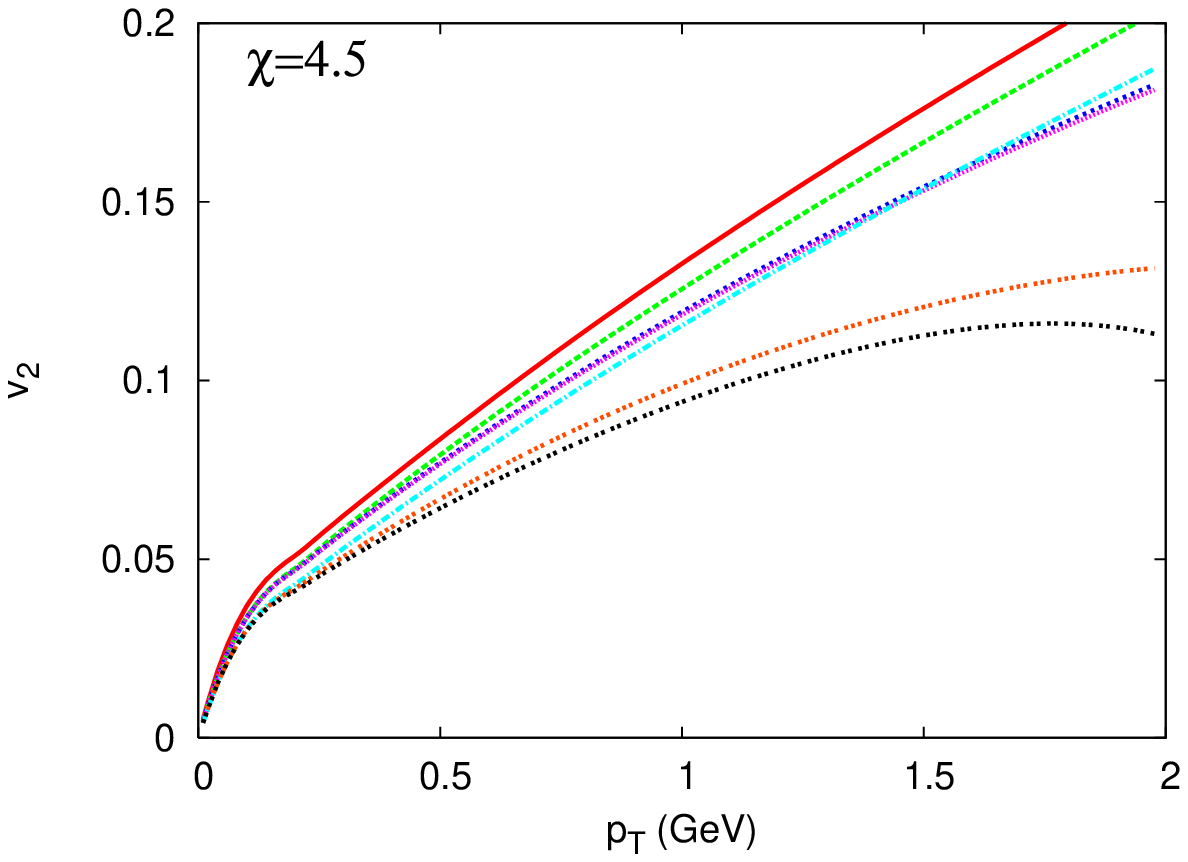}
\caption{(Color online) Differential $v_2$ spectra for Au-Au collisions at b=6.5 fm.  The resulting curves are generated in the same way as described for the $p_T$ spectra in fig.~\ref{fig:ptsp}}
\label{fig:v2sp}
\end{figure}

Figure~\ref{fig:v2sp} shows the differential elliptic flow using the
same parameter set from the $p_T$ spectrum.  The solid red curves
shows the ideal spectrum and, as expected, a larger elliptic flow is
generated for $\chi=4.5$ compared to $\chi=3$ since a larger fraction
of the space-time volume is described by hydrodynamics.

The viscous correction to the equations of motion causes only a small
change in the elliptic flow as seen by comparing the results at finite
viscosity using $f_o$ only with the ideal case.  For $\chi=3$ the
change is almost negligible.  For $\chi=4.5$ deviations are on the
order of 2\% at $p_T=2$ GeV.   

Including the viscous corrections to the distribution function can
bring about large changes in the elliptic flow.  We show the
corrections due to the auxiliary variable by $\delta f_\pi$ and those
from the gradients by $\delta f_G$ and we expect the two results to
agree.  When the two results start to diverge the gradient expansion is
no longer valid and a kinetic description is really required.

Based on our discussion in section \ref{sec:fo} the viscosity is what
sets the freezeout surface.  For $\eta/s=0.2$ we find that $\chi=3$
(upper figure).  In this case the viscous corrections are large but
can only be trusted up to $p_T \approx 1$ GeV.  We also show for
comparison the spectra for $\eta/s=0.05$ which can be trusted past 2
GeV.  For $\eta/s=0.05$ we take $\chi=4.5$ for reasons discussed in
section \ref{sec:fo}.  Again, the viscous correction decreases the
elliptic flow as a function of $p_T$.  Also shown are the spectra for
$\eta/s=0.2$ and the corrections are larger. In both cases the spectra
can be trusted past $p_T$ = 2 GeV.   

\section{Conclusions}

In summary we now make several conclusions regarding the effects of
shear viscosity on heavy ion collisions.  

We first recall the setup. The paper is restricted to an ideal gas
equation of state $p=\frac{1}{3}\epsilon$  and sets the initial
non-equilibrium fields to the value expected from the navier stokes
equations $\pi^{ij} = -\eta \llangle \partial^i u^j\rrangle$. The
initial distribution of entropy density follows the distribution of
participants. (This could be changed to a Color Glass Condensate model
initial conditions \cite{Hirano:2005xf}.) The paper simulates a fluid model based on \cite{Ottinger}
which is similar but differs from that of Israel and Stewart.  However
all models should ultimately agree on the magnitude of viscous
corrections provided the viscosity is sufficiently small.  

Several technical notes warrant discussion here. An algorithm for
a reliable solution of the viscous model was developed by Pareschi \cite{Pareschi}.  For small enough relaxation
times the auxiliary fields $\pi^{ij}$ should relax to the form
expected from the Navier-Stokes equation $\pi^{ij}\simeq-\eta \llangle
\partial^i u^j \rrangle$. Generically, relaxation models for viscosity
have long time parameters (the shear viscosity $\eta$ in this case)
and short time parameters. In the model considered here,
$\alpha$ is the 
short time parameter while  in the 
Israel-Stewart theory this short time parameter is $\eta/[(\epsilon +
p) \tau_{\pi}]$. These short time parameters can be constrained by the
$f$-sum rule \cite{KadanoffMartin, Teaney_corr, PTcorr}. In
general the results should not depend on these short time parameters.

We now summarize our physical results.  The integrated viscous
corrections to the flow are small.  This was seen in both the
hydrodynamic fields and also in the differential and integrated
elliptic flow when the thermal distribution function was restricted to
the ideal form.  (The remainder of this paragraph discusses only 
results with this restriction.) For the integrated $v_2$ this is seen in
the upper plot of fig.~\ref{fig:A2} where $A_2$ is shown for ideal runs and
viscous runs at $\eta/s=0.05$ and $0.2$. Corrections due to the
modified flow pattern are also small in the  differential $v_2$
spectrum as seen in fig.~\ref{fig:v2sp} by comparing the ideal and
viscous runs (again with $f_o$ only.) Although there is the possibility
for the elliptic flow to be modified from variations in the freezeout
surface across different runs this was minimized by freezing out on
contours of constant $\chi$.  One can see from fig.~\ref{fig:fosurf}
that the space-time freezeout contours are about the same at zero and
finite viscosity.  The fact that only small changes in the fields are
seen when including viscosity is not surprising.  The time scale of any
heavy ion collision is much shorter than the time needed for
dissipative effects to integrate and become large. 

\begin{figure}[hbtp]
\centering
\includegraphics[scale=0.85]{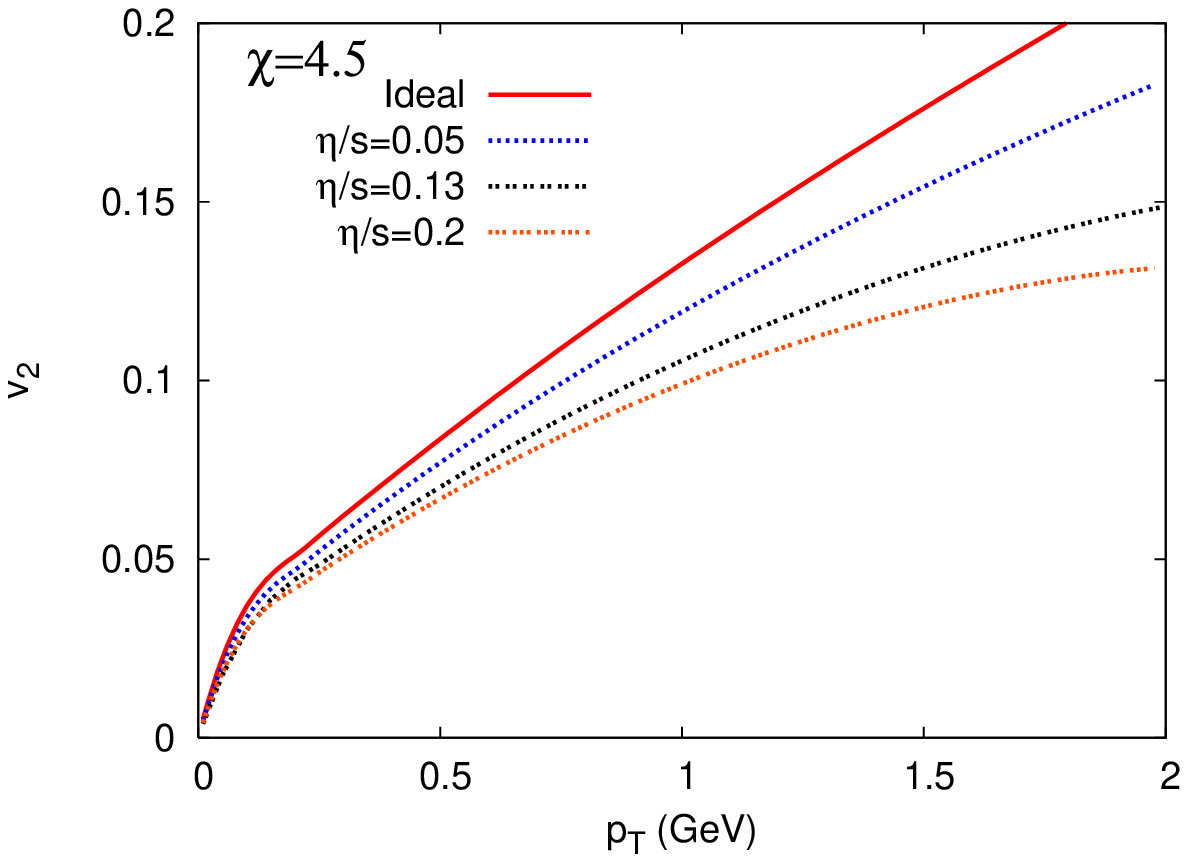}
  \vspace{9pt}
\includegraphics[scale=0.85]{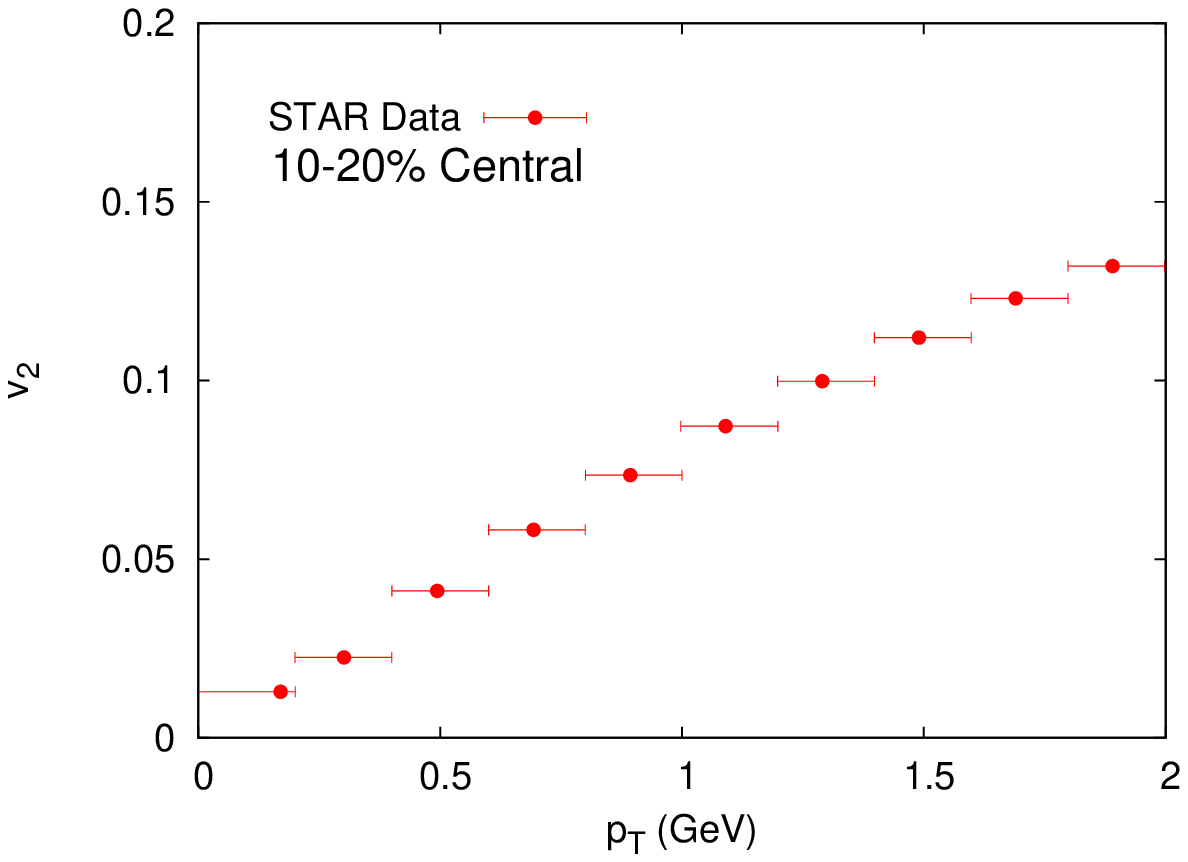}
\caption{(Color online) Top: Summary plot showing $v_2$ for massless particles for simulations using ideal hydro and $\eta/s=0.05, 0.2$.  Bottom: Charged hadron $v_2$ data using the standard reaction plane method as measured in Au-Au collisions at $\sqrt{s}=200$ GeV for a centrality selection of 10\% to 20\% \cite{Adams:2004bi}.}
\label{fig:v2sum}
\end{figure}

Even though viscosity does not modify the flow strongly we have shown
that there are still large corrections to the particle spectra due to
off-equilibrium corrections to the ideal particle distribution
function.  Any bounds for the viscosity (at least from this paper)
would have to come from the $v_2$ spectra.  As Lindblom \cite{Lindblom}
and earlier work by others \cite{KadanoffMartin}  has clarified, any
observable computed from the auxiliary fields $\pi^{ij}$ must agree
with the same observable generated by the physical gradients
$-\eta\llangle \partial^i u^j\rrangle$. When
deviations are seen  the viscous corrections can no longer be  trusted.
For a freezeout surface set by $\chi=4.5$ the viscous corrections agree
with gradients up to 2 GeV for viscosities as large as $\eta/s=0.2$ as
seen in figure \ref{fig:v2sp}.  It is therefore safe to use only the
auxiliary variable when generating spectra for this particular
parameter set.  In figure~\ref{fig:v2sum} we show a summary plot of the
differential elliptic flow.  We now show one additional curve for
$\eta/s=0.133$ yielding $(\eta/p)\partial_\mu u^\mu=0.6$ for this
particular choice of  freezeout surface.  We believe that this choice
of parameters is the closest physical scenario.  The lower plot of
figure \ref{fig:v2sum} shows the measured elliptic flow 
as measured by the STAR collaboration \cite{Adams:2004bi}.  We do not intend to
make a comparison, but simply would like to keep the data in mind.
Nevertheless since this simulation was performed with a massless gas
which has the largest elliptic flow, it seems difficult to imagine that
the $\eta/s \gsim 0.35$ will ever fit the data even if the initial 
conditions are modified along the lines of Ref.~\cite{Hirano:2005xf}.

Before a realistic comparison with data can be made the QGP/hadronic
phase transition must be taken into account.  In the vicinity of the
phase transition it is possible that the shear viscosity may become
very large.  Also, a more realistic model for the hadronic gas would be
the hard sphere model where $\eta\sim\frac{T}{\sigma_0}$.  This would
adjust at what point the simulation freezes out and would therefore
effect spectrum.  There is most likely a finite bulk viscosity due to
the fluctuations of the QGP and hadron concentrations in the mixed
phase or from chemical off-equilibrium in the hadronic phase
\cite{Kharzeev:2007wb}.  A final issue that should be taken into consideration
is that particles of different mass could possibly freezeout 
on different surfaces.

\chapter{Dilepton production from a viscous QGP}

\section{Introduction}

There is a general consensus that the early matter produced at RHIC behaves as a near perfect fluid \cite{Ollitrault:2006va}.  This conclusion was born out of the success of ideal hydrodynamic descriptions \cite{Teaney:2001av,Kolb:2003dz} of both hadron transverse momentum spectra and elliptic flow measurements up to 1.5-2 GeV/c.  Although it is too early to draw any definitive conclusions most likely the deviations from ideal hydrodynamic behavior can be ascribed to dissipative effects. This has already been suggested in some of the recent works on dissipative hydrodynamics \cite{Baier:2006gy,Romatschke:2007jx,Romatschke:2007mq,Chaudhuri:2007qp,Song:2007fn,Dusling:2007gi,Song:2007ux,Bozek:2007qt}.

In addition to hadronic observables which interact strongly and therefore depend only on the final state of the medium, electromagnetic probes are emitted throughout the entire space-time evolution reaching the detector without any final state interactions.  In terms of computing observables there is a big difference, since the resulting transverse momentum and elliptic flow spectra depend only on the final freezeout hypersurface, whereas the resulting dilepton yields depend on the full space-time volume.  A consistent description of heavy-ion phenomenology should use the same space-time evolution for both hadronic spectra and dilepton observables.

In this work we calculate the first viscous correction to dilepton emission from quark anti-quark annihilation in a dissipative medium.  The kinematic region when a thermal description is reliable is found by requiring that the viscous corrections are small.  When the viscous corrections become large a kinetic description is really required.  The viscous rates are then integrated over the space-time history of a hydrodynamic simulation of RHIC collisions.  We show how shear viscosity modifies the transverse momentum and invariant mass spectrum.  We find that the inverse slope of the transverse mass spectrum is sensitive to both the thermalization time as well as the shear viscosity and can therefore be used in order to learn about the early stages of heavy-ion collisions.  Finally a comparison is made with dileptons produced from a free-streaming quark-gluon plasma.   

\section{Dilepton Rates}

The rate of dilepton emission from a quark-gluon plasma due to q\={q} annihilation is given in the Born approximation as
\begin{equation}
\frac{dN}{d^4q}=\int\frac{d^3{\bf k}_1}{(2\pi)^3}\frac{d^3{\bf k}_2}{(2\pi)^3} f(E_1,T) f(E_2,T) v_{12} \sigma(M^2) \delta^4(q-k_1-k_2)\,,
\label{eq:KT}
\end{equation}
where $q=(q_0,{\bf q})$ is the virtual photon's four momentum and $M^2=(E_1+E_2)^2-({\bf k}_1 +{\bf k}_2)^2$ is the photon's invariant mass.  Throughout this work we consider massless quarks; therefore $E_{1,2}=\sqrt{{\bf k}_{1,2}^2+m_q^2}\approx|{\bf k}_{1,2}|$. The function $f(E,T)$ is the quark or anti-quark momentum distribution function, which in thermal equilibrium is given by $f(E,T)=1/(1+e^{E/T})$.

In the above expression $v_{12}$ is the relative velocity of a quark anti-quark pair and $\sigma(M^2)$ is the q\={q} cross section.  Both expressions are well known from the literature \cite{WongBk} and are given by $v_{12}=\frac{M^2}{2 E_1 E_2}$ and  $\sigma(M^2)=\frac{16\pi\alpha^2 e_q^2N_c}{3M^2}$.  The integral over the quarks' momentum can be done analytically with the result
\begin{equation}
\frac{dN}{d^4q} = -\frac{\alpha^2}{12 \pi^4} (N_c \sum_{u,d,s}e_q^2) f_b(q_0,T)\left[1+\frac{2T}{|{\bf q}|}\ln(\frac{n_+}{n_-})\right]\,,
\end{equation}
where $n_\pm=1/(e^{(q_0\pm|{\bf q}|)/2T}+1)$ and $f_b(q_0,T)=1/(e^{q_0/T}-1)$.

\section{Viscous Correction to the Dilepton Rates}

In order to account for dissipative effects in the dilepton emission rate we include the first viscous correction to the quark and anti-quark's distribution function in eq. \ref{eq:KT}.  This approach neglects any space-time inhomogeneities and assumes that the distribution functions relax to their dissipative forms much quicker than the medium evolves.  Ideally, one could solve the Baym-Kadanoff equations which would take non-equilibrium effects into account.  We note that the leading order born q\={q} rates do not contain pinch singularities which suggests that at least to leading order one may neglect space-time inhomogeneities \cite{Gelis:2001xt}.  This approximation allows us to calculate the dilepton emission rates locally in a space-time volume $d^4x$. 

As shown in \cite{Teaney:2003kp,Arnold:2000dr,GrootBk} viscosity modifies the ideal distribution function.  The resulting correction for fermions is
\begin{equation}
f(k)\to f(k)+\frac{C_1}{2 (\epsilon + p) T^2}f(k)[1-f(k)]k^\alpha k^\beta\pi_{\alpha\beta}\,,
\label{eq:vdist}
\end{equation}
where $\pi_{\alpha\beta}=\eta\langle \nabla_\alpha u_\beta \rangle\equiv\eta( 
\nabla_\alpha u_\beta+\nabla_\beta u_\alpha-\frac{2}{3}\Delta_{\alpha\beta}\nabla_\rho u^\rho)$ and $\eta$ is the shear viscosity not be confused with the space-time rapidity $\eta_s$.  The coefficient $C_1$ can be computed analytically for a massless fermion gas and is given by $C_1=14\pi^4/1350\zeta(5)\approx0.97$. Substituting the above result into the born annihilation rate (eq.~\ref{eq:KT}) and keeping terms up to first order in 
$\eta/s$ (quadratic in momentum) one obtains:
\begin{eqnarray}
\frac{dN}{d^4q}&=&\frac{4 N_c\alpha^2 e_q^2}{3(2\pi)^5}\int \frac{d^3{\bf k}_1 d^3{\bf k}_2}{E_1E_2}\delta^4(q-k_1-k_2) \times \nonumber\\
&&\left[ f(k_1)f(k_2) + \left(\frac{C_1}{2 (\epsilon + p) T^2}f(k_1)[1-f(k_1)]f(k_2)k_1^\alpha k_1^\beta+k_1\leftrightarrow k_2\right)\pi_{\alpha\beta}\right] \nonumber\\
&=&\frac{4 N_c\alpha^2 e_q^2}{3(2\pi)^5}\int \frac{d^3{\bf k}_1 d^3{\bf k}_2}{E_1E_2}\delta^4(q-k_1-k_2)\times\nonumber\\
&&\left[ f(k_1)f(k_2)+\frac{C_1}{(\epsilon+p) T^2}f(k_1)[1-f(k_1)]f(k_2) k_1^\alpha k_1^\beta\pi_{\alpha\beta}\right]
\label{eq:KTv1}
\end{eqnarray}

In simplifying the above result we have used the fact that the permutation of $k_1\leftrightarrow k_2$ gives the same contribution after integration.  We write the final result as the sum of the ideal and viscous correction
\begin{equation}
\frac{dN}{d^4q}= I_1(q) + \frac{C_1}{(\epsilon+p)T^2}I_2^{\alpha\beta}(q)\pi_{\alpha\beta}\,,
\end{equation}
with
\begin{eqnarray}
I_1&=&-\frac{N_c\alpha^2 e_q^2}{12 \pi^4}f_b(q_0)\left[1+\frac{2T}{|{\bf q}|}\ln(\frac{n_+}{n_-})\right]\,,\nonumber\\
I_2^{\alpha\beta}&=&\frac{4 N_c\alpha^2 e_q^2}{3 (2\pi)^5}\int\frac{d^3{\bf k}_1}{E_1 E_2}f(E_1)[1-f(E_1)]f(E_2)k_1^\alpha k_1^\beta\delta(E_1+E_2-q_0)\,.\nn
\end{eqnarray}

Since $I_2^{\alpha\beta}$ is a second rank tensor depending only on $u^\alpha$ and $q^\alpha$ it can be decomposed as
\begin{equation}
I_2^{\alpha\beta}=a_0 g^{\alpha\beta}+a_1 u^\alpha u^\beta+a_2 q^\alpha q^\beta + a_3(u^\alpha q^\beta + u^\beta q^\alpha)\,.
\end{equation}

The final result will contain the term $I_2^{\alpha\beta}\pi_{\alpha\beta}$.  By making use of the identities $u^\alpha\pi_{\alpha\beta}=g^{\alpha\beta}\pi_{\alpha\beta}=0$ only the term with coefficient $a_2$ will be non-vanishing.  $a_2$ is found by using the identity $a_2=P_{\alpha\beta}I_2^{\alpha\beta}$ where the projection operator in the local rest frame of the medium is
\begin{equation}
P_{\alpha\beta}=\frac{1}{2|{\bf q}|^4}[(3q_0^2-|{\bf q}|^2)u_{\alpha}u_{\beta}
-6q_0u_{\alpha}q_{\beta}+3q_{\alpha}q_{\beta}+|{\bf q}|^2g_{\alpha\beta}]\,.
\end{equation}

We now quote the final result for the first viscous correction to the born dilepton annihilation rates:
\begin{equation}
\frac{dN}{d^4q}=-\frac{N_c\alpha^2 e_q^2}{12\pi^4}\left[f_b(q_0,T)[1+\frac{2T}{|{\bf q}|}\ln(\frac{n_+}{n_-})]-\frac{C_1}{2(\epsilon+p)T^2} b_2(q_0,|{\bf q}|)q^\alpha q^\beta\pi_{\alpha\beta}\right]
\label{eq:KTvis}
\end{equation}
where we have defined
\begin{eqnarray}
b_2(q_0,|{\bf q}|)&=& \frac{1}{|{\bf q}|^5}\int_{E_-}^{E_+}f(E_1,T)f(q_0-E_1)(1-f(E_1))\nn&\times&\left[ (3q_0^2-|{\bf q}|^2)E_1^2-3q_0E_1M^2+\frac{3}{4}M^4\right]dE_1
\label{eq:b2}
\end{eqnarray}
and $E_{\pm}=\frac{1}{2}(q_0\pm|{\bf q}|)$.  For large invariant masses $(M/T \gg 1)$ one can replace the Fermi distribution with the classical Maxwell-Boltzmann distribution.  In the viscous correction to the distribution function this amounts to substituting $f_f(1-f_f)\to f_{MB}$.  In this limit an analytic expression can be found for the viscous correction to the dilepton rates.  In the limit that $(u\cdot q)/T \gg 1$ the resulting expression is given as
\begin{equation}
\frac{dN}{d^4q}=\frac{N_c\alpha^2 e_q^2}{12\pi^4}e^{-q_0/T}\left[1+\frac{C_1}{3(\epsilon+p)T^2}q^\alpha q^\beta\pi_{\alpha\beta}\right]\,,
\label{eq:appx}
\end{equation}
where as before $C_1\approx 0.97$.  We find that the above result holds at the accuracy of a few percent for $M \geq 3$ GeV at $T=400$ MeV.

A feature of the viscous correction is that it does not modify the invariant mass spectrum.  This is seen by looking at either of the above forms of the viscous correction (eq.~\ref{eq:appx} or \ref{eq:KTvis}).  In going from $d^4q$ to $dM^2$ the resulting integral will be a second rank tensor depending on $u^\alpha$ only.  The most general form the result can take is a linear combination of terms proportional to $g^{\alpha\beta}$ and $u^\alpha u^\beta$, which both vanish when contracted with $\pi_{\alpha\beta}$.  

Before performing the full space-time evolution we illustrate the effect of the viscous correction using
a simple model for the gradients.  We consider a 1D Bjorken expansion without transverse flow.  The viscous component of the stress-energy tensor can be easily computed and is given as 
\begin{equation} 
q^{\alpha}q^{\beta}\langle\nabla_\alpha u_\beta\rangle=\frac{2}{3\tau}q_{\perp}^2-\frac{4}{3\tau}m_{\perp}^2 \sinh^2(y-\eta_s)\,.
\label{eq:stressBj}
\end{equation}

\begin{figure}
\centering
\includegraphics[width=0.75\textwidth]{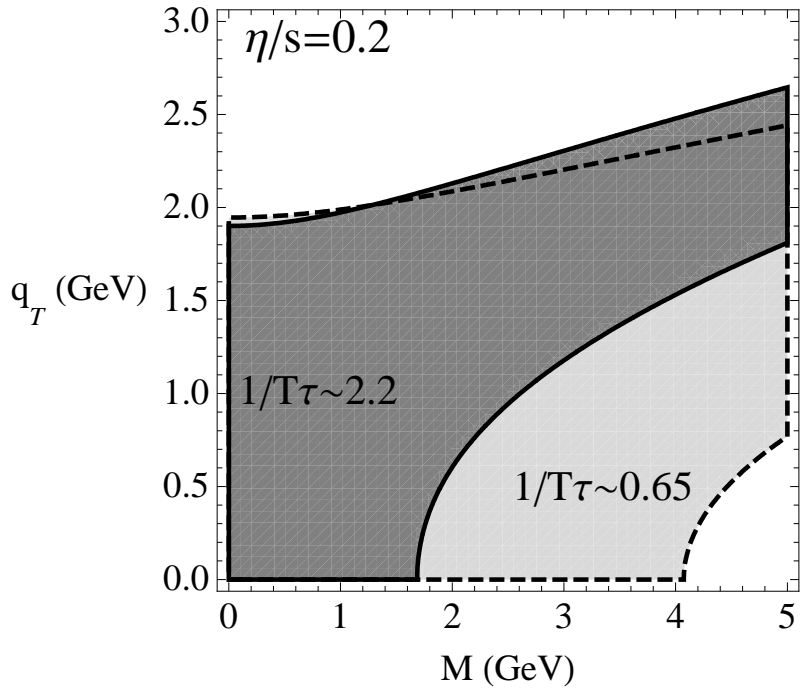}
\caption{\label{fig:map} Kinematic regions where the viscous correction is less then order one.  More precisely, the boundary is set by the condition $\lvert \delta f/f_0\rvert \leq 0.8$.}
\end{figure}

By substituting the above result into eq. \ref{eq:appx} an analytic expression can be found for the dilepton yields in the limit that $M/T\gg 1$.  After performing the integration over $\eta_s$ the result is
\begin{equation}
\frac{dN}{dM^2 dq_\perp^2 dy}=\frac{N_c\alpha^2 e_q^2}{12\pi^3}K_0(x)\left(1+\frac{2 C_1}{9\tau T}\left(\frac{\eta}{s}\right)\left[\left(\frac{q_\perp}{T}\right)^2-2\left(\frac{m_\perp}{T}\right) \frac{K_1(x)}{K_0(x)}\right]\right)\,,
\label{eq:Be}
\end{equation}
where $K_\nu(x)$ is the modified Bessel function evaluated at $x\equiv m_\perp/T$.  Fig.~\ref{fig:map} shows the kinematic regions where the viscous correction is small\newline ({\em i.e.} $dN_{vis}/dN_{ideal} \leq 0.8$) and therefore dictates when using a thermal description of dilepton production is suitable.  The criterion that dictates when hydrodynamics is applicable can be written as
\begin{equation}
\left(\frac{\eta}{s}\right)\times\frac{1}{\tau T} \ll 1
\end{equation}
and can therefore be separated into a condition on the medium, $\eta/s$, and a condition on the experimental setup, $1/(\tau T)$.  Throughout this work we always set $\eta/s=0.2$.  The region surrounded by the solid line is for $1/(\tau T)\approx 2.2$ corresponding to a temperature of 450 MeV at $\tau=0.2$ fm/c.  The region surrounded by the dotted line is for $1/(\tau T)\approx 0.65$ corresponding to a temperature of 300 MeV at $\tau=1$ fm/c.  At earlier times the viscous correction is larger and the allowed region is smaller.

The results shown in fig.~\ref{fig:map} should only be taken qualitatively.  Transverse flow alleviates the situation, opening up the boundaries shown above.  Also, the result presented was in the limit $M/T >> 1$ where an analytical result was obtained.  Fig.~\ref{fig:map} is still useful, since it still serves as a qualitative picture where the viscous corrections become large even after including flow and relaxing the classical limit.  Outside of the kinematic boundaries a thermal description may no longer be reliable.

\begin{figure}
\centering
\includegraphics[width=0.85\textwidth]{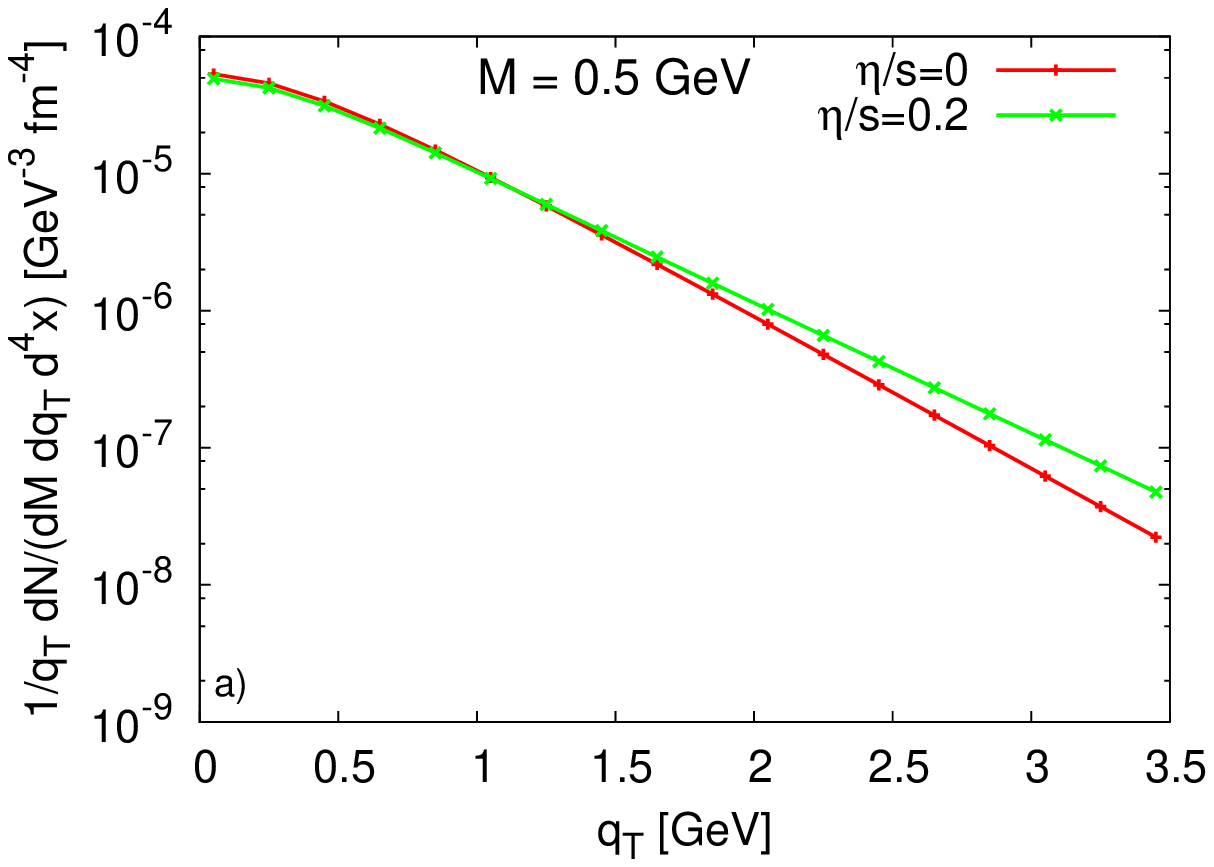}
\vspace{9pt}
\includegraphics[width=0.85\textwidth]{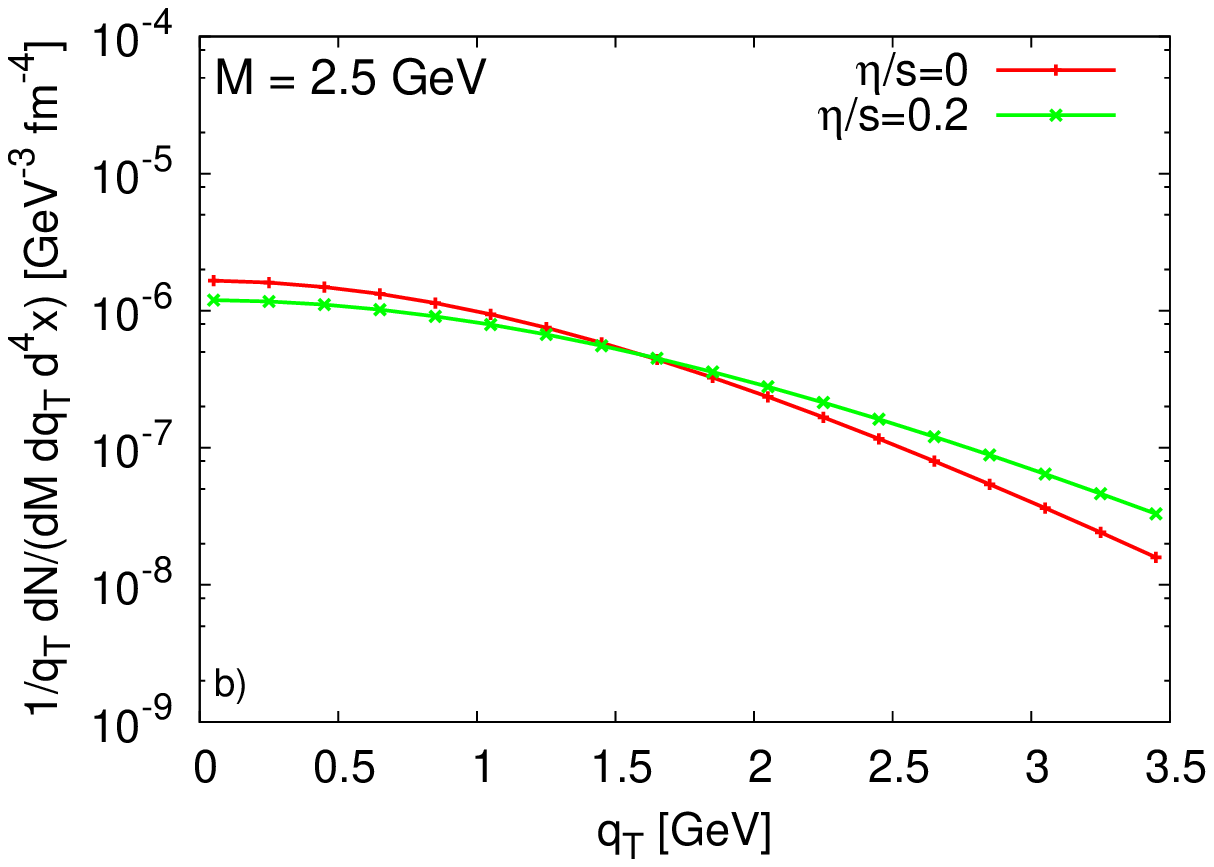}
\caption{\label{fig:qt}(Color online) Dilepton transverse momentum spectra
 for $T=0.4$ GeV and $\eta/s=0.2$ at $\tau=1$ fm/c for a boost invariant expansion with no transverse flow.}
\end{figure}

Throughout the remainder of this work we now resort to eq.~\ref{eq:KTvis}, which is evaluated numerically, in order to compute the dilepton yields accurately at all masses.  Fig.~\ref{fig:qt} shows the dilepton spectrum generated for a temperature T=0.4 GeV at proper time $\tau=1$ fm/c and using a viscosity to entropy ratio of $\eta/s=0.2$. 

Figures of the invariant mass spectrum are not shown because, as discussed earlier, the spectrum is unmodified when including the viscous correction.  Looking at the transverse momentum spectrum a {\em hardening} of dileptons is seen that is reminiscent of the single particle spectrum in \cite{Teaney:2003kp}.  The magnitude of the viscous correction is dictated by both $\eta/s$ as well as the proper time.  At earlier times the shear between the longitudinal and transverse directions is larger resulting in bigger corrections at smaller proper times due to the $1/\tau$ factor in eq.~\ref{eq:stressBj}.

\section{Evolution Model}

In order to model the space time evolution of the collision region we use the results of \cite{Dusling:2007gi}, which is summarized in this section.  The hydrodynamic model is a 1+1 dimensional boost invariant expansion with initial conditions tuned in order to simulate Au-Au collisions at RHIC energies ($\sqrt{s}=200$ GeV).  Dissipative corrections to the ideal hydrodynamic expansion is treated using a second order relaxation scheme first proposed by \"{O}ttinger and Grmela \cite{OG}.

The hydrodynamical model uses an ideal gas equation of state $p=\frac{1}{3}\epsilon$.  The relationship between energy density and temperature is chosen in order to mimic the entire phase region through the mixed phase of an ideal $N_f=3$ QGP equation of state with a $1^{st}$ order phase transition.  This happens to correspond to an ideal gas having an effective $DoF\approx16$.  At an initial time the entropy is distributed in the transverse plane according to the distribution of participants for a Au-Au collision.  Then one parameter, $C_s$, is adjusted to set the initial temperature and total particle yield.  The value $C_s=15$ closely corresponds to the results of full hydrodynamic simulations and corresponds to an initial temperature of $T_0=420$ MeV for $\tau_0=1$ fm/c.

Detailed plots of the hydrodynamic solution with and without viscosity is shown in \cite{Dusling:2007gi}.  For modest values of the shear viscosity ($\eta/s \lsim 0.3$) dissipative effects did not integrate to give large changes to the ideal hydrodynamic solution.  
The net effect of a finite viscosity was twofold.  First, the longitudinal pressure is initially reduced causing a slower decrease of energy density per unit rapidity at early times.  The reduction of longitudinal pressure is accompanied by a larger transverse pressure which drives larger transverse velocities.  The larger velocities then cause the energy density to deplete faster at later times.

Even though the changes to the ideal hydrodynamic result is small a full viscous simulation is still needed in order to have access to the velocity gradients which enter into the dissipative corrections of the quark and anti-quark distribution functions.

The hydrodynamic model is started at $\tau_0=0.2$ fm/c in order to account for some of the pre-equilibrium production of dileptons which will contribute at larger masses.  Dileptons are produced as long as the temperature of the medium is greater than a critical temperature taken as $T_c=0.170$ GeV.  We do not look at dileptons produced during a mixed phase or hadronic phase in this work.  At any space-time point the hydrodynamic model provides the three terms of the stress tensor ($\pi^{rr}, r^2\pi^{\phi\phi}$ and $\tau^2\pi^{\eta\eta}$).  Actually for an azimuthaly symmetric hydrodynamic simulation only two terms of the viscous stress tensor are independent since the third term could have been found by making use of the tracelessness of the shear tensor as well as its orthogonality to $u^\mu$ \cite{Heinz:2005bw}.  The equation for $q^{\alpha}q^{\beta}\pi_{\alpha\beta}$ used in eq.~\ref{eq:KTvis} is given as   
\begin{eqnarray}
q^{\alpha}q^{\beta}\pi_{\alpha\beta}&=&q_{\perp}^2\cos^2(\theta)\pi^{rr} +q_{\perp}^2\sin^2(\theta)r^2\pi^{\phi\phi} + m_{\perp}^2\sinh^2(\eta_s) \tau^2\pi^{\eta\eta} \nonumber\\ && + m_{\perp}^2\cosh^2(\eta_s) v^2\pi^{rr} - 2m_{\perp}\cosh(\eta_s) q_{\perp}\cos(\theta)v\pi^{rr}\,,
\end{eqnarray}
where $\theta\equiv\phi^q-\phi^v$ is the relative angle between the virtual photon's momentum ($q_\perp$) and the fluid cell's radial velocity. 

\begin{figure}
\centering
\includegraphics[width=0.85\textwidth]{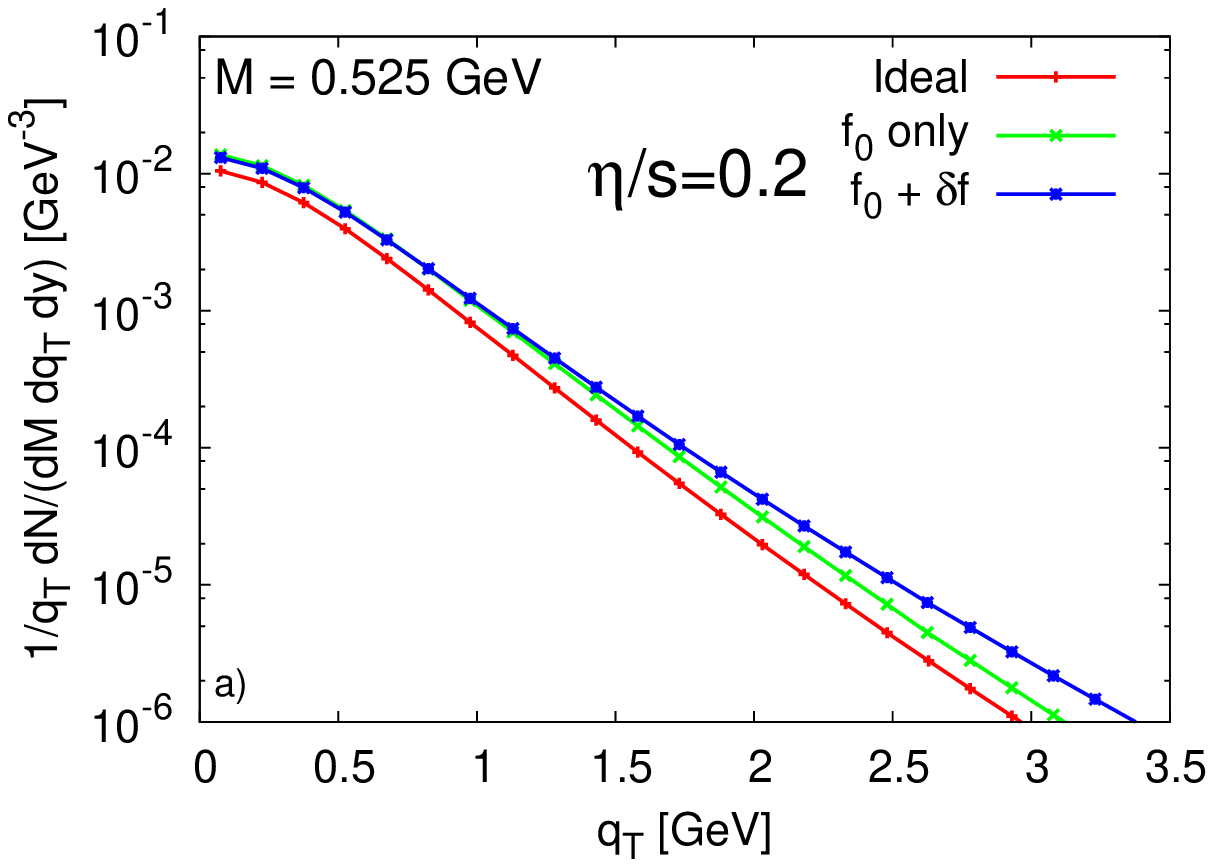}
\vspace{9pt}
\includegraphics[width=0.85\textwidth]{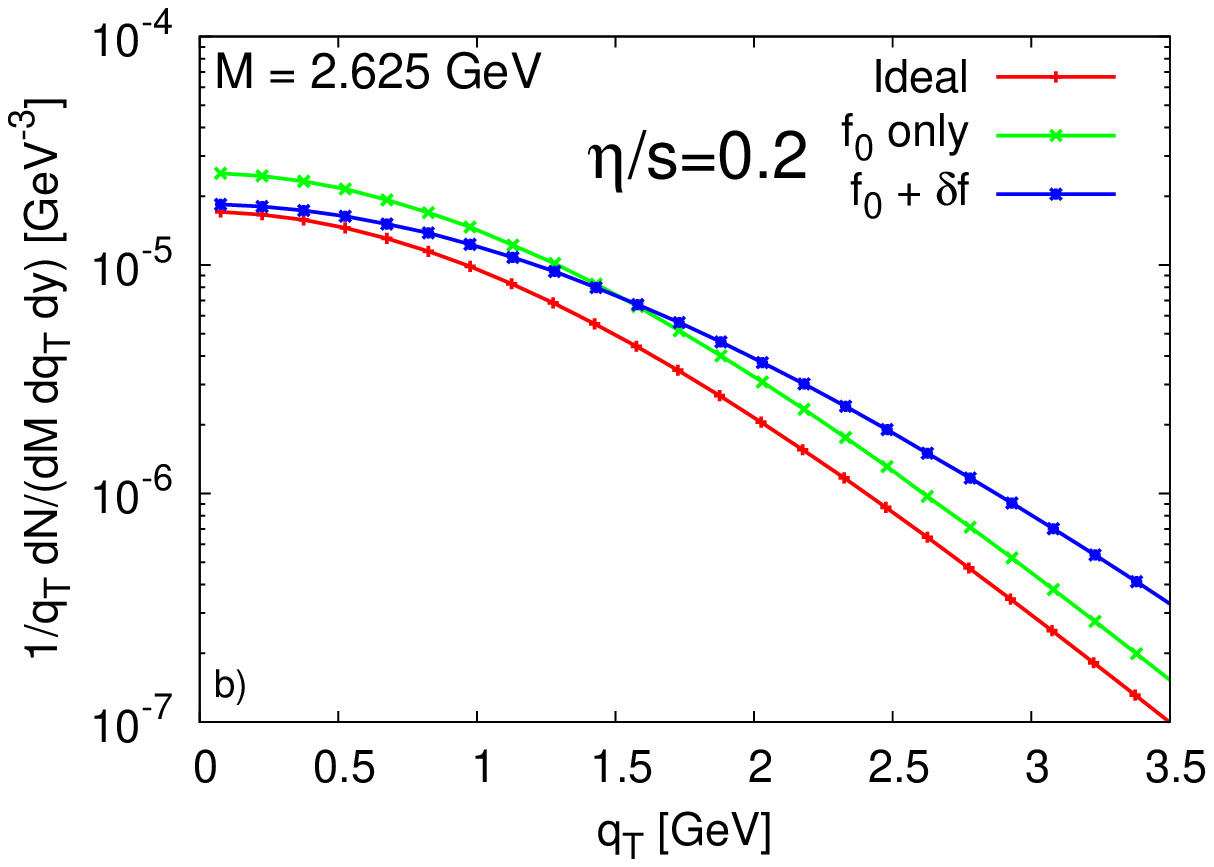}
\caption{\label{fig:qtINT}(Color online) Dilepton transverse momentum spectra
 after the full space-time integration of a boost invariant expansion with arbitrary transverse expansion and azimuthal symmetry.}
\end{figure}

Fig.~\ref{fig:qtINT} shows the resulting transverse momentum spectrum after the full space-time integration at two invariant mass points: $M=0.525$ GeV (top) and $M=2.625$ GeV (bottom).  First the red curve shows spectra generated from an ideal hydrodynamic simulation ($\eta/s=0$).  Next the green curve shows the spectra generated from a viscous simulation having $\eta/s=0.2$ but without including the viscous correction to the distribution function.  This curve therefore shows the effect that viscosity has on modifying the ideal hydrodynamic equation of motions.  We find that a finite viscosity leads to a slight increase in the overall yield.  This is due to the fact that a finite viscosity causes the energy density to deplete more slowly at early times.  This effect therefore brings about an effective increase in the lifetime of the simulation above the critical temperature. We find $\approx 30\%$ increase in the low mass region and $\approx 50\%$ increase in the higher mass region.  

Finally, the blue curves in fig. \ref{fig:qtINT} show the viscous result including the viscous correction to the distribution function.  We find that the magnitude of the viscous correction increases with the invariant mass.  This was similarly observed in fig.~\ref{fig:map} where the range in $q_\perp$ having viscous corrections of order less than one (as shown by shaded regions) decreased in size with increasing mass.  The simulation results are discussed in more detail in the next section.    

\section{Discussion}

In order to further understand the viscous corrections, the effective temperature ($T_{eff}$) of the dilepton spectrum from the full space-time integration is constructed as a function of invariant mass.  The effective temperature is found by fitting the transverse mass spectrum at a given mass to the following expression, 
\begin{equation}
\frac{dN}{dM^2 m_\perp dm_\perp dy} \propto e^{-m_\perp/T_{eff}}\,.
\end{equation}

In this work the fit is done in the transverse momentum region of $0.5\leq q_\perp \text{(GeV)} \leq 2.0$.  As expected, we find that the transverse mass spectra does not exactly fit the above form.  Actually, other ranges in $q_\perp$ could have been chosen where the fit works better.  However, the results are qualitatively the same and therefore a different choice in $q_\perp$ range will not change the discussion that follows.  If a quantitative comparison were to be made with data, it would be more appropriate to compare to the actual $q_\perp$ spectra instead.  Regardless, $T_{eff}$ still serves as a useful quantity since it probes the average temperature of the medium as well as the radial flow profile and viscous correction.  Looking at fig.~\ref{fig:qtINT} we expect the viscous correction to increase the effective temperature, with larger corrections at higher masses.    

\begin{figure}
\centering
\includegraphics[width=0.8\textwidth]{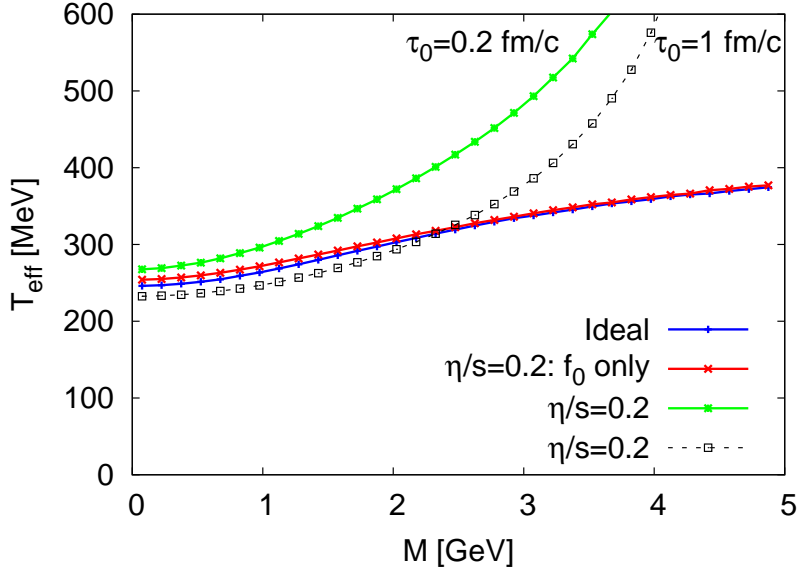}
\caption{\label{fig:Teff}(Color online) Effective temperature as a function of invariant mass.}
\end{figure}

In fig.~\ref{fig:Teff} the effective temperature is shown as a function of invariant mass.  The solid blue curve labeled {\em ideal} shows $T_{eff}$ for an ideal ($\eta/s=0$) hydrodynamic expansion started at proper time $\tau_0=0.2$ fm/c.  The shape of the curve is dictated by the underlying radial flow as well as the average temperature of the emission region.  At higher masses the dominate source of dileptons is from the higher temperature regions, which occur at earlier proper times.  This explains the slight rise in $T_{eff}$ with mass.  We now look at the solid red curve, which is generated from a hydrodynamic evolution having $\eta/s=0.2$.  In this case we do not include the viscous correction to the distribution function and the resulting modifications to the effective temperature are due to changes in the hydrodynamic evolution.  As discussed earlier, modifications from viscosity to the hydrodynamic solution are small and we therefore don't expect to see large deviations from the ideal case.  This is indeed the case.

We now focus the discussion on the role of the viscous correction to the distribution function.  The green curve in fig.~\ref{fig:Teff} shows the effective temperature of dileptons coming from a viscous medium having $\eta/s=0.2$ from a simulation started at a proper time of $\tau_0=0.2$ fm/c.  The result is that the effective temperature increases greatly as a function of invariant mass.  From the magnitude of the correction, the upper bound of the domain of hydrodynamics is found to be at most $M_{max}\approx 2.0$ GeV for this parameter set.  There are two reasons why the viscous correction increases with mass.  First, there is the explicit mass dependence in the viscous correction itself.  This is easiest to see by looking at the approximate form, eq.~\ref{eq:Be}.  The second reason is because the high mass contribution is mainly produced in the early, high temperature stages of the evolution.  Looking at eq.~\ref{eq:stressBj} the viscous correction grows like $1/\tau$ at early times.  In order to see the effect of the early emission a final simulation is done (dashed-black curve) where the hydrodynamic evolution is started at $\tau_0=1$ fm/c.  In this case the viscous corrections are more modest and $M_{max}\approx 4.5$ GeV.          

It is therefore a combination of both the thermalization time as well as the magnitude of $\eta/s$ that dictates when a hydrodynamic description is reliable.  Since the effective temperature rises so quickly with mass, as long as there is non-vanishing viscosity, there will always be a mass region outside of the region of a hydrodynamic description.  From eq.~\ref{eq:Be} one can extract an approximate condition for the mass.  Since most of the particle yield is at low $q_\perp$ we should guarantee that the viscous correction is small at $q_\perp=0$.  Furthermore at high and intermediate masses the ratio of Bessel functions is approximately one.  Then the condition that the viscous correction must be less than of order one can be expressed as
\begin{equation}
M_{max} \approx \frac{2 \tau_0 T_0^2}{\eta/s}\,.
\end{equation}

When the viscous corrections to the spectra become large a kinetic approach is required.  One can ask whether the viscous correction at early times {\em mock up} the effects of off-equilibrium production that would be taken into account by a full kinetic theory.  In order to test this hypothesis $T_{eff}$ spectra is calculated from a free streaming non-interacting gas of quarks \cite{Gyulassy:1997ib,Kapusta:1992uy}.  We should point out that our treatment is very similar to the recent work of \cite{Mauricio:2007vz}.  In this model the initial parton distribution is taken as thermal with the temperature chosen in order to reproduce the thermal dilepton number given by the hydrodynamic simulation.  Starting with the thermal initial condition at $\tau=0.2$ fm/c the total dilepton yield is found by integrating the free streaming result \ref{eq:A10} up to a final time of $\tau=1.0$ fm/c.  The details of this calculation is given in the appendix.  We now discuss the results.

We consider two scenarios.  The first is running the hydrodynamic simulation starting at $\tau_0=0.2$ fm/c until $T_c$.  The second scenario runs the free streaming model from $0.2\leq \tau \text{(fm/c)} \leq 1.0$.  Then at $1.0$ fm/c the hydrodynamic evolution is started and ran until $T_c$.  We should stress that the second model is not very realistic since the free streaming model is not asymptotic with the hydrodynamic evolution at $\tau=1$ fm/c.  A future work might use a more realistic model for the evolution of the distribution function then the proposed free streaming case.  For example, one could start with an initially anisotropic distribution which evolves to its thermal form from multi-quark scattering \cite{Xu:2005wj}, at which point a hydrodynamic evolution is started.  However, we expect the true result to lie between the two scenarios used in this work.

\begin{figure}
\centering
\includegraphics[width=0.85\textwidth]{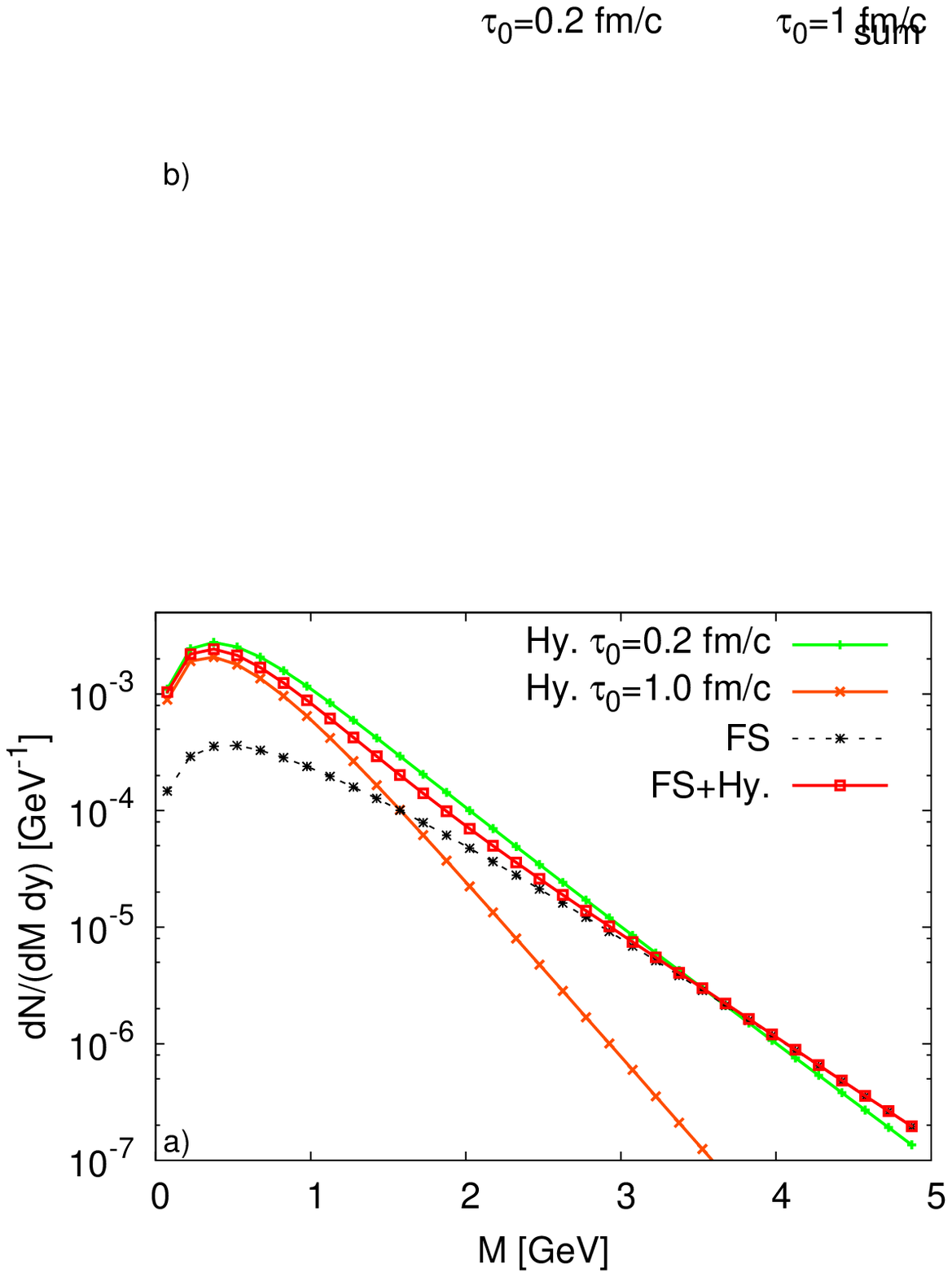}
\vspace{9pt}
\includegraphics[width=0.85\textwidth]{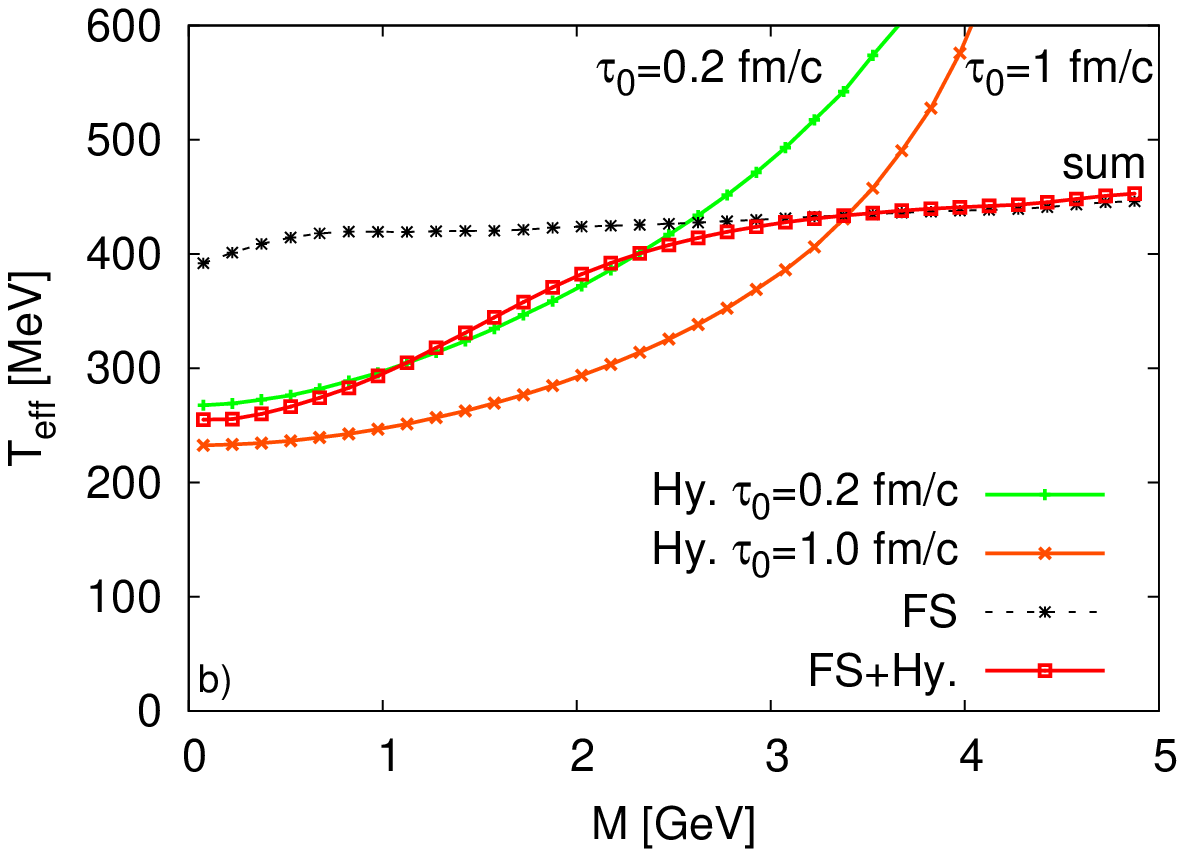}
\caption{\label{fig:summary}(Color online) Top: Dilepton invariant mass spectra. Bottom: Effective temperature as a function of invariant mass.}
\end{figure}

Figure~\ref{fig:summary} shows both the invariant mass spectrum (top) and effective temperature (bottom) for the two scenarios outlined above.  We note that the invariant mass spectrum is generated by integrating over {\em all} $q_\perp$.  The curves to compare are the hydrodynamic simulation started at $\tau_0=0.2$ (labeled {\em Hy. $\tau_0=0.2$ fm/c}) and the sum of the hydrodynamic simulation started at $\tau_0=1$ and the free-streaming (labeled {\em FS+Hy.}).  We first note that the invariant mass spectrum is qualitatively the same for the two scenarios and it would not be possible to discern between the two scenarios from experimental data.  Qualitative differences do appear in the $T_{eff}$ spectrum.  First we find that the free-streaming with hydro solution mimics the early time hydro only solution at low masses.  However, at high masses the two result diverge when the viscous correction can no longer be trusted.  While the early hydrodynamic solution increases greatly with mass the free streaming solution flattens off at higher mass.  We therefore argue that through a detailed analysis of $q_\perp$ spectra one could hopefully extract information on the thermalization time, viscosity to entropy ratio and thermalization mechanism in heavy-ion collisions.      

Some comments are in order on how viscosity may modify the dilepton emission from the hadronic phase, which comprises of a larger part of the overall yield in the mass region considered here.  A more detailed analysis will be presented in a future work.  As the medium evolves the viscous pressure decreases rapidly so one may expect dissipative effects to be smaller.  However the viscosity becomes larger as the temperature decreases.  For a pion gas in the chiral limit $\eta/s=\frac{15}{16\pi}\left(\frac{f_\pi}{T}\right)^4$ and it therefore becomes a dynamical question on how large the viscous corrections become in the hadronic phase. 

\section{Conclusions}

In conclusion, we have calculated the first viscous correction to dilepton production from leading order q\={q} annihilation.  The rates are then integrated over the space-time history of a viscous hydrodynamic simulation of RHIC collisions.  We argue that a thermal description is only reliable for invariant masses less than $\approx(2 \tau_0 T_0^2)/(\eta/s)$ and above this a kinetic description is required.  For $\eta/s=0.2$ and $\tau_0=1$ fm/c this corresponds to $M\lsim 4.5$ GeV.  We have shown that viscosity does not change the invariant mass distribution but strongly modifies the transverse momentum distribution and can therefore be used to extract information on both the viscosity to entropy ratio as well as the thermalization time.  Finally, we have also made comparisons with an initially free streaming QGP.  Qualitative differences in transverse momentum are seen, which could again possibly be used to learn about the thermalization mechanism.

\section*{Appendix: Dilepton yields from a free streaming QGP}
\addcontentsline{toc}{section}{Appendix: Dilepton yields for a free streaming QGP}
\chaptermark{Appendix}

In this appendix the dilepton yields are derived for a free streaming boost invariant expansion.  The starting point is the collision-less Boltzmann equation
\begin{equation}\label{eq:bz}
p^{\mu}\partial_{\mu}f(p,x)=0\,,
\end{equation}
where $f(p,x)$ will be considered as the phase-space distribution for the quark and anti-quark.  Under the assumption of boost invariance as well as homogeneity in the transverse plane the Boltzmann equation can be written as
\begin{equation}
\partial_\tau f-\frac{\tanh\chi}{\tau}\partial_\chi f=0\,,
\label{eq:fs}
\end{equation}
where $\chi=y-\eta_s$.  The initial condition is such that the quark distribution is isotropic and starts from local thermal equilibrium, $f(p,\tau=\tau_0)=\frac{1}{e^{p_0/T}+1}$.  One can write $p_0=u\cdot p=p_\perp \cosh(\chi)$ by using the assumption of boost invariance and homogeneity in the transverse plane.  We note that even for quarks out of equilibrium it still might be useful to use the equilibrium form of the distribution function where $T$ is instead considered as an effective temperature describing the initial state.  The solution of eq.~\ref{eq:fs} at any time $\tau$ is
\begin{equation}
f(p,x)=\frac{1}{e^{\frac{p_{\perp}}{T}\sqrt{1+\sinh^2(\chi)\left(\frac{\tau}{\tau_0}\right)^2}}+1}\,.
\end{equation}

With the explicit form of the distribution function available one can calculate the dilepton rates using the same kinetic theory expression used before (see eq.~\ref{eq:KT})
\begin{equation}
\frac{dN}{d^4q}=\int d^4x\int\frac{d^3p_1}{(2\pi)^3}\frac{d^3p_2}{(2\pi)^3}
f(p_1,x)f(p_2,x)v_{12}\sigma\delta^{(4)}(p_1+p_2-q)\,.
\end{equation}

Making use of the expressions for the relative velocity and cross section as quoted earlier the above equation can be expressed in the following form
\begin{equation}
\frac{dN}{d^4q}=B\int d^4x\int d^4p_1d^4p_2\delta(p_1^2)\delta(p_2^2)\delta^{(4)}(p_1+p_2-q)f(p_1,x)f(p_2,x)\,,
\end{equation}
where $B=\frac{32\pi\alpha^2 e_q^2}{(2\pi)^6}$.

First let us quote some well known identities:
\begin{eqnarray}
d^4x&=&\tau d\tau d\eta_s d^2x_{\perp}=\pi R^2\tau d\tau d\eta_s \nonumber\\
d^4p&=&\frac{1}{2}dp^2dy_pp_{\perp}dp_{\perp}d\phi_p \nonumber\\
d^4q&=&M dM q_{\perp}dq_{\perp} dy d\phi \nonumber\\
\delta^{(4)}(P-q)&=&4\delta(P^2-M^2)\delta(y_p-y)\delta(P_{\perp}-q_{\perp})\delta(\phi_P-\phi) \nonumber
\end{eqnarray}
where $P^{\mu}=p_1^{\mu}+p_2^{\mu}$ and $y$ and $\phi$ are rapidity and angle in the 
transverse plane.  We place a subscript $P$ on quantities to indicate they are derived from $P^{\mu}$.  The free streaming dilepton rate can now be expressed as
\begin{eqnarray}
\frac{dN}{d^4q}= B\int d^4x
\int dy_1p_{1,\perp}dp_{1,\perp}d\phi_1dy_2p_{2,\perp}dp_{2,\perp}d\phi_2f(p_1,x)f(p_2,x)
\times \nonumber\\
\delta(P^2-M^2)\delta(P_{\perp}^2-q_{\perp}^2)\delta(y_P-y)\delta(\phi_p-\phi)
\end{eqnarray}

Since the distribution function is boost invariant the integral over $\eta_s$ is trivial due to the delta function, $\delta(y_P-y)$.  By defining $y_{\pm}=y_1\pm y_2$ and $\phi_{\pm}=\phi_1\pm\phi_2$ the delta functions can be rewritten as
\begin{eqnarray}
\delta(P^2-M^2)&=&\frac{1}{2p_{1,\perp}p_{2,\perp}}\delta(\cosh y_--\cos\phi_--\frac{M^2}{2p_{1,\perp}p_{2,\perp}}) \nonumber \\
\delta(P_{\perp}^2-q_{\perp}^2)&=&\frac{1}{2p_{1,\perp}p_{2,\perp}}\delta(\cos\phi_-+\frac{p_{1,\perp}^2+p_{2,\perp}^2-q_{\perp}^2}{2p_{1,\perp}p_{2,\perp}}) \nonumber
\end{eqnarray}

After rewriting the integration variables as $dy_1dy_2=\frac{dy_+dy_-}{2}$ and $d\phi_1d\phi_2=\frac{d\phi_+d\phi_-}{2}$ the integral over $\phi_-$ and $y_-$ can be done explicitly yielding 
\begin{eqnarray}
\frac{dN}{MdMdyq_{\perp}dq_{\perp}}&=&4\pi^2R^2 B\nn&\times&\int\tau d\tau
\int dy_+ \frac{p_{1,\perp}dp_{1,\perp}p_{2,\perp}dp_{2,\perp}}{(2p_{1,\perp}p_{2,\perp})^2}\frac{1}{\lvert\sin\phi_-\rvert}\frac{1}{\sinh y_-}\nn&\times&f(p_1,x)f(p_2,x) 
\end{eqnarray}
where 
\begin{eqnarray}
\lvert\sin\phi_-\rvert&=&\frac{\sqrt{((p_{1,\perp}+p_{2,\perp})^2-q_{\perp}^2)
(q_{\perp}^2-(p_{1,\perp}-p_{2,\perp})^2)}}{2p_{1,\perp}p_{2,\perp}} \nonumber\\
\sinh y_-&=&\frac{\sqrt{(M^2+q_{\perp}^2-(p_{1,\perp}+p_{2,\perp})^2)(M^2+q_{\perp}^2-(p_{1,\perp}-p_{2,\perp})^2)}}{2p_{1,\perp}p_{2,\perp}}\nn
\end{eqnarray}

The delta function in the above equation enforces the following constraints
\begin{eqnarray}\label{eq:constraint}
\left\lvert\frac{p_{1,\perp}^2+p_{2,\perp}^2-q_{\perp}^2}{2p_{1,\perp}p_{2,\perp}}\right\rvert&\leqslant& 1 \nonumber \\
\frac{M^2+q_{\perp}^2-p_{1,\perp}^2-p_{2,\perp}^2}{2p_{1,\perp}p_{2,\perp}}&\geqslant& 1
\end{eqnarray}

Let us make a further shift of variables, $p_{\pm}=p_{1,\perp}\pm p_{2,\perp}$. The
constraints (\ref{eq:constraint}) then take particularly simple form.  The final expression is
\begin{eqnarray}\label{eq:fsrate}
\frac{dN}{dM^2dydq_{\perp}^2}&=&\pi R^2 \frac{N_c \alpha^2 e_q^2}{48\pi^4} \int\tau d\tau \int_{-\infty}^{+\infty} dy_+\int_{q_{\perp}}^{\sqrt{M^2+q_{\perp}^2}}dp_+\int_{-q_{\perp}}^{q_{\perp}}dp_-\nn
&\times&\frac{1}{\sqrt{(M^2+q_{\perp}^2-p_+^2)(M^2+q_{\perp}^2-p_-^2)}}\frac{1}{\sqrt{(p_+^2-q_{\perp}^2)(q_{\perp}^2-p_-^2)}}\nn
&\times& (p_+^2-p_-^2) f\left(p_1,\tau\right)f\left(p_2,\tau\right)
\label{eq:A10}
\end{eqnarray}
where
\begin{eqnarray}
f\left(p_1,\tau\right)&=&\left[ 1+\exp\left(\frac{p_++p_-}{2T}\sqrt{1+\left(\frac{\tau}{\tau_0}\right)^2\sinh\left(\frac{y_++y_-}{2}\right)}\right)\right]^{-1}\nonumber\\
f\left(p_2,\tau\right)&=&\left[ 1+\exp\left(\frac{p_+-p_-}{2T}\sqrt{1+\left(\frac{\tau}{\tau_0}\right)^2\sinh\left(\frac{y_+-y_-}{2}\right)}\right)\right]^{-1}\nonumber\\
y_-&=&\sinh^{-1}\left[\frac{2\sqrt{\left(M^2+q_\perp^2-p_+^2\right)\left(M^2+q_\perp^2-p_-^2\right)}}{\left(p_+^2-p_-^2\right)} \right]
\end{eqnarray}

\chapter{Heavy Ion Phenomenology}

\section{Di-muons at the CERN SPS collider}

\subsection{Introduction}

It is expected that above a critical temperature, $T_c\approx 170$ MeV, QCD 
undergoes a chiral phase transition where the relevant degrees of freedom 
change from mesons and baryons to a phase of strongly coupled quarks and 
gluons, the strongly coupled quark-gluon plasma (sQGP). This new phase of 
matter is being searched for in a number of ultra-relativistic heavy ion 
facilities.  There are a number of current observations in favor of the sQGP, 
ranging from hydrodynamical flow (soft probes) to jet quenching (hard probes). 
However, most of these observations are blurred by the fact that the
space-time evolution of the sQGP is short and its conversion to hadronic
matter is involved. Since the latter dominates the final stage of the
evolution, it is producing competing signals that interfere with those from
the sQGP. In this respect, dilepton and photon emissions are interesting
probes of the collision region as neither interact strongly with the medium 
produced in these collisions, thus they probe the {\em early} stages of the 
collision.  This is in contrast to hadronic observables which thermalize 
along with the collision region thus providing information only on the 
late (or freeze-out) stage of the collision. 

Making quantitative predictions of the production rates of dileptons and 
photons is difficult for a number or reasons.  Since the temperature produced 
in typical heavy-ion collisions is in the range of 200-300 MeV which is 
about the QCD scale factor, $\Lambda$, the differential cross sections can not
be computed in a weak-coupling expansion.  Another uncertainty is detailed 
knowledge of the evolution of both hadronic matter and quark gluon phase 
produced in heavy-ion collisions.  In addition there is also a background of 
dileptons from other processes not occurring in the quark-gluon plasma such 
as hadronic decays. 

In the past there have been a number of experiments probing photons and 
dileptons created in hadronic collisions.  One of the most recent experiments 
was the CERES (NA45) taking place as the CERN SPS collider which looked for 
dielectrons.  It was found that the dielectron production exceeded the 
theoretical expectations for {\em conventional} processes in both hadronic 
and QGP matter~\cite{HungShuryak96}, especially in the mass region 
$0.3\leq M \text{(GeV)} \leq 0.6$~\cite{CERES95}. A number of theoretical
analyses were put forward to explain this excess based on effective
Lagrangians with medium modification~\cite{Rapp,LiGale98} and dropping
vector meson masses~\cite{BrownRho}. Model independent emission rates
constrained by the strictures of broken chiral symmetry and data were
unable to account for the excess rate reported by NA45
\cite{paper1,paper2,paper3}. However, the large statistical and systematic 
errors reported by NA45 in exactly the excess region, did not allow for a 
definitive conclusion as to the theoretical nature of the emissivities.

In this letter we revisit these issues in light of the recently reported
dimuon data from the NA60 collaboration using In-In collisions at 
158 Gev/Nuc~\cite{sanja05}. These data have far better statistics, which gives much better
constraints on any medium modification to the vector mesons \cite{Hees06, Renk06}.  We use the model independent analysis
in~\cite{paper1,paper2,paper3} to analyze these data, whereby the
emissivities are constrained by broken chiral symmetry in a dilute
hadronic medium, and by non-perturbative QCD in the sQGP. The collision
expansion and composition are extracted from an underlying hydrodynamical 
evolution set to reproduce the CERN SPS conditions.

\subsection{Dilepton Emission Rates}

The rate of dilepton emission per unit four volume for particles in 
thermal equilibrium at a temperature T is related to the thermal 
expectation value of the electromagnetic current-current 
correlation function~\cite{McLerran, Weldon}.  For massless leptons 
with momenta $p_1$ and $p_2$, the rate per unit invariant momentum 
$q=p_1+p_2$ is given by:

\begin{equation}
\label{eq:rate}
\frac{dR}{d^4q}=\frac{-\alpha^2}{3\pi^3 q^2}\frac{1}
{1+e^{ q^0/T}}\text{Im}{\bf W}^F(q)
\end{equation}

where $\alpha=e^2/4\pi$, T is the temperature and 

\begin{equation}
\label{eq:WF}
{\bf W}^{F}(q)= i\int{d^4x}\text{ } 
e^{iq\cdot x}\text{Tr}\left[e^{-({\bf H}-\mu_B{\bf N}-\Omega)/T} 
T^* {\bf J}^\mu(x) {\bf J}_\mu(0) \right]
\end{equation}
where $e{\bf J}_\mu$ is the hadronic part of the electromagnetic current, 
{\bf H} is the hadronic Hamiltonian, $\mu_B$ is the baryon chemical 
potential, {\bf N} is the baryon number operator, and $\Omega$ is the 
Gibbs energy.  The trace is over a complete set of hadron states.

In order to take into account leptons with mass $m_l$ the right-hand side 
of Eq.~\ref{eq:rate} is multiplied by

\begin{equation}
\label{eq:lep_mass}
(1+\frac{2m^2_l}{q^2})(1-\frac{4m^2_l}{q^2})^{1/2}
\end{equation}
To compare the theoretical dilepton production rates with those 
observed in heavy ion collisions, the rates must be integrated over 
the space-time history of the collision region and then finally 
integrated over the dilepton pair's transverse momentum and rapidity 
in order to compare with the yields measured by the NA60 collaboration.  
The final expression for the rates is given as:
\begin{eqnarray}
\label{eq:acc}
\frac{dN}{dM}&=&2\pi M\int{dy}\int{dq_\perp}\cdot 
q_\perp\times Acc(M,q_\perp,y)\nn&\times&\int_{\tau_0}^{\tau_{f.o}}
{\tau d\tau}\int_{-\infty}^{\infty}{d\eta}\int_{0}^{r_{max}}{rdr}\int_{0}^{2\pi}{d\theta} \frac{d^4R}{d^4q d^4x}\left(M,|\vec{q}|, 
T,\mu_B,x\right)\nn
\end{eqnarray}
where $M=\sqrt{q^2}$ is the dilepton invariant mass, y is the dilepton 
pair rapidity, $\eta$ is the spatial rapidity, $q_\perp$ is the dilepton pair transverse momentum (with $\theta$ defined as the angle between $q_\perp$ and the fluid element's velocity), x is 
the hadron fraction, r is radial coordinate (with $r_{max}$ set by the freeze-out temperature), and
$Acc(M,q_\perp,y)$ is the experimental acceptance taking into account 
that the CERES detector covers a limited rapidity in the interval 
$y=2.9-4.5$ in the lab frame.  

The integration over $\eta, r, \theta$ and $\tau$ was done over the full hydrodynamic simulation of the collision region as described below.  $|\vec{q}|$ can be found by considering the two invariants $q^\mu q_\mu=M^2$ and $u^\mu q_\mu$ constructed from the dilepton momentum and fluid 4-velocity which can be expressed as:
\begin{align}
\label{eq:mtm_4vel}
&q^\mu = \left(M_\perp \cosh(y),q_\perp,M_\perp \sinh(y) \right) \nonumber\\
&u^\mu = \left(\gamma_\perp \cosh(\eta), \gamma_\perp v_\perp, \gamma_\perp \sinh(\eta) \right)
\end{align}
giving
\begin{equation}
\label{eq:qmag}
|\vec{q}| = \left[-M^2+\left(\gamma_\perp M_\perp \cosh(\eta)-u_\perp q_\perp\cos(\theta)\right)^2\right]^{1/2}
\end{equation}
where $u_\perp=\gamma_\perp v_\perp$, $\gamma_\perp=\frac{1}{\sqrt{1-v_\perp^2}}$, and $v_\perp$ is the transverse fluid velocity which is taken from the hydrodynamic simulation.

The acceptance function has a complicated dependence on $M, q_\perp$ and y, but since our rates are y-independent we have used an acceptance with $M$ and $q_\perp$ dependence built to specifications provided by the NA60 collaboration \cite{PC}.  Without detailed hadronic data available (such as transverse mass spectra and HBT analysis) a careful consideration of hadronic input, such as freeze-out temperature, cannot be made.  Therefore there is a large uncertainty in the overall normalization of the yields, which depends strongly on $T_{f.o.}$.  In addition, the assumption of boost invariance can also affect the normalization as the acceptance probes very specific rapidities.  The approach taken here is to normalize our results to the excess data in the peripheral centrality windows which fixes the normalization in the central bins.

\subsection{Spectrum above T$_C$}

At temperatures T$>$T$_C$ lattice calculations have predicted that 
the relevant degrees of freedom consists of (strongly) interacting
quarks and gluons.  In order to compute the dilepton production rates as 
one would expect from a conventional phase of quark-gluon plasma we use 
the Born q\={q} annihilation term \cite{cleymans, bellac} which for massless 
quarks is

\begin{equation}
\label{eq:Bornqqbar}
\text{Im} {\bf W}^R=\frac{1}{4\pi}\left( N_C 
\sum_{q=u,d,s}e^2_q \right) q^2 \left[ 
1+\frac{2T}{|\Vec{q}|}\ln(\frac{n_+}{n_-}) \right]
\end{equation}
where $N_C$ is the number of colors, $e_q$ the charge of the quarks, 
and $n_\pm=1/(e^{(q_0\pm|\Vec{q}|)/2T}+1)$.  It should be mentioned 
that~\ref{eq:Bornqqbar} reduces to the well-known vacuum result 
$(\text{Im} W^R=\frac{-q^2 N_C}{4\pi}\sum_q e_q^2)$ at T=0 and that the 
finite temperature rate is always smaller then the T=0 rate due to 
Pauli blocking.  

\begin{figure}
\centering
\includegraphics[scale=.85]{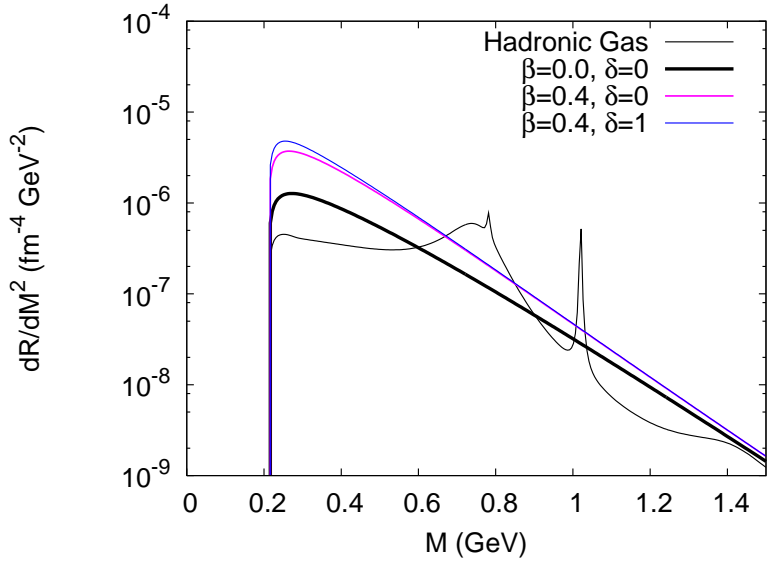}
\caption{Integrated dimuon rates from the plasma phase for T=150 MeV.  
The thick solid line shows the perturbative q\={q} annihilation rates 
while the this solid lines show the results for 
non-vanishing A$^2_4$, B$^2$ and E$^2$ condensates and for 
only a non-vanishing A$^2_4$ condensate.  
For comparison the integrated 
hadronic rate at T=150 MeV and $\mu_B=225$ MeV is also shown, which will 
be discussed in the next section.}
\label{fig:dimuon_rates_qgp}
\end{figure}

It has also been seen in lattice simulations that near the critical 
temperature $T_c$ there are still substantial chromoelectric and 
chromomagnetic condensates present leading to additional 
non-perturbative effects.  It was shown in \cite{LWZH98} that the 
enhancement to the dilepton rates in a plasma with non-vanishing 
chromoelectric and chromomagnetic condensates can be given by
\begin{eqnarray}
\label{eq:nonpert}
\text{Im} {\bf W}^R&=&\frac{1}{4\pi}\left( N_C 
\sum_{q=u,d,s}e^2_q \right) \left[ q^2 
\left< \frac{\alpha_s}{\pi}A^2_4 \right>-
\frac{1}{6}\left< \frac{\alpha_s}{\pi}E^2\right>+
\frac{1}{3}\left< \frac{\alpha_s}{\pi}B^2\right>  \right]\nn&\times& 
\left( \frac{4\pi^2}{T|\Vec{q}|}\right) 
\left( n_+(1-n_+)-n_-(1-n_-) \right)
\end{eqnarray}
where the values of the above condensates in~\ref{eq:nonpert} can only 
be estimated using non-perturbative calculations such as lattice QCD.  
The net result is a substantial enhancement (as seen in 
Fig.~\ref{fig:dimuon_rates_qgp} for the case of dimuons) of the dilepton 
production rates below an invariant mass of $\approx 0.4$GeV.  For the 
remainder of the paper we refer to the perturbative plasma rates as those 
given by~\ref{eq:Bornqqbar} and the non-perturbative plasma rates as those 
given the sum of equations~\ref{eq:Bornqqbar} and~\ref{eq:nonpert} using 
$\delta\equiv \left<\frac{\alpha_s}
{\pi}E^2\right>/(200 MeV)^4=\left<\frac{\alpha_s}
{\pi}B^2\right>/(200 MeV)^4=0$ and 
$\beta\equiv \left<\frac{\alpha_s}{\pi}A^2_4\right>/T^2=0.4$ 
which is the lower of the non-perturbative curves in 
Fig.~\ref{fig:dimuon_rates_qgp}.  An explanation of the choice 
of $\beta=0.4$ can be found in \cite{LWZH98}.

%pl_had_muon_muB225_T150_drdM2
%drdM2plot.gp

\subsection{Spectrum below T$_C$}

Even though there are various approaches to calculating production 
rates, they differ in the way in which the current-current correlation 
function in Eq.~\ref{eq:rate} is approximated and evaluated.  
The approach taken here is to use a chiral reduction formalism in 
order to reduce the current-current correlation function in~\ref{eq:WF} 
into a number of vacuum correlation functions which can be constrained 
to experimental $e^+e^-$ annihilation, $\tau$-decay, two-photon fusion 
reaction, and pion radiative decay experimental data.

For temperatures, T $\leq m_\pi$ and for nucleon densities, 
$\rho_N \leq 3\rho_0 $ the trace in Eq.~(\ref{eq:WF}) 
can be expanded in pion and nucleon states.  
Keeping terms up to first order in meson and nucleon 
density gives \cite{paper3}
\begin{equation}
\label{eq:exp}
\text{Im} {\bf W}^F(q)=-3q^2
\text{Im} {\bf \Pi}_V(q^2)+\frac{1}{f^2_a}
\int{da}{\bf W}^F_1(q,k)+\int{dN}{\bf W}^F_N(q,p)
\end{equation}
with phase space factors of
\begin{equation}
\label{eq:phase_space_N}
dN=\frac{d^3p}{(2\pi)^3}\frac{1}{2E_p}\frac{1}{e^{(E_p-\mu_B)/T}+1}
\end{equation}
and
\begin{equation}
\label{eq:phase_space_a}
da=\frac{d^3k}{(2\pi)^3}\frac{1}{2\omega_k^a}\frac{1}{e^{\omega_k^a/T}-1}
\end{equation}
with nucleon and meson energies of E$_p=\sqrt{m^2+p^2}$ and  
$\omega_k^a=\sqrt{m_a^2+k^2}$ respectively.

The first term in~\ref{eq:exp} is the transverse part of the isovector 
correlator $\langle 0|T^*{\bf VV}|0 \rangle$ which can be determined 
experimentally from electroproduction data and gives a result 
analogous to the resonant gas model.  At low and intermediate invariant 
mass the spectrum is dominated by the $\rho(770$ MeV) and $\rho'(1450$ MeV).

The term linear in meson density (the second term in Eq.~\ref{eq:exp}) can 
be related to experimentally measured quantities via the three flavor 
chiral reduction formulas \cite{CRF}.  It is shown in~\cite{paper1,paper3} 
that the dominant contribution comes solely from the part involving 
two-point correlators which gives:

\begin{align}
\label{eq:lin_in_meson}
&{\bf W}^F_1(q,k)=\frac{12}{f_\pi^2}q^2\text{Im} 
{\bf \Pi}_V^I(q^2)+\frac{12}{f_K^2}q^2\text{Im} 
\left( {\bf \Pi}_V^I(q^2)+\frac{3}{4}{\bf \Pi}_V^Y(q^2) \right)\nonumber\\
&-\frac{6}{f_\pi^2}(k+q)^2\text{Im} {\bf \Pi}_A^I 
\left( (k+q)^2\right)+(q\to-q)\nn&-\frac{6}{f_K^2}(k+q)^2\left[\text{Im} 
{\bf \Pi}_A^V\left( (k+q)^2\right)+\text{Im} {\bf \Pi}_A^U\left( 
(k+q)^2\right) \right] + (q\to -q)\nonumber\\
&+\frac{8}{f_\pi^2}\left( (k\cdot q)^2-m_\pi^2 
q^2\right) \text{Im} {\bf \Pi}_V^I(q^2)\times\Re 
\Delta_R^\pi(k+q)+(q\to-q)\nonumber\\
&+\frac{8}{f_K^2}\left( (k\cdot q)^2-m_K^2 q^2\right) 
\text{Im} \left({\bf \Pi}_V^I(q^2)+\frac{3}{4}{\bf \Pi}_V^Y\right)
\times\Re \Delta_R^K(k+q)+(q\to -q)
\end{align}
Where $\Delta_R^a$ is the retarded meson propagator given by 
$1/(q^2-m_a^2+i\epsilon)$ and ${\bf \Pi}_A$ is the transverse part of 
the iso-axial correlator $\langle 0|T^*{\bf j}_A{\bf j}_A|0 \rangle$.  The 
spectral functions appearing in Eq.~(\ref{eq:lin_in_meson}) can be 
related to both $e^+e^-$ annihilation as well as $\tau$-decay data 
as was compiled in \cite{Huang95}.  As already shown in \cite{paper1} 
the spectral function can be directly related the form factor, 
${\bf F}_V$, via the KSFR relation where ${\bf F}_V$ is 
parameterized in the common Breit-Wigner form where the 
resonance parameters and decay constants are taken from 
empirical data.  Included in the data are contributions to 
the spectral function from the $\rho, \omega, \phi, a_1, K_1$ 
and some of their radial excitations (see Table I in \cite{paper3}).

\begin{figure}
\centering
\includegraphics[scale=.85]{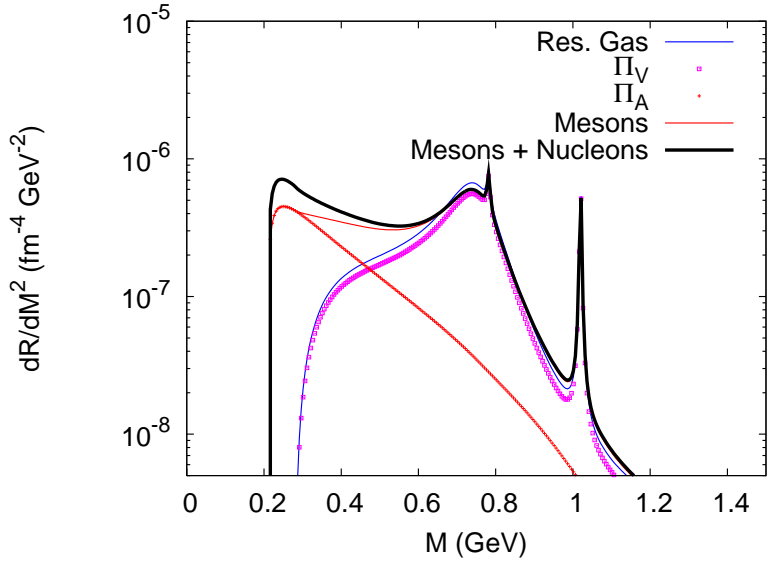}
\caption{The total integrated dimuon rates from a hadronic gas at T=150 MeV.  
The curve labeled Res. Gas shows the analogue of the resonant gas contribution (the first term in 
Eq. \ref{eq:exp}).  The points labeled ${\bf \Pi}_V$ and ${\bf \Pi}_A$ 
give the contributions from all of the respective spectral functions 
in equations \ref{eq:exp} and \ref{eq:lin_in_meson}.  The thin line 
labeled meson is the total rate given by a hadronic gas without 
nucleons.  The thick solid line gives the total dimuon rate when 
nucleons  (shown here for $\mu_B=225$ MeV) are taken into account.} 
\label{fig:dimuon_rates_had}
\end{figure}

It can be seen in Fig.~\ref{fig:dimuon_rates_had} that the term 
linear in meson density decreases the rates from the resonance 
gas contribution for the mass region above the two pion threshold.  
However below the two pion threshold the only contribution to the 
rates come from the ${\bf \Pi}_A$ terms in Eq.~\ref{eq:lin_in_meson}.  
This is because the axial spectral density is integrated over 
all momentum in the thermal averaging (Eq.~\ref{eq:exp}), 
which weakens the $(k+q)^2$ factor in Eq.~\ref{eq:lin_in_meson} 
allowing the $1/q^2$ term in Eq.~\ref{eq:rate} to dominate at low $q^2$. 

The final term in Eq.~(\ref{eq:exp}) which is proportional to 
the nucleon density is the spin-averaged forward Compton 
scattering amplitude of virtual photons off a nucleon.  
Experimentally, data is only available for values of $q^2\leq0$, 
so while the photon rate which requires $q^2=0$ can be determined 
by use of the optical theorem the contribution to the dilepton 
rates must be determined by chiral constraints.  Broken chiral 
symmetry dictates uniquely the form of the strong interaction 
Lagrangian (at tree level) for spin-$\frac{1}{2}$ particles.  
Perturbative unitarity follows from an on-shell loop-expansion 
in $\frac{1}{f_\pi}$ that enforces current conservation and 
crossing symmetry.  To one-loop, the $\pi N$ contribution is 
parameter free.  The large contribution of the $\Delta$ to the 
Compton amplitude near threshold is readily taken into account by 
adding it as a unitarized tree term to the one-loop 
result~\cite{paper2,SZ99}.  The enhancement in the dimuon rates 
due to a non-vanishing baryon density can be seen in 
Fig.~\ref{fig:dimuon_rates_had} where the solid curve shows the 
total dimuon spectra with an enhancement as large as a factor of 
two in the invariant mass region of $2m_\mu \leq M \text{(GeV)} \leq 0.6$.

\subsection{Hydrodynamic Evolution}

As mentioned earlier, in order to compare the theoretical dilepton production 
rates with those seen in heavy-ion collisions it is necessary to integrate 
these rates over the space time evolution of the collision region.  We 
consider a region localized in space-time consisting of thermal hadronic 
matter acting as a source of particles.  Equilibrium of the collision 
region is strictly a local property with different temperatures and 
baryon densities possible in different space-time domains.  

A computational hydrodynamic code was already developed by one of us and 
it has been modified to the conditions of the SPS collider for Indium 
on Indium collisions.  In this paper we only briefly outline the physics 
behind this code and show the results of the In-In collisions which has 
not been modeled before.  For technical details regarding the hydrodynamic 
calculations the reader is referred to the prior works by one of 
us \cite{Teaney01}. 

\subsubsection{Hydrodynamics}

The hydrodynamic equations for a relativistic fluid consist of
the local conservation of energy and momentum, which can be written in 
compact form as $\partial_u T^{\mu\nu}=0$, as well as local 
charge conservation $\partial_\mu J^\mu_i=0$ where 
$T^{\mu\nu}=(\epsilon+p)U^\mu U^\nu-pg^{\mu\nu}$ is the 
energy-momentum tensor with $\epsilon$ the energy density, $p$ 
the pressure, $U^\mu=\gamma(1,{\bf v})$ is the proper velocity of 
the fluid, and $J^\mu_i$ is any conserved current ({\em e.g.} 
isospin, strangeness and baryon number in the case of strong interactions).

The same space-time evolution scenario as first proposed by 
Bjorken \cite{bjorken83} is assumed where the equation of motion 
can be described by the Bjorken proper time $\tau=\sqrt{t^2-z^2}$ 
and the spatial rapidity $y=\frac{1}{2}\ln\frac{t+z}{t-z}$.  One of 
the main results, following from the assumption of a central-plateau 
structure in the rapidity distribution is that of boost invariance, 
stating that the initial conditions and thus the subsequent evolution 
of the system are invariant with respect to a Lorentz boost.  Thus a 
solution at any value of $y$ can be found by boosting the solution at 
$y=0$ to a new frame moving with velocity $v=\tanh (y)$ in the negative 
z-direction.  

With the assumption of boost invariance the equations of motion are a 
function of the transverse coordinates and the proper time $\tau$ only.  
After integrating over the transverse plane of the collision region one 
finds that $(dS_{tot}/d\eta)$, $(dn_B/d\eta)$, and the net transverse 
momentum per unit rapidity are all conserved.

\subsubsection{Equation of State}

In order to solve the equations of motion as given by the vanishing of 
the divergence of the energy-momentum stress tensor one must have 
an Equation of State (EoS) relating the local values of the pressure, 
energy density, and baryon density ($n_B$).  The approach taken here 
is to consider an EoS with a variable latent heat in the $e/n_B$ plane 
where the following two derivatives hold along a path of constant $n_B/s$:

\begin{equation}
\label{eq:speed_sound}
\left(\frac{dp}{de}\right)_{n_B/s}\equiv c^2_s
\end{equation}

\begin{equation}
\label{eq:dsoverde}
\left(\frac{ds}{de}\right)_{n_B/s} = \frac{s}{p+e}
\end{equation}

If the speed of sound is defined everywhere along with the entropy of 
one arc in the $e,n_B$ plane the above derivatives can be integrated 
in order to determine $s(e,n_B)$.  From the entropy all other 
thermodynamic variables, such as T and $\mu_B$, can be found as needed. 

We consider a hydrodynamic evolution that consists of three phases, a hadronic phase, 
a QGP phase, and a mixed phase.  The hadronic phase is taken to be made 
of ideal gas mixtures of the lowest SU(3) multiplets of mesons and 
baryons.  All intensive thermodynamic quantities including $p, e, s,$ 
and $n_B$ can be found as a sum of that quantity's contribution 
from each specie in the gas consisting of a simple Bose or 
Fermi distribution.  The hadronic phase is taken up to a temperature 
of $T_C \leq 170$ MeV and has a squared speed of sound of 
approximately $1/5c^2$.  For temperatures above $T_C$ only the 
squared speed of sound, $c^2_s$ is specified.  For the mixed phase 
it is taken almost at zero ($c_s^2 = 0.05c$). For the QGP phase the 
degrees of freedom are taken to be massless and the speed of sound is 
accordingly $c_s^2 = 1/3c$. The extension of this analysis to the
sQGP is beyond the scope of this work.

\subsubsection{Initial Conditions}

The initial conditions of the collision consist of setting the 
entropy and baryon density proportional to the number of 
participating nucleons in the transverse plane at some 
initial proper time $\tau_0 = 1$ fm/c.  Since both the 
entropy and baryon number per unit rapidity are conserved 
the final yields of pions and nucleons are proportional to 
the number of participants.  The number of participants 
were calculated by use of a Glauber model and the initial 
entropy and baryon densities were fixed by two constants 
$C_s$ and $C_{n_B}$, which respectively are the entropy 
and net baryon number produced per unit spatial rapidity 
per participant.  These constants were fixed to the 
conditions at the CERN SPS collider in order to fit the 
total yield of charged particles and the net yield of 
protons.  Table~\ref{tab:param} summarizes the input 
parameters used in the hydrodynamic calculations. In order to address the centrality of the collision the 
impact parameter was chosen in order to reproduce the 
number of participants as reported in \cite{Usai}. 

\begin{table}
\centering
\begin{tabular}{l|c}
\hline
Parameter & Value \\
\hline
$c^2_{mixed}$	& 0.05c	\\
$c^2_{QGP}$	& 0.33c	\\
T$_C$	& 170 MeV\\
T$_{f.o.}$	&	130 MeV\\
$\tau_0$	&	1.0 fm/c\\
$n_B/s$	& 0.0238\\
$C_s$	&	8.06\\
$C_{n_B}$	&	0.191\\
\hline
\end{tabular}
\caption{\label{tab:param} Parameters used in the hydrodynamic 
simulation of In-In collisions.}
\end{table}

\subsection{Results for In-In Collisions at CERN SPS}

\begin{figure} 
\centering
\includegraphics[scale=1.25]{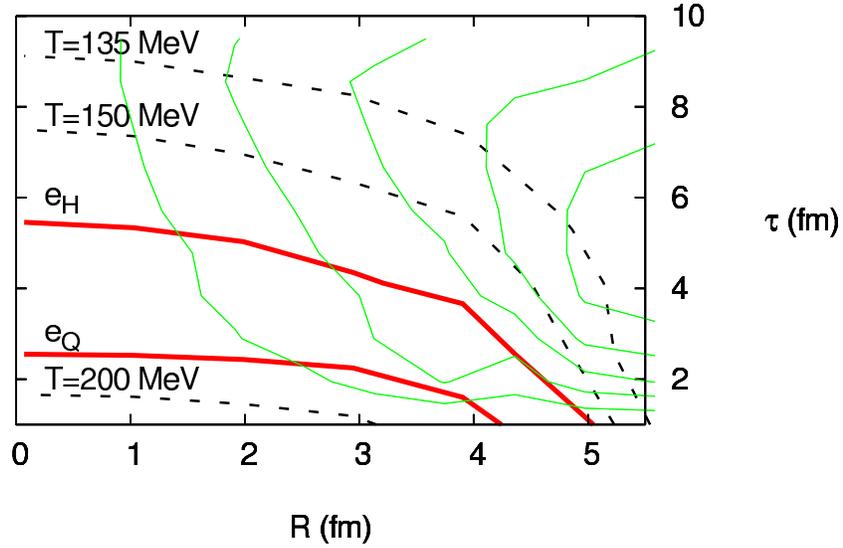}
\hspace{10pt}
\caption{The hydrodynamic solution for semi-central In-In collisions 
at the SPS collider.  The thin lines show contours of constant 
transverse fluid rapidity $(v_\perp=tanh(y_\perp))$ with 
values of 0.1,.02,..,0.5.  The dashed lines show contours of 
constant temperature with values of (working radially outward) 
T=200 MeV, T=150 MeV and T=135 Mev.  The $e_Q=1.70$ GeV/fm$^3$ 
and $e_H=0.50$ GeV/fm$^3$ contours represent the phase changes 
from QGP to mixed and from mixed to hadronic matter respectively.}
\label{fig:RvsTauSC}
\end{figure}

\begin{figure}
\centering 
\includegraphics[scale=0.8]{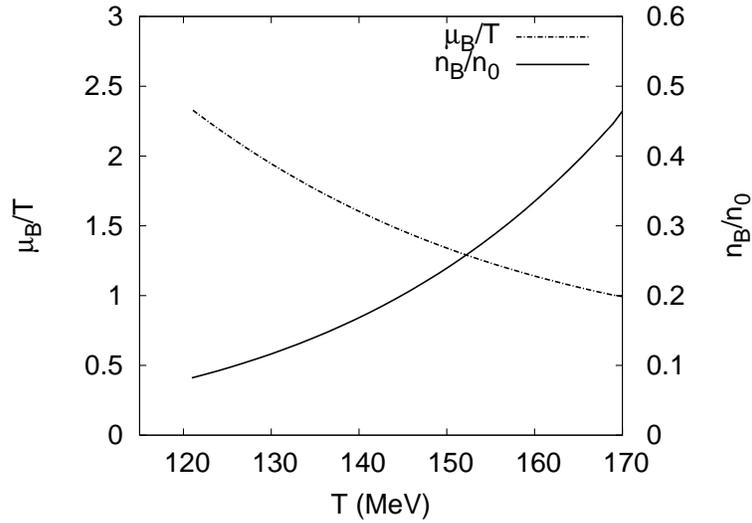}
\caption{Dependence of the baryon chemical potential, 
$\mu_B$, on the temperature for the hadronic phase.}
\label{fig:muBvsT}
\end{figure}

The hydrodynamic result for In-In Semi-Central collisions is shown 
in Fig.~\ref{fig:RvsTauSC}.    The two thick lines labeled $e_Q$ and 
$e_H$ represent contours of constant energy density showing the 
transition from the plasma phase to the mixed plasma and hadronic 
phase and the transition from the mixed to the purely hadronic 
phase respectively.  It can be seen that the QGP phase takes up 
a much smaller space-time volume then the hadronic phase, however 
the rates still appear in the spectrum as the high temperatures 
in this region enhance the rates by an order of magnitude.  
The effect of nucleons depends on the baryon chemical potential 
throughout the evolution.  This is plotted as a function of temperature 
for the pure hadronic phase in Fig.~\ref{fig:muBvsT}. 

\subsection{Results and Discussion}

\begin{figure}[hbtp]\label{fig:result}
\centering
\includegraphics[scale=.5]{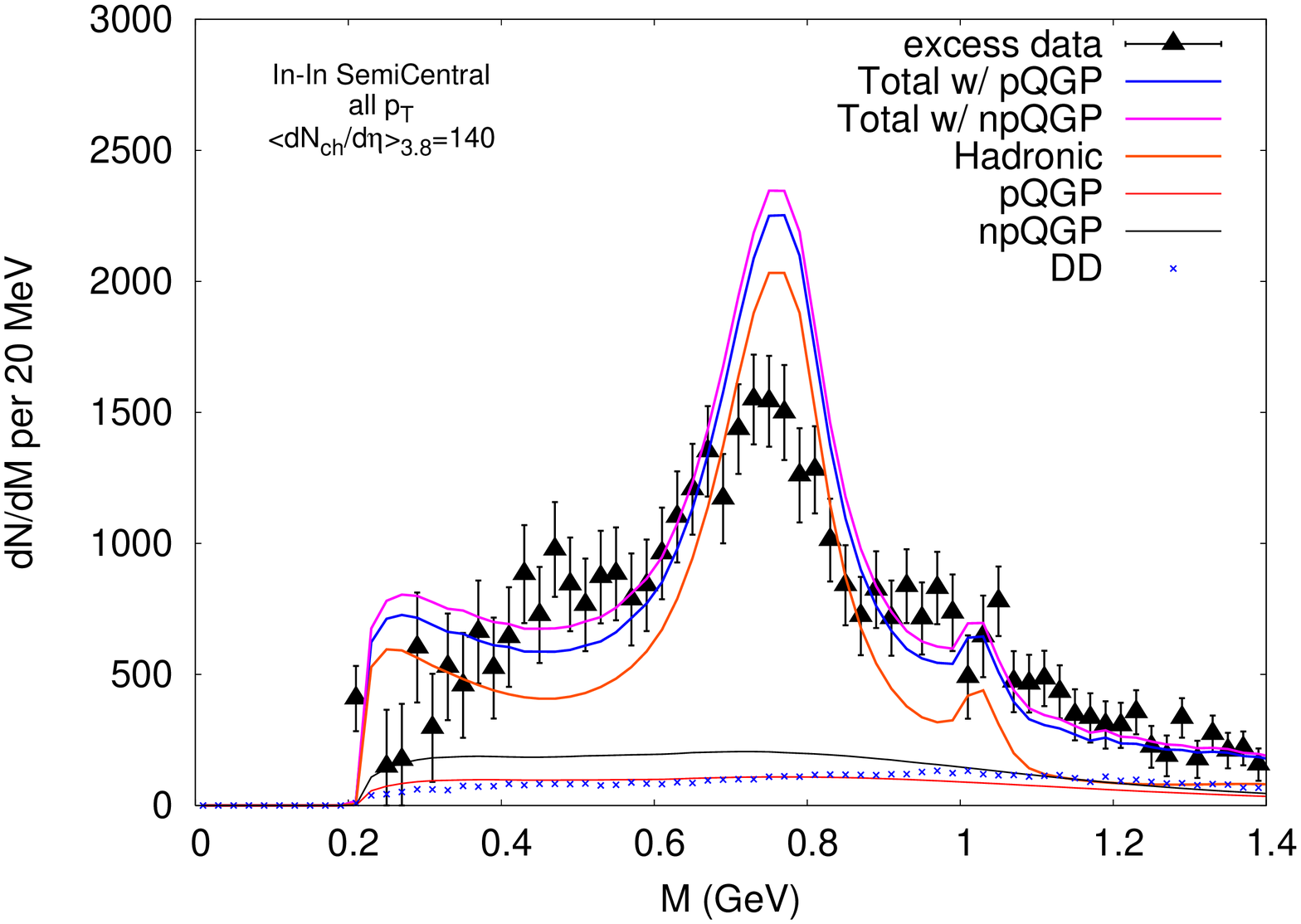}
  \vspace{9pt}
\includegraphics[scale=.5]{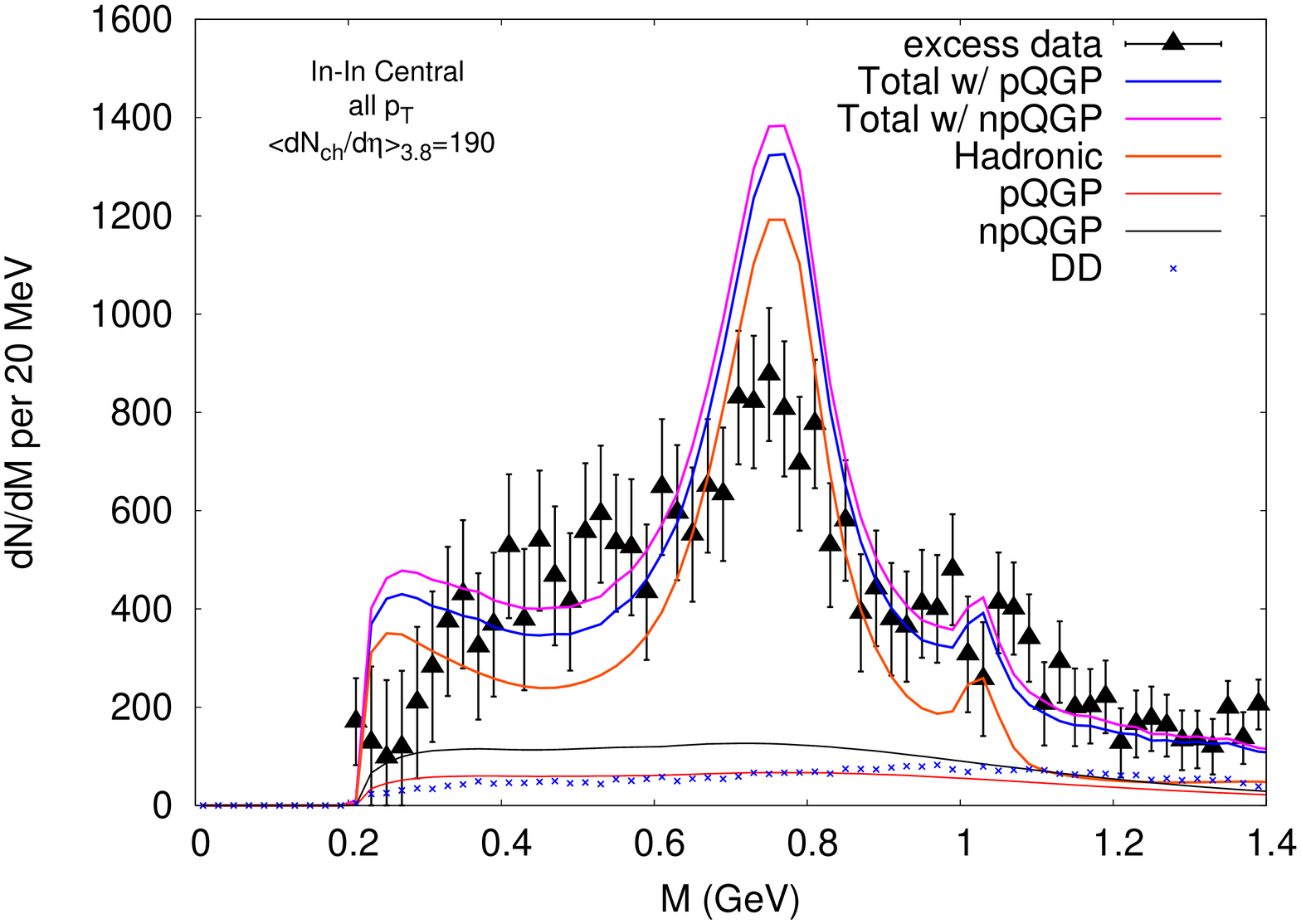}
  \caption{ NA60's excess dimuon data compared to our thermal yields which include contributions from either the perturbative or non-perturbative QGP phase, the hadronic phase and the D\={D} contribution.  Shown for all four centrality windows.}
  \label{result}
\end{figure}

Our final dimuon yields for the four centrality windows is shown in 
Fig.~\ref{fig:result} where it is compared to the excess data measured by the 
NA60 collaboration.  In all four figures we show the total dimuon yield, which includes contributions from the hadronic phase, either the perturbative or non-perturbative plasma phase, as well as the D\={D} contribution as provided by the NA60 collaboration.  For all cases we also show separately the perturbative and non-perturbative plasma contributions to the overall yield.  It can be seen to be almost negligible in the peripheral data.  For the central data, where there is a larger plasma contribution we also show curves showing separately the hadronic contribution.
 
Even though it can 
be seen that the theoretical rates are able to describe most 
of the features of the spectrum, a number of things should be 
noted before a direct comparison is made.  The rates below M=0.4 GeV 
should not be taken literally since they are obtained by 
saturating the total measured yield in that region 
by $\eta$ Dalitz decays only, thereby lowering the 
excess close to the dimuon threshold.  Actually, by reducing 
the $\eta$ yield by 10\% the data has much better agreement with 
the theory for M$\leq0.4$ GeV.  The charm decay data was analyzed 
and provided by NA60, and since the contribution from charm 
decays is not subtracted from the excess data it must added to 
our rates for comparison with experiment.  The excess spectra 
which is shown in the figure is created from subtracting the 
cocktail (omitting the $\rho(770)$) from the total observed data.  
This would erase any $\omega$ or $\phi$ peak at the vacuum positions.  
Since our hadronic rates don't modify either the position or 
width of the $\omega$ or $\phi$ it can be very difficult to 
distinguish any residual $\omega$ or $\phi$'s from the cocktails'.  

It can be seen right away that the dimuon yields are reproduced in the peripheral centrality windows.  This is expected as the matter is dilute 
and any medium modification to the spectral densities will be accounted for in the virial expansion (Eq.~\ref{eq:exp}).   In the central bins it can be seen that the shape of the spectrum changes as one goes to more central collisions.  Even though the general shape of the spectrum is reproduced by our rates, our rates slightly over-predict the yield at the $\rho$ peak by about $\approx 50\%$ for semi-central and by $\approx 60\%$ for the central data.  Even though our rates agree fairly well with the remaining data away from the $\rho$ peak, there is still room for enhancement in the low mass region, $0.4 \leq \text{M (GeV)}  \leq 0.6$. 

We should finally mention what happens when the non-perturbative 
QGP rate is used instead of the perturbative result.  Similar 
to the perturbative QGP results in Fig.~\ref{fig:result} 
the non-perturbative plasma rate is about a factor or two larger in the low mass region.  Even though this does help to explain
some of the excess in the low mass region, especially in the more central data, the space time volume of the plasma phase is too small
to have a large effect.

\begin{figure}[hbtp]\label{fig:result2}
\centering
\includegraphics[scale=.5]{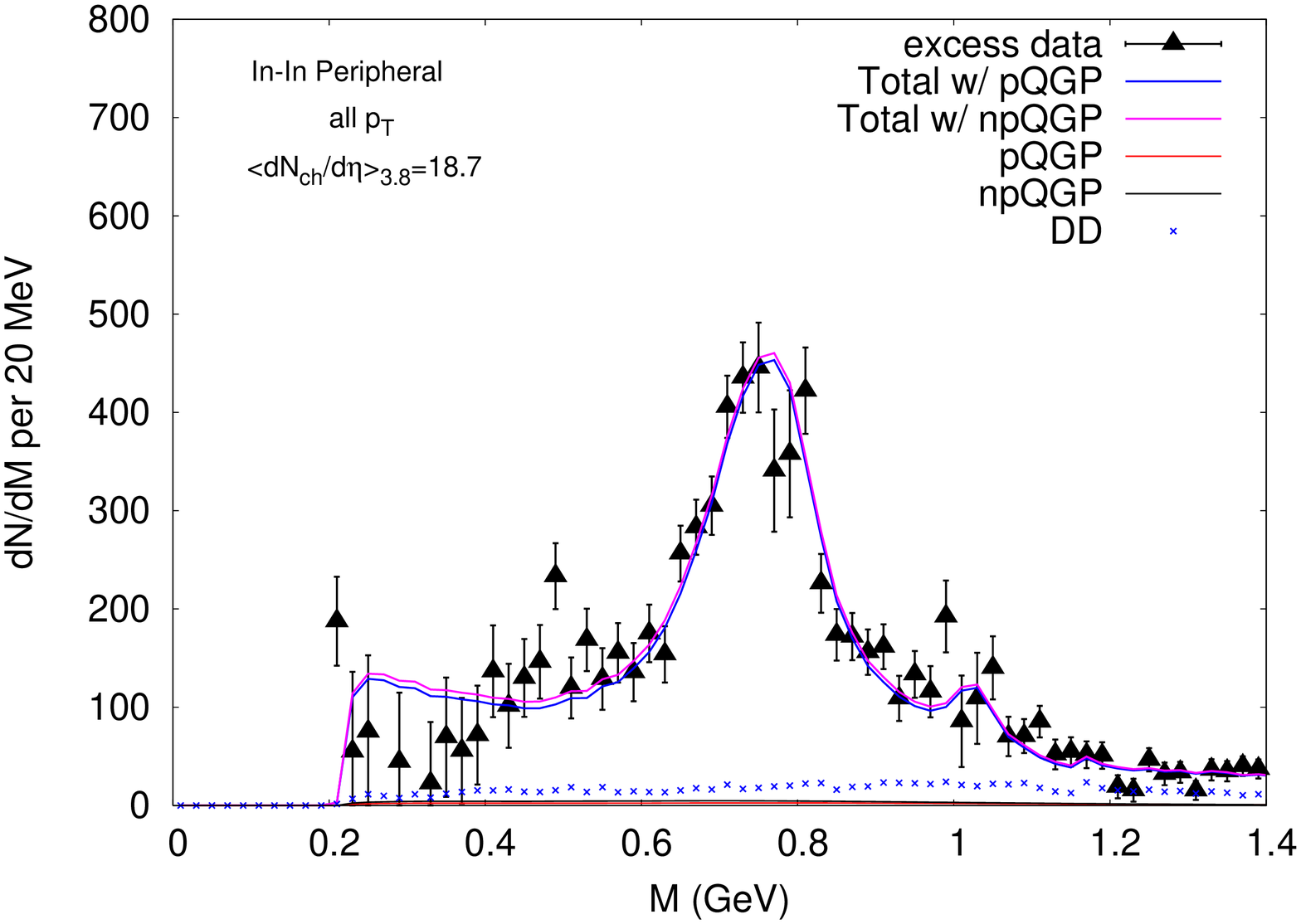}
  \vspace{9pt}
\includegraphics[scale=.5]{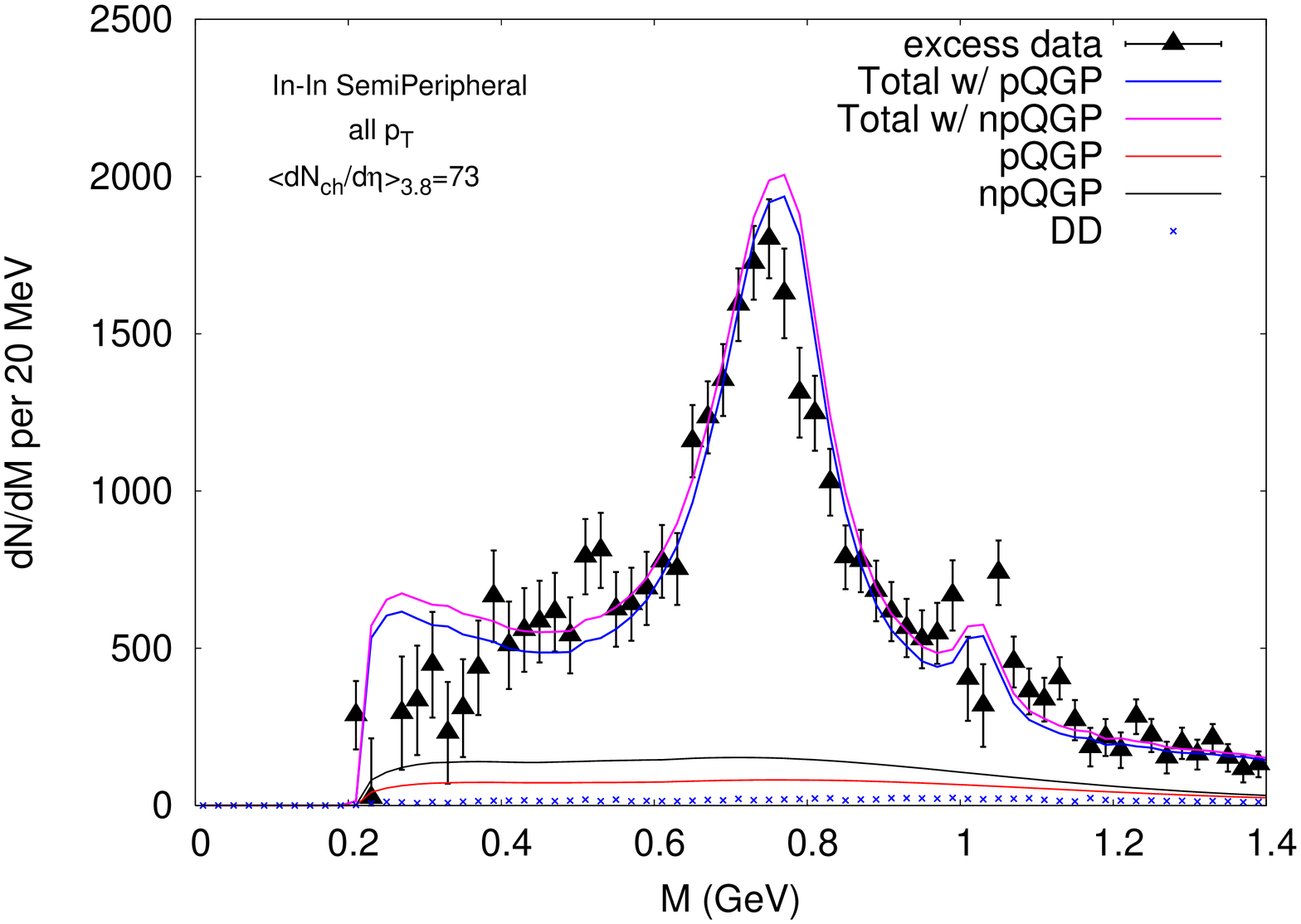}
\end{figure}  

\subsection{Conclusions}

Using a parameterization of the results given by a hydrodynamic model 
of the collision region at the CERN SPS collider, the NA60 dimuon 
spectrum was reproduced using a pure thermal model assuming that 
there exists a sQGP phase above $T_C$ with an interacting hadronic 
phase persisting until freeze-out.  The dimuon spectrum from the sQGP 
phase originates primarily from q\={q} annihilation with non-perturbative 
effects due to non-vanishing gluon condensates.  After hadronization it 
is assumed that there remains a dilute hadronic gas in which the dimuon 
rates can be constrained entirely from chiral symmetry arguments and 
experimental data.  The combination of these two rate equations, 
after being folded over the space-time evolution of the collision region, 
are able to explain most of the excess dimuon data, especially in the more peripheral collisions where our 
assumptions about diluteness hold.  In the more central data, where the assumption of diluteness may 
breakdown, it is necessary to investigate how higher order terms in the virial expansion modify the spectrum.

\section{Di-electrons at RHIC}

In this section we now discuss the implications for the recent measurements by the PHEINIX collaboration at RHIC \cite{:2007xw,Toia:2007yr}.  Being a continuation of the work from the previous section (see also \cite{Dusling:2007su}) we do not go through all the details of the model for the hydrodynamic evolution. Table~\ref{tab:paramAu} summarizes the relevant parameters of the hydrodynamic simulation used in this section.  
\begin{table}[h]
\centering
\begin{tabular}{l|c}
\hline
Parameter & Value \\
\hline
EOS	& Lattice Motivated	\\
T$_C$	& 190 MeV\\
T$_{f.o.}$	&	120 MeV\\
$\tau_0$	&	0.2 fm/c\\
impact parameter (b)	&	0 fm\\
\hline
\end{tabular}
\caption{\label{tab:paramAu} Parameters used in the hydrodynamic simulation of Au-Au collisions.}
\end{table}

\begin{figure}
\centering
\includegraphics[scale=.4]{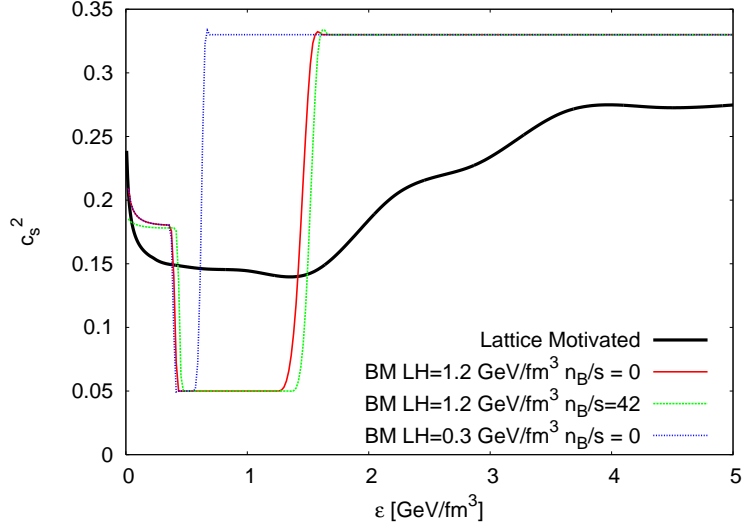}
\caption{Summary of the equations of states used in this work.  The lattice motivated EOS was taken from \cite{Laine:2006cp} and is used in the analysis of the RHIC data in this section.  The BM EOS consists of a first order phase transition having a variable latent heat (shown for LH=1.2 GeV/fm$^3$ and 0.3 GeV/fm$^3$.)  }
\label{fig:EOS}
\end{figure}

There are some modifications that should be addressed.  First, we now start the hydrodynamic evolution at $\tau_0=0.2$ fm/c in order to account for some of the pre-equilibrium dilepton production in the QGP stage of the evolution.  This will increase the yields in the high mass region above the $\phi$.  It was already shown in the prior section (see also \cite{Dusling:2008xj}) that non-equilibrium effects modify the dilepton $p_T$ spectrum but not the invariant mass spectrum.  A second modification is the use of a more realistic equation of state motivated (taken from \cite{Laine:2006cp}) by recent lattice measurements.  In figure \ref{fig:EOS} we show the squared speed of sound versus energy density for four different equation of states.  The solid black curve is the lattice motivated EOS.  The other three curves are bag model (BM) equation of states ({\em i.e.} first order phase transition) with variable latent heat.  We show the BM EOS having a latent heat of 1.2 GeV/fm$^3$ and 0.3 GeV/fm$^3$.  Also shown is the EOS having latent heat 1.2 GeV/fm$^3$ with a fixed baryon number to entropy density ($n_B/s=42$). 

\begin{figure}
\centering
\includegraphics[scale=.4]{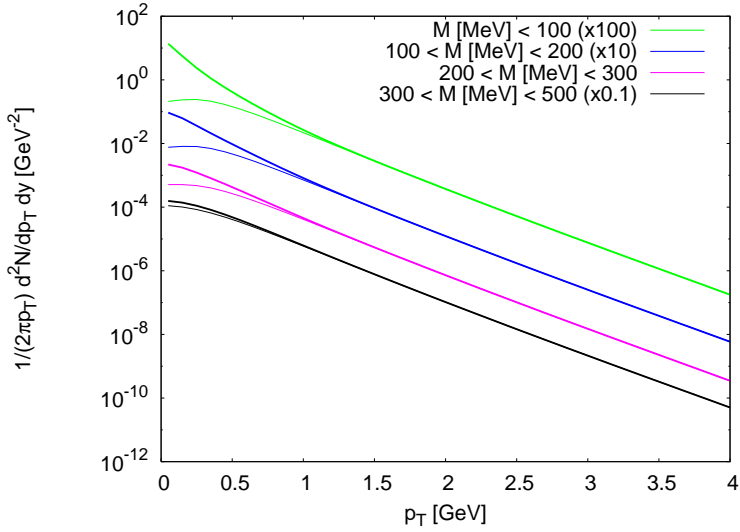}
\caption{Differential $p_T$ spectra in different mass windows.  The thin lines include the zeroth and first order, $W^F_\pi$, contribution in pion density while the thick lines also included the second order, $W^F_{\pi\pi}$, contribution.}
\label{fig:spectraW2}
\end{figure}

We now discuss the results of the space-time integration.  In fig.\ref{fig:spectraW2} we show the differential $p_\perp$ yields in four different mass windows.  The thin lines show the zeroth and first order contribution, $W_\pi$, and the solid lines include the additional two pion contribution, $W_{\pi\pi}$.  We find a large enhancement from the two pion contribution at low $p_\perp$ which can be described as arising from Bremsstrahlung type processes.  

\begin{figure}
\centering
\includegraphics[scale=.4]{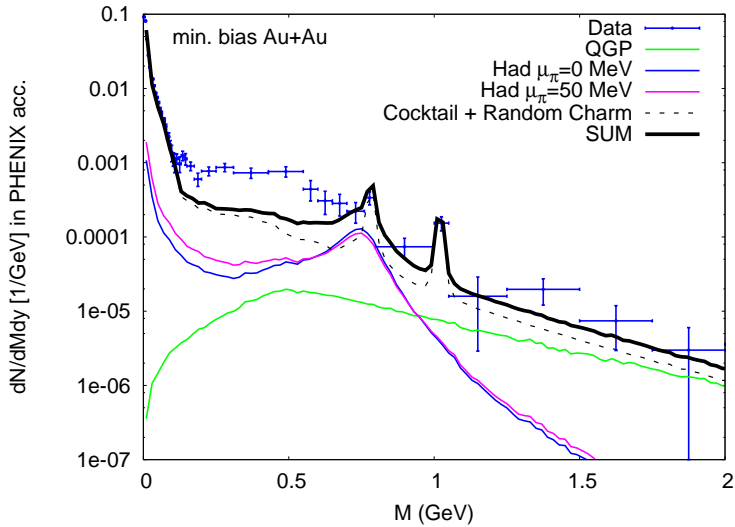}
\caption{The data points show the measured di-electron spectrum from PHENIX.  The dotted line shows the hadronic cocktail provided by PHENIX including the charm contribution.  The solid curve labeled 'SUM' includes the cocktail, QGP and hadronic gas at $\mu_\pi=50$ MeV.}
\label{fig:massAcc}
\end{figure}

\begin{figure}
\centering
\includegraphics[scale=.4]{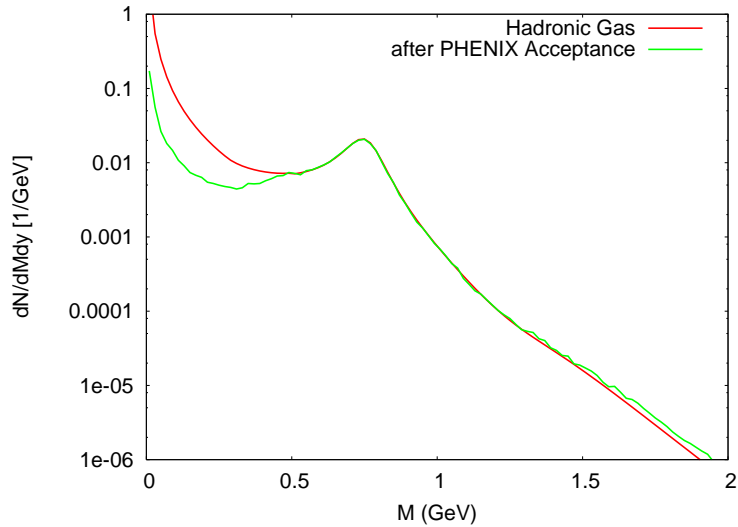}
\caption{Hadronic dilepton yields before and after the PHENIX acceptance.}
\label{fig:massW2compare}
\end{figure}

Figure \ref{fig:massAcc} shows the invariant mass spectrum after going through the PHENIX acceptance.  Shown separately are the contributions form the QGP and hadronic phases.  Two scenarios are considered.  The first is a hadronic phase with zero pion chemical potential.  The second is the same hadronic evolution with a constant pion chemical potential of $\mu_\pi=50$ MeV enhancing the rates at low mass throughout the entire evolution.  The solid black line shows the sum of the contributions from the cocktail, QGP and hadronic phase at $\mu_\pi=50$ MeV.  We find that after the acceptance cuts the one and two pion final state contributions are not able to explain the low mass enhancement seen in the data.   
In order to understand these cuts further we show in figure~\ref{fig:massW2compare} the hadronic rates before and after the acceptance cuts.  The yields in this figure are normalized at the $\rho$ peak.  It is clearly seen that the acceptance at PHENIX reduces the rates by about a factor of three or more for $M < 0.5$ GeV.  To a good approximation, it turns out that the $W^F_{\pi\pi}$ does not contribute at all after the acceptance cuts.  Even though we find a large enhancement at low $p_\perp$ as seen in fig.~\ref{fig:spectraW2}, the {\em sweet spot} for the PHENIX detector is for dilepton pairs having $m_\perp \geq 0.4$ GeV while our $2\pi$ enhancement feeds in below this kinematic region.
\begin{figure}[h!]
\centering
\includegraphics[scale=.4]{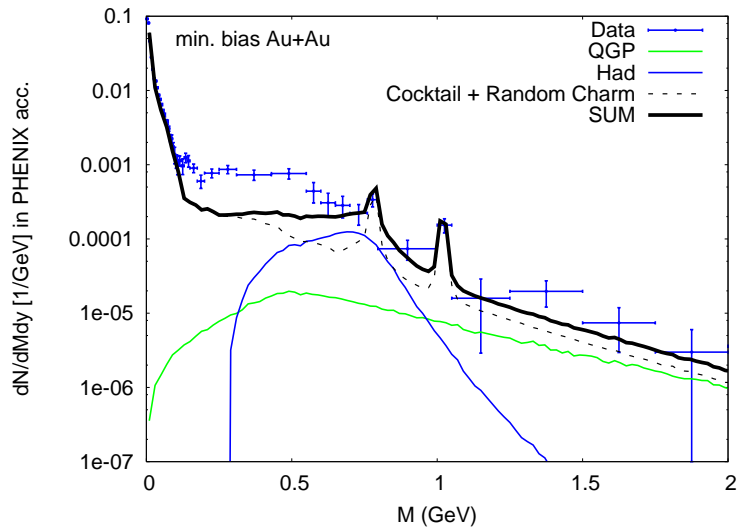}
\caption{PHENIX di-electron results compared to thermal $2\pi$ annihilation with a collisionally broadened $\rho$.}
\label{fig:massCB}
\end{figure}
In order to further the comparison with other works we also show the yields from $2\pi$ annihilation proceeding through an intermediate $\rho$ meson with a medium modified propagator.  The calculation of the in-medium width is discussed in section \ref{sec:coll}.  In figure~\ref{fig:massCB} we show the hadronic rates from only $2\pi$ annihilation with collisional broadening of the $\rho$ propagator.

Our results are consistent with the microscopic calculation of \cite{Bratkovskaya:2008bf} when using a collisionally modified rho showing that we at least have good constraints on the hydrodynamic evolution.  Of course, we cannot explain any of the low mass emission since there is no thermal emission below the $2\pi$ threshold in this model.  

In summary, thermal hadronic emission helps explain the excess above the cocktail near the $\rho$ mass.  However, the low mass ($M\leq0.6$ GeV) dilepton excess remains elusive.   

\section{Role of Viscosity}

In this section we briefly discuss the effect that shear viscosity has on the dilepton spectrum.  The motivation for looking at the viscous correction is twofold.  First, there is the recently conjectured universal lower bound on the shear viscosity to entropy ratio, $\eta/s=1/4\pi\approx 0.08$ \cite{Kovtun:2004de}.  Therefore, one should see how viscosity modifies the current results which rely on kinetic equilibrium.  Second, the empirical hadronic data on elliptic flow and $p_\perp$ spectra seem to support a small but non-zero shear viscosity.  Since many of the simulations of the dilepton data at RHIC and CERN rely on kinetic equilibrium it will be interesting to see how off-equilibrium corrections modify the result.  Including viscosity will also help make contact with microscopic kinetic theory calculations.

We will look at viscosity in the context of In-In collisions at NA60.  Recently NA60 has measured the inverse slope ($T_{eff}$) of the $p_\perp$ spectra as a function of invariant mass.  Many of the theoretical calculations, including ours, were unable to account for the large rise in $T_{eff}$ as a function of mass up to the about 1 GeV.  After 1 GeV there is a sudden drop in the effective temperature which is claimed by the NA60 collaboration to have origins from early emission before the build up of collective flow thus resulting in its lower effective temperature.

\begin{figure}[hbtp]\label{fig:TeffLH}
\centering
\includegraphics[scale=.45]{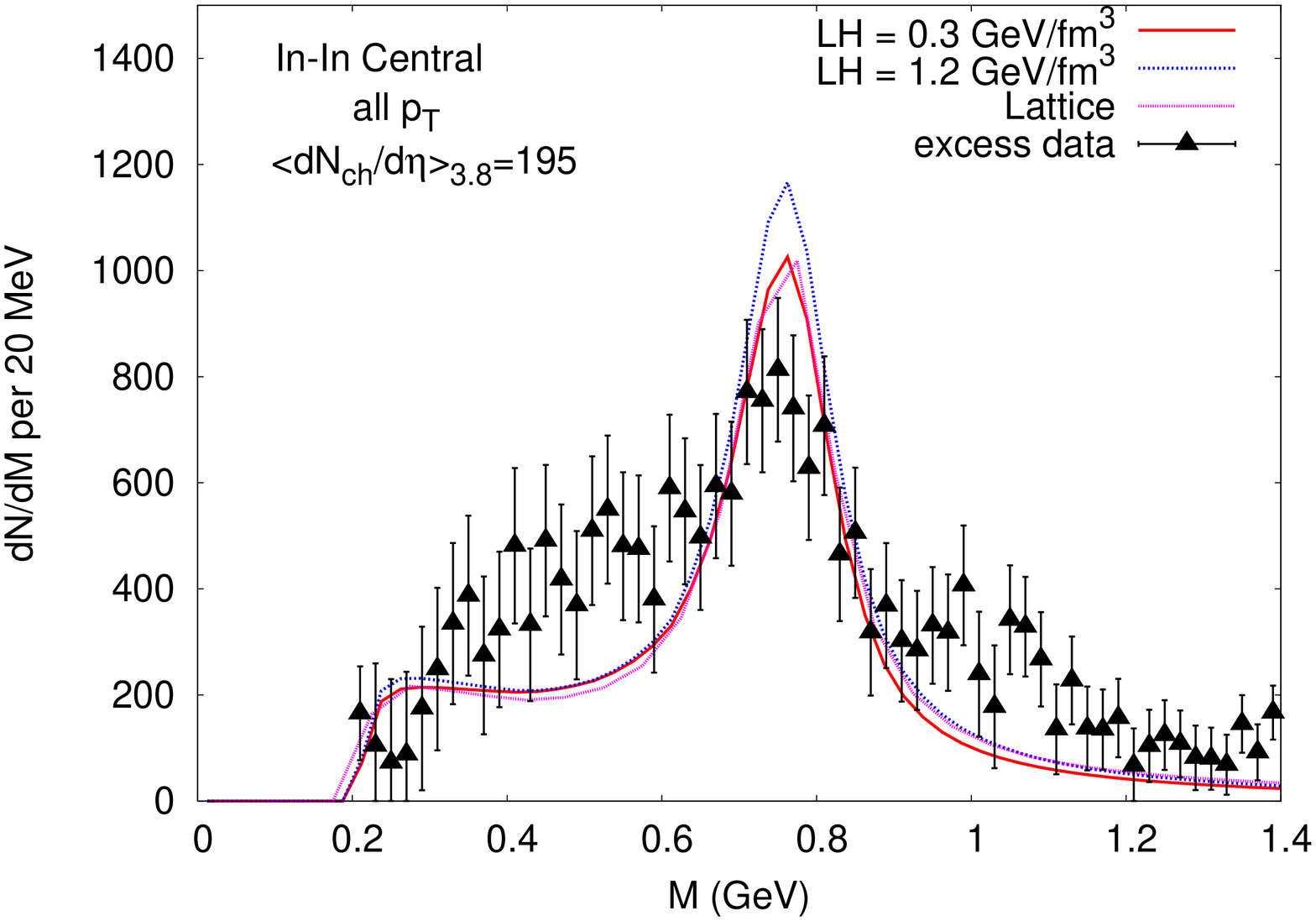}
  \vspace{9pt}
\includegraphics[scale=.45]{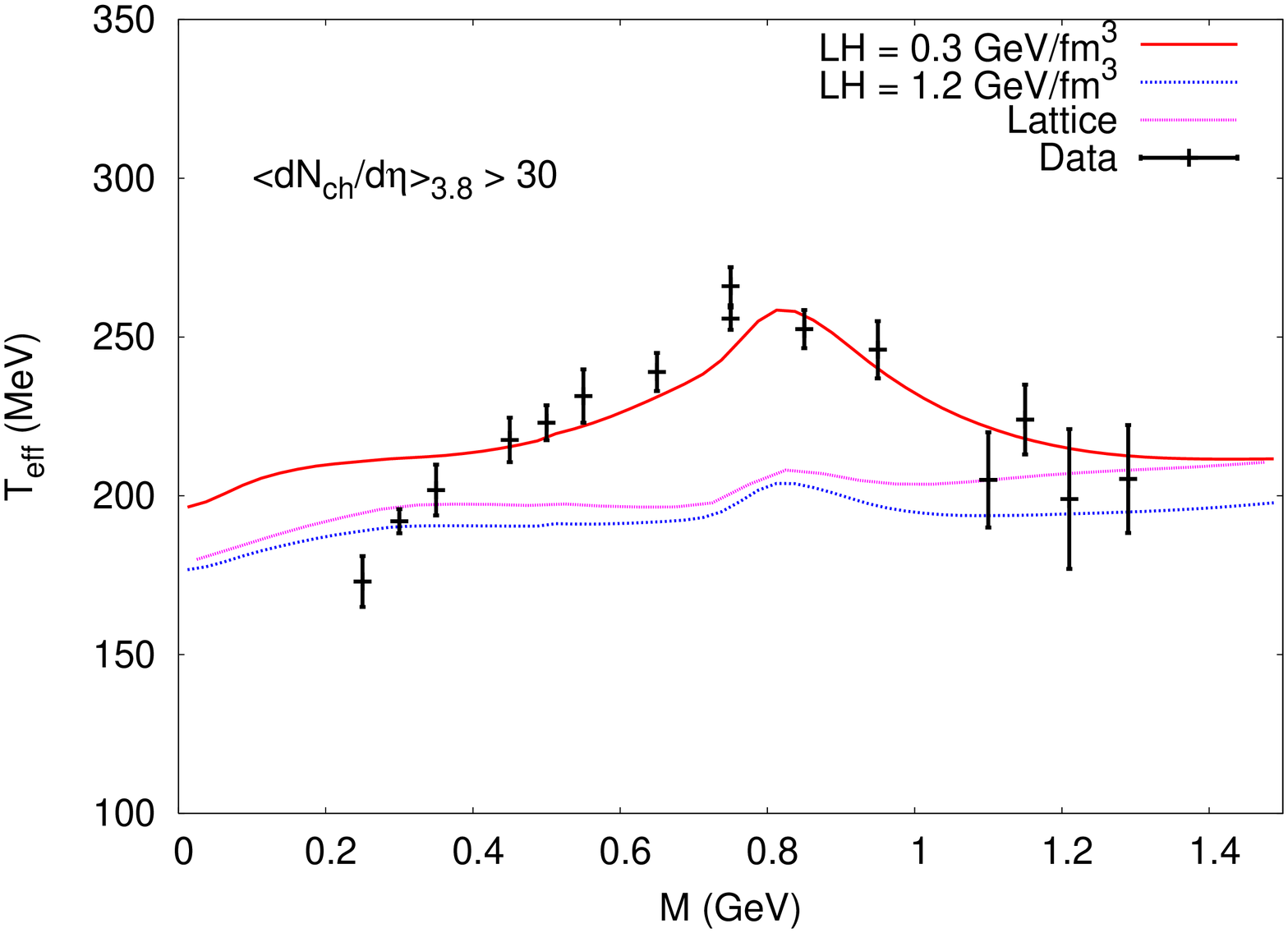}
\caption{The above figure shows the effect of changing the latent heat on the invariant mass spectra (top) and effective temperature (bottom) for central In-In collisions.  Both hydrodynamic evolutions use $T_c=165$ MeV, $\tau_0=0.6$ fm/c and $T_{f.o.}=130$ MeV. Also shown for comparison is the result using the Lattice EOS.}
\end{figure}

First let us look at our previous calculation (chapter 5.1) and see the effect of the changing the latent heat of the phase transition.  Our previous calculation used a latent heat of 1.2 GeV/fm$^3$.  Figure~\ref{fig:TeffLH} compares the results for a latent heat of 1.2 GeV/fm$^3$ and a much smaller latent heat of 0.3 GeV/fm$^3$.  In terms of the mass spectra we find that the larger latent heat allows for more dilepton emission.  This is simply because the space-time volume of the phase transition region is much larger.  However, a large difference is seen in the $T_{eff}$ results.  We find that LH=0.3 leads to much higher temperature which is consistent with the data.  There are two reasons for this large increase.  First, a smaller latent heat is effectively a much harder EOS (see fig.~\ref{fig:EOS}).  Second, differences in the space-time evolution lead to a larger fraction of the yield coming from the freeze-out contribution versus the hadronic contribution in the LH=0.3 EOS.

Even though LH=0.3 does a decent job in explaining the $T_{eff}$ data it will be hard to also explain hadronic observables in the same framework.  It was shown in \cite{Teaney:2001av} that a latent heat of 0.8 GeV/fm$^3$ does the best job of fitting hadronic observables at the SPS.  Second, this equation of state is far from consistent with the lattice motivated EOS of fig.~\ref{fig:EOS}.  It turns out that the LH=0.8 and the lattice EOS lead to very similar effective temperatures in the dilepton spectra.      

With the above in mind we now examine how viscous corrections modify the dilepton spectra.  We start by summarizing how shear viscosity modifies the particle / dilepton spectra by separating its role into two main effects.
\begin{itemize}
\item Shear viscosity modifies the ideal equations of motion.  For central collision its main effect is to increase the transverse flow of the medium.  This effect integrates over time and the largest change in the transverse flow profiles are seen at the latest times.  The higher flow velocity will manifest itself by producing harder $p_\perp$ spectrum of the produced particles at freezeout.
\item Shear viscosity modifies the particle distribution function.  The viscous correction to the distribution function schematically goes as ~$1/T\tau$.  Therefore the viscous correction starts off large and becomes smaller as the system evolves in proper time.  For realistic freezeout out surfaces it is found that the $p_T$ spectrum hardens when including the viscous correction to the distribution function.
\end{itemize}
One can now easily see how viscosity modifies the effective temperature.  At early times the viscous correction to the distribution function causes an increase in $T_{eff}$ while at later times it is instead the larger transverse flow which increases $T_{eff}$.  This first effect will be most relevant during the QGP phase and the later effect most relevant during the hadronic phase.

Let us now look at a quantitative example.  We consider the hydrodynamic evolution of central In-In collisions starting at $\tau_0=1$ fm/c.  The QGP phase has a shear viscosity to entropy ratio of $\eta/s=0.08$ and we consider a hadronic phase both with ($\eta/s=0.75$) and without ($\eta/s=10^{-6}$) shear viscosity present.  We also consider two scenarios for the dilepton production in the hadronic phase.  In one case we consider the hadronic reaction $\pi\pi\to\rho\to\gamma^*\to e^+e^-$ where the width of the intermediate $\rho$ meson is collisionally broadened.  In this case one can include the viscous corrections to the rate in the same manner as was done for the q\={q} case in chapter 4.  We also do the calculation using the rates derived from the chiral reduction formalism in Chapter 2.  However, as these rates relied on the use of kinetic equilibrium it is not trivial to include the viscous correction to the particle distribution function.  Not including the viscous correction to the particle species is of course inconsistent, however, it is found that in the hadronic phase the largest viscous effects come from changes in the flow profile.  Regardless, one can think of this result as showing the contribution from {\em flow} effects alone in the hadronic phase.

\begin{figure}[hbtp]\label{fig:TeffC}
\centering
\includegraphics[scale=.45]{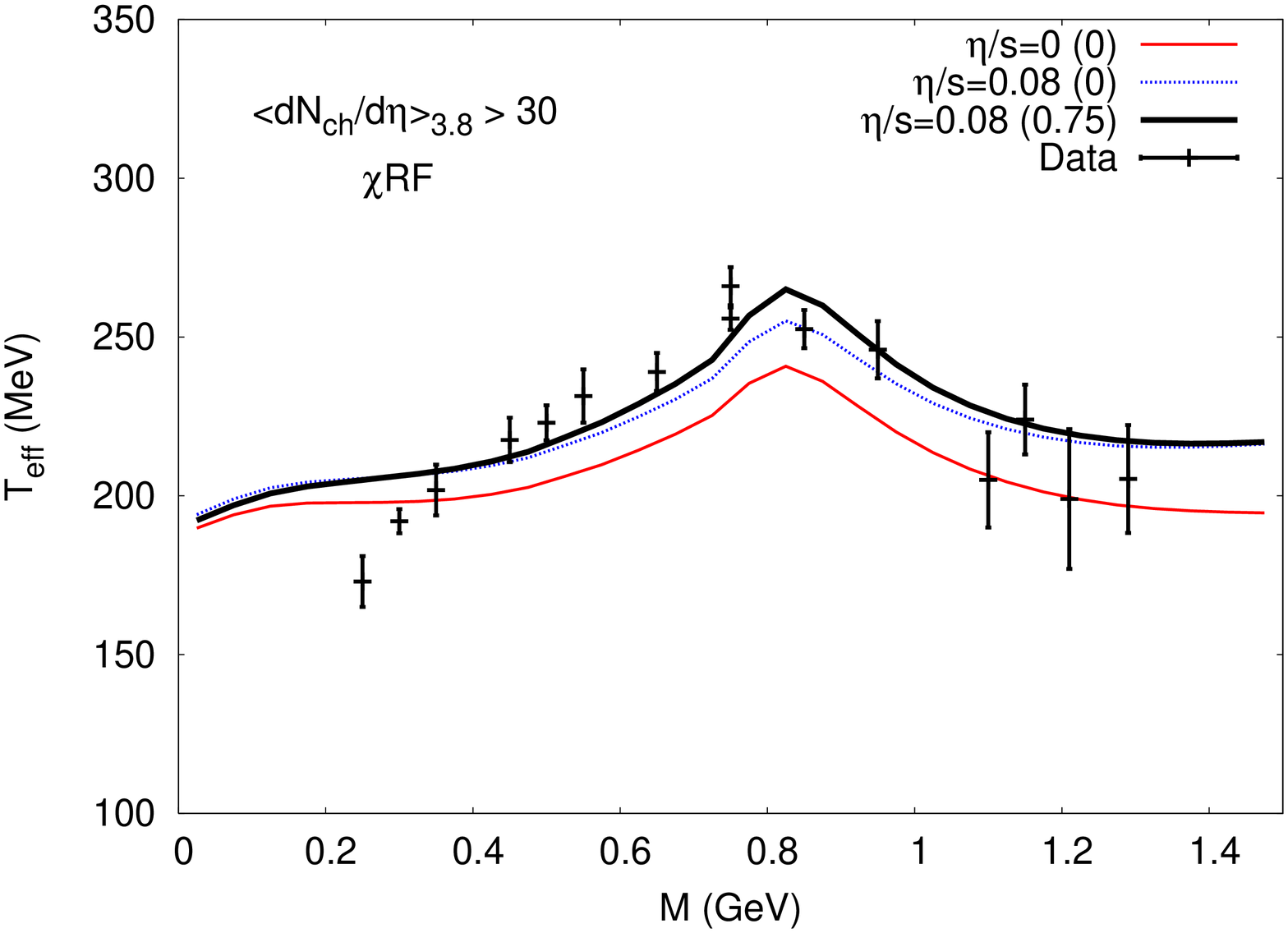}
  \vspace{9pt}
\includegraphics[scale=.45]{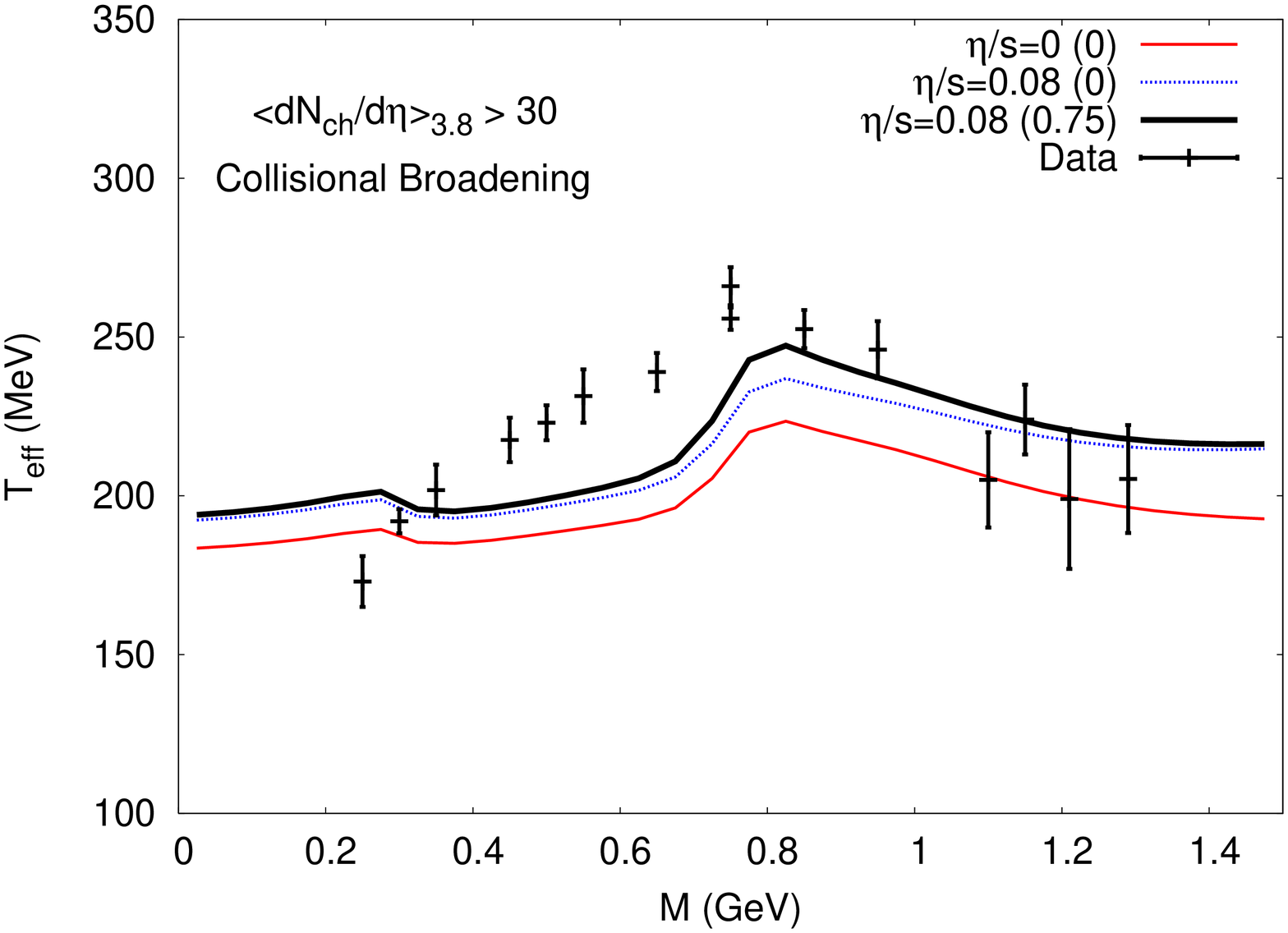}
\caption{Effective temperature as a function of mass when the hadronic phase is constrained by the chiral reduction formula (top) or by thermal production through a medium modified rho (bottom).  In each label the number in parenthesis is the viscosity in the hadronic phase.  The upper curve in both plots is for $\eta/s_{qgp}=0.08$ and $\eta/s_{had}=0.75$.  The middle curves are $\eta/s_{qgp}=0.08$ and $\eta/s_{had}=0$.  The lower lying curves are the results of the ideal simulation.  All the hydrodynamic evolutions use  $T_c=160$ MeV, $\tau_0=1.0$ fm/c, $T_{f.o.}=135$ MeV and a LH=0.3 GeV/fm$^3$. }
\end{figure}

In fig.~\ref{fig:TeffC} we show the effective temperature as a function of invariant mass.  We still use a LH of 0.3 GeV/fm$^3$ but modify the freezeout temperature such that the maximum temperature agrees with the data at the rho mass.  By comparing with the ideal curves we can see how viscosity modifies the $p_\perp$ slopes in the two hadronic production scenarios. It is seen that even a modest viscosity in the QGP phase brings about a large increase in the temperature.  Based on phenomenological cross sections \cite{Itakura:2007mx} the shear viscosity in the hadronic phase is much larger.  The increase in effective temperature is more modest since the gradients are smaller at later proper times.  It is important to note that for $M\geq 1.0$ GeV the hadronic viscosity does not modify the temperature since this region is dominated by q\={q} annihilation in our model.  At least in this model it appears that $\eta/s=0.08$ already saturates the experimental data.  Any larger and the resultant temperature would be above the data for $M\geq 1.0$ GeV.

Now we look at how the initial thermalization time effects the above result.  In order for the temperature to be consistent at the rho mass we have to increase the freezeout out temperature and LH of the model.  Since changes in the LH and freezeout temperature don't effect the qgp temperature too strongly this wont effect the discussion to follow.  The hydrodynamic evolution is now started at $\tau=0.5$ fm/c.  Focusing on $M\geq 1$ GeV the effective temperature of the ideal QGP phase is consistent with the data.  However, when including even the minimal viscosity of $\eta/s=0.08$ the temperature above M=1 GeV is no longer consistent with the data. 
\begin{figure}[hbtp]\label{fig:TeffTau}
\centering
\includegraphics[scale=.45]{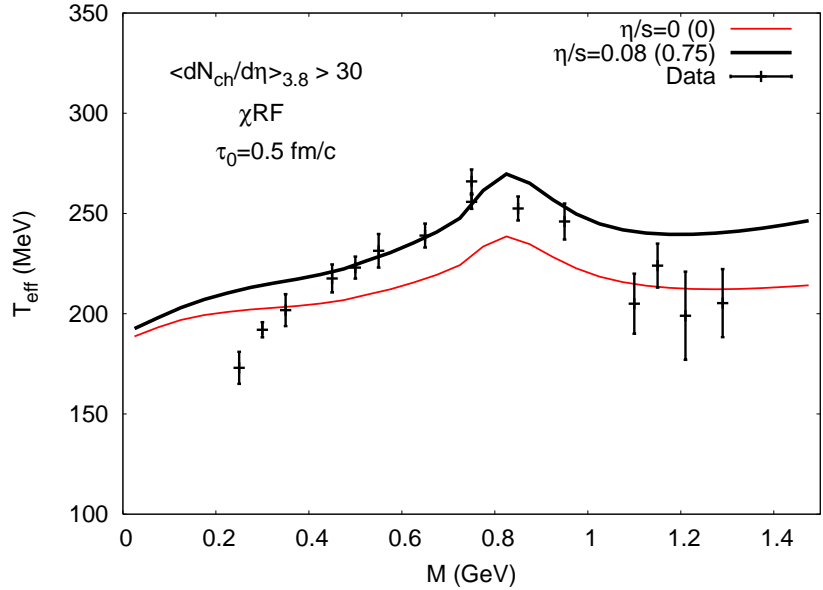}
\caption{Effective temperature as a function of mass when the hadronic phase is constrained by the chiral reduction formula.  In each label the number in parenthesis is the viscosity in the hadronic phase.  The upper curve is for $\eta/s_{qgp}=0.08$ and $\eta/s_{had}=0.75$.  The lower lying curve is the result of the ideal simulation.  The parameters of the hydrodynamic evolution are  $T_c=160$ MeV, $\tau_0=0.5$ fm/c, $T_{f.o.}=140$ MeV and a LH=0.5 GeV/fm$^3$. }
\end{figure}
If we assume that the QGP phase must have a minimum viscosity of $\eta/s=0.08$ then we can conclude that within our model a thermalization time of $\tau_0=0.5$ fm/c is inconsistent with the data while a thermalization time closer to $\tau_0=1$ fm/c is supported by the data. 

\begin{figure}[hbtp]\label{fig:massC}
\centering
\includegraphics[scale=.45]{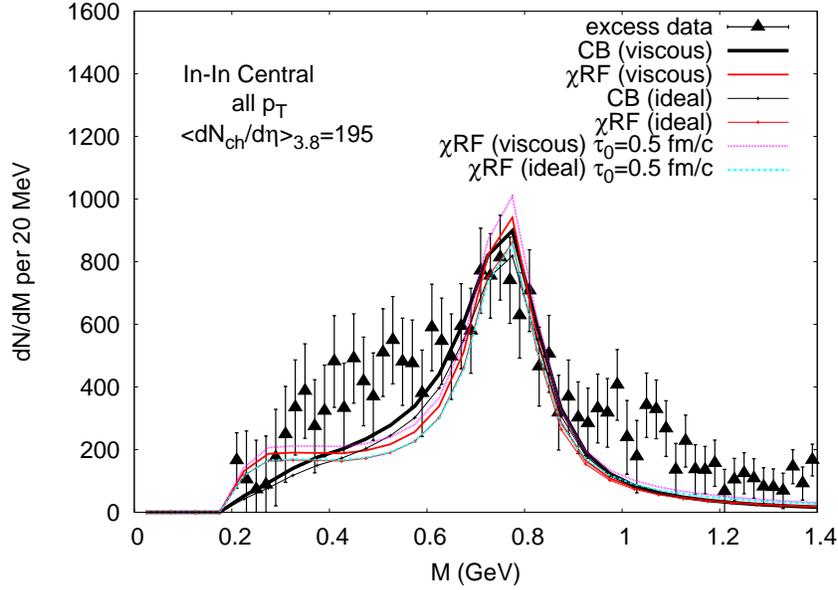}
\caption{Summary plot of the invariant mass spectra compared to the NA60 data.  The first four curves show the ideal and viscous ($\eta/s_{qgp}=0.08, \eta/s_{had}=0.75)$ using both the chiral reduction formula ($\chi$RF) and collisional broadening (CB) with the parameter set described in fig.~\ref{fig:TeffC}.  The last two curves show the viscous and ideal curves using the parameter set of fig.~\ref{fig:TeffTau}.} 
\end{figure}

Finally, it should be discussed how the viscosity and thermalization time change the mass spectra.  It was already shown in chapter 4 that the viscous correction to the distribution function leaves the mass spectra unchanged.  However, changes in the flow profiles could possible change the spectra.  Figure~\ref{fig:massC} shows a summary of the mass spectra comparing a few different evolution models used in the prior examples.  To first order the mass spectra remain unchanged. 

\def\gsim{\mbox{~{\protect\raisebox{0.4ex}{$>$}}\hspace{-1.1em}
        {\protect\raisebox{-0.6ex}{$\sim$}}~}}
\def\lsim{\mbox{~{\protect\raisebox{0.4ex}{$<$}}\hspace{-1.1em}
        {\protect\raisebox{-0.6ex}{$\sim$}}~}}
\chapter{Conclusions and Outlook}

Dilepton emission rates from a hadronic gas at finite temperature are constrained by broken chiral symmetry in a density expansion.  The rates are expressed in terms of measured $e^+e^-$ annihilation and $\tau$ decay data.  We focus on reactions having one pion and two pions in the final state.  This brings about a mixing of the vector and axial spectral densities and results in a large enhancement of the rates below the $2\pi$ threshold.  For the one pion state (corresponding to processes of the type $I\to F\pi + e^+e^-$) this enhancement dominates at intermediate momentum ($\vec{q}\gsim 0.5$ GeV) and low mass ($M \leq 2m_\pi$).  The two pion final state ($I\to F \pi\pi + e^+e^-$) dominates at low momentum ($\vec{q}\lsim 0.5$ GeV) and low mass.

This result has direct consequences for dilepton phenomenology.  Both NA60 and PHENIX see an increase in yield at low $p_\perp$ and low mass and we have shown that the $2\pi$ final state contribution significantly enhances the yields in this kinematic region.  Even though the enhancement can be thought of as arising due to Bremsstrahlung type processes it is a direct consequence of the way chiral symmetry is broken. 
 
The second part of this work simulates a viscous relativistic hydrodynamic model of heavy ion collisions.  We solve a second order theory that relaxes on short time scales to the Navier-Stokes stress tensor.  Then the viscous correction to hadronic $p_\perp$ spectra and differential elliptic flow is calculated taking into account off-equilibrium corrections to the distribution function.  In comparison to the ideal hydrodynamic results we find that shear viscosity decreases the differential elliptic flow and {\em hardens} the single particle spectra at larger transverse momentum.  Both of these results are welcomed by the data. 

Next, the shear viscous correction to dilepton production from q\={q} annihilation is computed.  Starting from kinetic theory the quark's distribution function is corrected with its off-equilibrium counterpart.  The resulting rates are integrated over the space-time evolution of the QGP phase using the above viscous hydrodynamic simulation.  While the changes to the invariant mass spectra are small the viscous correction to the transverse momentum spectra becomes larger at higher invariant mass and/or $p_\perp$.  

Given the success of ideal hydrodynamic models in their explanation of hadronic spectra a consistent description of electromagnetic observables should use the same space-time evolution.  Therefore, in the last part of this thesis the dilepton rates are integrated over the space-time evolution of the hydrodynamic evolution model.  This is a large improvement over other models that employ parameterizations or phenomenologically motivated blast-wave type fits for the evolution.  The results are then compared to the measured dilepton spectra at NA60 and PHENIX.  

At NA60 the chiral reduction approach yields fairly good agreement with the invariant mass data.  The agreement is better for more peripheral collisions with some discrepancy in the more central bins suggesting that the system might not be dilute enough for the density expansion.  Part of the suppression at the $\rho$ pole and the enhancement at lower mass which is seen in the data is a direct consequence of broken chiral symmetry.

In regards to NA60's $T_{eff}$ data, we have shown that the measured effective temperature can be obtained by either decreasing the latent heat, including viscosity in the QGP/Hadronic phases or some combination there of.  If the mass region above 1 GeV is indeed dominated by QGP emission we have shown how the thermalization time and $\eta/s$ of the QGP can be extracted from the experimental $p_\perp$ data. 

Next we investigated the mass spectrum measured by PHENIX.  Using the same comprehensive emission rates that worked on the percent level for NA60 the low mass enhancement seen at RHIC cannot be reconciled.  This discrepancy can be traced back to the PHENIX acceptance which rejects all di-electron pairs having $m_\perp \lsim 0.4$ GeV.   

The future prospect for this work is to extract precision data, such as the EOS or transport coefficients, from the experimental heavy ion data.  By fitting a combination of hadronic observables ($p_\perp$ spectra, elliptic flow) and electromagnetic observables (dilepton $T_{eff}$ and mass spectra) within the same viscous hydrodynamic framework one could potentially extract quantities such as the thermalization time, latent heat, and shear viscosity to entropy ratio. 

%%%%%%%%%%%%%%%%%%%%%%%%%%%%%%%%%%%%%%%%%%%%%%%%%%%%%%%%%%%%%%%%%%%%%%%%%%%%%%%%

%%%%%%%%%%%%%%%%%%%%%%%%%%%%%%%%%%%%%%%%%%%%%%%%%%%%%%%%%%%%%%%%%%%%%%%%%%%%%%%%

%\bibliographystyle{unsrtnat}
\renewcommand{\baselinestretch}{1}
\normalsize
\clearpage
\end{document}